# DEVELOPMENT OF IMPEDANCE AND CAPACITANCE BASED SENSORS FOR THE ESTIMATION OF ADULTERANT INGREDIENTS IN DIFFERENT BIO-CONSUMABLES

Thesis submitted for the degree of

Doctor of Philosophy (Sc.)

In

# Electronic Science

by

# Chirantan Das

Department of Electronic Science

University of Calcutta

Kolkata, India

2020

*To*

***Ma & Baba..***

# Abstract

Electrical Impedance Spectroscopy (EIS) technique is found to be an excellent candidate for bio-sensing and food quality monitoring applications due to its rapid, robust, cost-effective and point-of-care approach. The present research work investigates the implementation of EIS technique supported by several optical spectroscopic techniques such as Ultraviolet-Visible (UV-Vis) and Fourier Transform Mid Infrared (FT-MIR) to detect and quantify several toxic adulterants in foods and bio-consumables. A comprehensive understanding on the background theory related to the study has been developed to analyze the overall polarization of the system and the effect of frequency on complex permittivity of such system have been observed. In the current work, the technique is applied to adulterated saccharides, honey and turmeric samples through a prototype sensing device. All the corresponding measurements have been performed by dipping a custom-made parallel plate conductivity cell with unity cell constant inside the solution under test. EIS study exhibited a steady variation of electrical parameters such as impedance, capacitance, conductance and current values with increasing adulterant percentage in the solution. Variation in such properties due to adulteration provides a systematic sensor plot through which one can determine their percentage of adulteration in unknown adulterated samples. Co-efficient of sensitivity has been extracted from the EIS data for adulteration study in terms of one of the measured parameters. Also, the results of UV-Vis and FT-MIR studies have been used for comparative analyses which corroborate with the EIS results, wherever applicable. In spite of several achievements, the prototype sensing using conductivity cell has its limitations. Hence, effort has been made on the design and fabrication of sensor chips for detection and quantification of a variety of adulterants in milk samples. Several characterizations have been performed on the device, including spectroscopic ellipsometry measurements, to obtain different teflon layer thicknesses for hydrophobic layer optimization and lowering of contact angle hysteresis. Contact angle measurements have been performed to check the surface wettability of the liquid sample for different hydrophobic layer thicknesses. Categorical determination of polar, non-polar and ionic adulterants in milk employing EIS technique has been achieved through a phase angle-impedance mapping based sensor plot which shows distinct range of psi ($\Psi$) and delta ($\Delta$) values for all the adulterants considered. Thus, the overall study suggests the possibility of developing a rapid and precise method for the detection and quantification of different adulterants in foods and bio-consumables.

# Acknowledgement

I am glad to acknowledge, first and foremost, both my supervisors, Dr. Sanatan Chattopadhyay and Dr. Anupam Karmakar for their tremendous effort to lead the entire research project and their continual support at each and every minute step of this research. I also like to acknowledge them for their fruitful discussions with me to develop the comprehensive research ideas. And, without any hesitation, I like to express my gratitude to both of them for always standing by my side.

I would like to acknowledge the Department of Electronic Science, University of Calcutta, along with all its faculties and staffs for providing me the platform for pursuing the research work and also for allowing me to utilize its infrastructure.

I would like to acknowledge the Centre for Research in Nanoscience and Nanotechnology (CRNN), University of Calcutta, for providing all the necessary facilities and support for performing the entire research work.

I also acknowledge the Department of Polymer Science and Technology, University of Calcutta, for allowing me to utilize its infrastructure.

It's a pleasure for me to acknowledge the Centre of Excellence (COE) for the Systems Biology and Biomedical Engineering, University of Calcutta, funded by the World Bank through TEQIP Phase II, for providing me the fellowship to pursue my research and developing significant research infrastructure.

I would like to acknowledge the Sensor and Systems Development Group, University of Calcutta, funded by the UGC through UPE Phase II, for providing me the fellowship to pursue my research and developing significant research infrastructure.

I gladly express my gratitude to Prof. Krishnendu Acharya, Prof. Sudip Kundu, Dr. Chirasree Roychaudhuri and Dr. Dipankar Chattopadhyay for fruitful discussions with them on several research issues.

I am whole-heartedly grateful to all of my research colleagues, namely Subhadip Chakraborty, Rajib Saha, Subhrajit Sikdar, Dr. Basudev Nag Chowdhury, Dr. Avishek Das, Nirmal Kumar Bera, Anuraag Mukherjee, Subrata Mandal, Sandip Bhattacharya, Dr. Mainak Palit, Dr. Anindita Das, Somdatta Paul, Jenifar Sultana, Anannya Bhattacharya, Alivia Basak, Debopam Bhattacharya and Mr. Ashis Bhattacharya for their continuous support in the research project.

Also, mere thanks is not enough to describe the brotherhood we (Me, Subhadip, Rajib & Subhrajit) have shared for the last five years and I hope it will be everlasting.

Last but not the least, I express my gratitude to my mother Smt. Nilima Das, my father Sri. Dilip Kumar Das, my brother Chiranjib Das, my sister-in-law Sumita Das, my nephew Tintin and all other relatives for their unconditional support and sympathy to my involvement into the research work during the entire project.

# Contents

# Chapter 1

## Introduction: Emergence of different food adulterant sensors and its challenges            (1-10)

# Chapter 2

# Literature Review: Chronological development of Food adulterant sensors      (11-24)

# Chapter 3

# Impact of carbohydrate adulteration: Detection and quantification      (25-39)

# Chapter 4

# Applications of EIS for assessing the quality of honey (40-55)

# Chapter 5

# Estimation of metanil yellow in turmeric by capacitive measurements      

# Chapter 6

## Design and fabrication of on-wafer capacitive sensors for detecting different adulterants in milk by employing EIS technique      (72-101)

# Chapter 7

# Conclusions and Future scopes    (102-106)

# References    (107-136)

# List of Figures

Fig. 6.1: Flowchart to illustrate the systematic cleaning process of Si wafers before device fabrication.

Fig. 6.2: Flowchart illustrating the teflon coating process.

Fig. 6.3: (a) - (f) Plots showing the contact angle of milk sample for different coating spin speeds, (g) experimentally measured (symbols) and model fit (lines) spectroscopic ellipsometric parameters, $\Psi$ and $\Delta$, for the teflon film (4000 rpm spin speed) on $SiO_2$/Si device in the range of 200–800 nm wavelength at 60° angle of incidence and, (h) figure showing the variation of teflon film thickness and contact angle for different spin speeds.

Fig. 6.4: Process flow of sensor device fabrication and image of a real-time fabricated device with liquid sample placed on it.

Fig. 6.5: Plots of (a) impedance, (b) capacitance, (c) conductance and (d) variation of current at +5 V for different percentages of tap water content in milk samples collected from five different sources.

Fig. 6.6: Plots of (a) capacitance, (b) impedance, (c) conductance and (d) variation of current at +5 V for different percentages of detergent content in five milk samples.

Fig. 6.7: FT-MIR spectra of pure and adulterated milk with circled area indicating the generation of new peaks due to adulteration.

Fig. 6.8: Plots of capacitance, conductance and impedance of milk samples adulterated with polar adulterants - (a) Urea, (b) Melamine, (c) Starch, (d) Allantoin, (e) Cyanuric Acid and non-polar/ionic adulterants - (f) Benzoic Acid, (g) Sodium Bicarbonate and (h) Ammonium Sulphate.

Fig. 6.9: Molecular structures of polar adulterants- (a) Urea, (b) Melamine, (c) Starch, (d) Allantoin and (e) Cyanuric acid showing individual bond dipoles and direction of their overall dipole moments. Multi-coloured spheres represent distinct elements - **Red:** Oxygen; **Blue:** Nitrogen; **Grey:** Carbon; **White:** Hydrogen.

Fig. 6.10: Molecular structures of ionic and non-polar adulterants- (a) Sodium bicarbonate, (b) Ammonium sulphate and (c) Benzoic acid showing individual bond dipoles and direction of their overall dipole moments and schematic of (d) solvent-shared ion association and (e) fully solvated ion association resulting from ion dissociation in milk adulterated with ionic adulterants. Multi-coloured spheres represent distinct elements - **Red:** Oxygen; **Blue:** Nitrogen; **Grey:** Carbon; **White:** Hydrogen; **Purple:** Sodium; **Yellow:** Sulphur.

Fig. 6.11: Plot of the phase angle-impedance mapping for precise detection of eight adulterants in milk sample.

# List of Tables

# List of Symbols

## (Unless mentioned otherwise)

## Universal Constants

k            : Boltzmann constant

$\varepsilon_0$          : Permittivity of free space/vacuum

## Geometrical/Dimensional Parameters

d            : Distance between the electrodes

A            : Electrode area

## Material Parameters

$\varepsilon$           : Permittivity of the material

$\varepsilon^*$          : Frequency dependent complex permittivity of the system

$\boldsymbol{p}$          : Dipole moment

$\vec{P}$          : Polarization

n            : Number of dipoles per unit volume

$p_i$          : Ionic dipole moments

$p_d$          : Sucrose dipole moments

$n_{i1}$ and $n_{i2}$     : Two consecutive numbers of ionic dipoles per unit volume

$n_w$          : Total water dipoles per unit volume

$n_i$          : Quantity of ionic dipoles per unit volume

$n_d$          : Quantity of sucrose dipoles per unit volume

## Physical Parameters

$E$            : External field

$W$           : Potential energy

T             : Absolute temperature

$\vec{D}$             : Electric displacement vector

$\psi$             : Phase angle

$\Delta$             : Phase difference

$\Psi$             : Amplitude component

$R^2$            : Co-efficient of determination

$\beta$             : Co-efficient of sensitivity

S             : Standard error of regression

n             : Refractive index

k             : Extinction coefficient

$I$             : Intensity of light

$I_o$            : Initial intensity of the incident light

$T$             : Transmittance

a.u.           : Arbitrary unit

## Electrical Parameters

Z             : Impedance

C             : Capacitance

R             : Resistance

G             : Conductance

$C_{sol}$          : Solution capacitance

$C_{dl}$           : Double-layer capacitance

| | |
|---|---|
| $R_{sol}$ | : Solution resistance |
| $C_{stray}$ | : Parasitic capacitance |
| $L_{stray}$ | : Parasitic inductance |
| X | : Reactance |
| I-V | : Current-voltage |

## General Abbreviations

| | |
|---|---|
| EIS | : Electrical Impedance Spectroscopy |
| UV-Vis | : Ultraviolet-Visible |
| FT-MIR | : Fourier Transform Mid Infrared |
| FSSAI | : Food Safety and Standards Authority of India |
| USFDA | : United States Food and Drug Administration |
| EFSA | : European Food Safety Authority |
| NIR | : Near-Infrared |
| IR | : Infrared |
| GC | : Gas Chromatography |
| NMR | : Nuclear Magnetic Resonance |
| KBr | : Potassium Bromide |
| LOC | : Lab-on-a-Chip |
| DI | : De-Ionized |
| SEM | : Scanning electron microscopy |
| HPLC | : High-Performance Liquid Chromatography |
| LC | : Liquid Chromatography |
| MS | : Mass spectrometry |
| SPME-GC-MS | : Solid phase micro-extraction-gas chromatography-mass spectroscopy |
| LC-MS | : Liquid chromatography-mass spectrometry |
| ELISA | : Enzyme-linked immunosorbent assay |
| DNA | : Deoxyribonucleic acid |
| PCR | : Polymerase chain reaction |
| RAPD | : Randomly Amplified Polymorphic DNA |
| LAMP | : Loop mediated isothermal amplification |
| SCAR | : Sequence Characterized Amplified Region |
| CE-MS | : Capillary electrophoresis–mass spectrometry |
| PNA | : Peptide nucleic acid |
| EEC | : Electrical equivalent circuit |
| PDMS | : Polydimethylsiloxane |
| DMF | : Digital microfluidic |
| MEMS | : Micro-electromechanical system |
| $CaCO_3$ | : Chalk powder |
| $Na_2CO_3$ | : Sodium carbonate |
| IC | : Integrated circuit |
| SOC | : System-on-chip |
| w | : Adulterant concentration |
| IS | : Indian Standards |
| SS | : Sucrose Syrup |
| CS | : Corn Syrup |
| HFCS | : High Fructose Corn Syrup |

MS                : Maltose Syrup

KBr               : Potassium Bromide

FWHM              : Full width at half maximum

MY                : Metanil Yellow

FAO               : Food and Agriculture Organization of the United Nations

WHO               : World Health Organization

LIBS              : Laser Induced Breakdown Spectroscopy

VLSI              : Very large scale integration

$SiO_2$           : Silicon dioxide

$Al_2O_3$         : Alumina

Al                : Aluminum

Cu                : Copper

Cr                : Chromium

Pt                : Platinum

W                 : Tungsten

Au                : gold

# List of publications

## A. Journal publications

## Journals related to thesis:

1. <u>Chirantan Das</u>, Subhadip Chakraborty, Nirmal Kumar Bera, Dipankar Chattopadhyay, Anupam Karmakar, Sanatan Chattopadhyay. Quantitative estimation of soda ash as an adulterant in aqueous sucrose solution by employing electrical impedance and capacitance spectroscopy, *Measurement*, *148*, 106937 (2019).

2. <u>Chirantan Das</u>, Subhadip Chakraborty, Krishnendu Acharya, Nirmal Kumar Bera, Dipankar Chattopadhyay, Anupam Karmakar and Sanatan Chattopadhyay. FT-MIR supported Electrical Impedance Spectroscopy based study of sugar adulterated honeys from different floral origin, *Talanta*, 171, 327-334 (2017).

3. <u>Chirantan Das</u>, Subhadip Chakraborty, Krishnendu Acharya, Nirmal Kumar Bera, Dipankar Chattopadhyay, Anupam Karmakar and Sanatan Chattopadhyay. Impedimetric Approach for Estimating the Presence of Metanil Yellow in Turmeric Powder from Tunable Capacitance Measurement, *Food Analytical Methods*, 1-11 (2019).

4. <u>Chirantan Das</u>, Subhadip Chakraborty, Basudev Nag Chowdhury, Subhrajit Sikdar, Rajib Saha, Anuraag Mukherjee, Anupam Karmakar and Sanatan Chattopadhyay. A diagrammatic approach of impedimetric phase angle-modulus sensing for identification and quantification of various polar and non-polar/ionic adulterants in milk. (**Communicated**)

5. Subhadip Chakraborty, <u>Chirantan Das</u>, Rajib Saha, Avishek Das, Nirmal Kumar Bera, Dipankar Chattopadhyay, Anupam Karmakar, Dhrubajyoti Chattopadhyay, Sanatan Chattopadhyay. Investigating the quasi-oscillatory behaviour of electrical parameters with the concentration of D-glucose in its aqueous solution at room temperature by employing impedance spectroscopy technique. *Journal of Electrical Bioimpedance*, 6(1):10. (2015).

6. Subhadip Chakraborty, <u>Chirantan Das</u>, Nirmal Kumar Bera, Dipankar Chattopadhyay, Anupam Karmakar, and Sanatan Chattopadhyay. "Analytical modelling of electrical impedance based adulterant sensor for aqueous sucrose solutions." *Journal of Electroanalytical Chemistry*. 784: 133-139. (2017).

## Journals not related to thesis:

1. Kakali Ghoshal, Subhadip Chakraborty, <u>Chirantan Das</u>, Sanatan Chattopadhyay, Subhankar Chowdhury, Maitree Bhattacharyya. Dielectric properties of plasma membrane: A signature for dyslipidemia in diabetes mellitus. *Archives of biochemistry and biophysics*, 635, 27-36 (2017).

2. Tapas K. Ghosh, Sourav Sadhukhan, Dipak Rana, Gunjan Sarkar, <u>Chirantan Das</u>, Sanatan Chattopadhyay, Dipankar Chattopadhyay and Mukut Chakraborty. Treatment of recycled cigarette butts (man-made pollutants) to prepare electrically conducting material. *Journal of the Indian Chemical Society*, vol. 94, pp: 863-870. (2017).

3. Subhadip Chakraborty, <u>Chirantan Das</u>, Kakali Ghoshal, Maitree Bhattacharyya, Anupam Karmakar and Sanatan Chattopadhyay. Low Frequency Impedimetric Cell Counting: Analytical Modeling and Measurements, *IRBM*. (2019).

4. Subhadip Chakraborty, Sreyasi Das, <u>Chirantan Das</u>, Kaushik Das Sharma, Anupam Karmakar and Sanatan Chattopadhyay. On-chip Estimation of Hematocrit Level For Diagnosing Anemic Conditions by Impedimetric Techniques, *Biomedical Microdevices*, 22, 1-11.

## 5. B. Conference publications

## Conferences related to thesis:

1. <u>Chirantan Das</u>, Subhadip Chakraborty, Anupam Karmakar, Sanatan Chattopadhyay. On-chip Detection and Quantification of Soap as an Adulterant in Milk Employing Electrical Impedance Spectroscopy, *International Symposium on Devices, Circuits and Systems (ISDCS)*, IEEE, 2018.

2. <u>Chirantan Das</u>, Subhadip Chakraborty, Anupam Karmakar, Sanatan Chattopadhyay. Comparative study for the impedimetric detection and quantification of adulterants in different bio-consumables, *3rd International Symposium on Devices, Circuits and Systems (ISDCS)*, IEEE, 2020.

3. Subhadip Chakraborty, <u>Chirantan Das</u>, Anupam Karmakar, Sanatan Chattopadhyay. Analyzing The Quasi-oscillatory Nature Of Electrical Parameters With The Concentration Of Sucrose In Aqueous Solution At Room Temperature. ***Advance Materials Proceedings***, 2016.

## Conferences not related to thesis:

1. Avishek Das, <u>Chirantan Das</u>, Rajib Saha, Anupam Karmakar, Mainak Palit, Himadri Sekhar Dutta, Sanatan Chattopadhyay, Electrical Characterization of n-ZnO Nanowire/p-Si Hetero-junction Diode in Presence of Traps, 978-1-4673-9513-7/15, ***CODEC*** (2015).

2. Subhadip Chakraborty, <u>Chirantan Das</u>, Rajib Saha, Sreyasi Das, Raghwendra Mishra, Roshnara Mishra, Anupam Karmakar, Sanatan Chattopadhyay. Bio-dielectric Variation as a Signature of Shape Alteration and Lysis of Human Erythrocyte: An On-chip Analysis, ***International Symposium on Devices, Circuits and Systems (ISDCS)***, IEEE, 2018.

3. Anuraag Mukherjee, Subhadip Chakraborty, <u>Chirantan Das</u>, Anupam Karmakar and Sanatan Chattopadhyay. Study of Optical and Electrical Characteristics of chemically extracted Lotus and Taro Bio-Wax for Hydrophobic Surface Engineering. ***International Conference on Opto-Electronics and Applied Optics (Optronix)*** **IEEE**. (pp. 1-4) (2019).

4. Anuraag Mukherjee, Subhadip Chakraborty, <u>Chirantan Das</u>, Rajib Saha, Subrata Mandal, Anupam Karmakar and Sanatan Chattopadhyay. Fabrication and characterization of hydrophobic bi-layer high-k dielectric films for digital microfluidic applications. ***International Workshop on Physics of Semiconductor Device (IWPSD)***. (2019).

5. Alivia Basak, Subhadip Chakraborty, <u>Chirantan Das</u>, Anuraag Mukherjee, Rajib Saha, Anupam Karmakar, Sanatan Chattopadhyay Electrically isolated buried electrode biosensor for detecting folic acid concentration. ***3$^{rd}$ International Symposium on Devices, Circuits and Systems (ISDCS)***, IEEE, 2020.

6. Sandip Bhattacharya, Rajib Saha, Subhrajit Sikdar, Subrata Mandal, <u>Chirantan Das</u>, Sanatan Chattopadhyay. Investigation of density and alignment of ZnOnanowires grown by double-step chemical bath deposition (CBD/CBD) technique on metallic, insulating and semiconducting substrates. ***3$^{rd}$ International Symposium on Devices, Circuits and Systems (ISDCS)***, IEEE, 2020.

7. Subhadip Chakraborty, <u>Chirantan Das</u>, Rajib Saha, Anupam Karmakar, Sanatan Chattopadhyay, Arindam Chatterjee, Madhusudan Das. Dielectric study of kidney stones by fabricating an MIS structure: Material analysis and challenges. International Seminar cum Research Colloquium on MEMS based Sensors and Smart Nanostructured Devices (***MSSND 2019***).

## C. Book Chapter

1. Sanatan Chattopadhyay, Subhadip Chakraborty, <u>Chirantan Das</u>, Rajib Saha: Recent progresses on micro- and nano-scale electronic biosensors: A review. ***Nanospectrum: A Current Scenario***, Edited by S. Chakrabarti, P. Mukherjee, G. Khan, A. Adhikary, P. Patra, J. Bal, 2015, 19-40; Allied Publishers Pvt. Ltd, ISBN: 978-93-85926-06-8.

# Chapter 1

# Introduction: Emergence of different food adulterant sensors and its challenges

## 1.1 Overview

This chapter outlines the reasons for conducting this research, research problem raised from the existing knowledge and the organization of this thesis.

Section 1.2 describes the background and motivation of this investigation.

Section 1.3 describes the types of adulteration and the different foods which are prone to such adulteration.

Section 1.4 illustrates the objective to pursue this research.

Section 1.5 describes the different challenges that a society faces due to such types of adulterations.

Section 1.6 summarizes how to overcome such challenges by enlisting a variety of adulterant detection techniques in diversified domains.

Section 1.7 illustrates different characterization tools used in this current work.

Section 1.8 presents the orientation and organization of the thesis.

## 1.2 Background and Motivation

Food safety monitoring is considered to be an important aspect for dealing with threats to public health and well-being of the mass population. Globally, this has emerged to be one of the most significant public concerns which is directly related to the alteration of food habits, growth of mass supplying establishments and acute globalization of food supply chains [**1,2**]. Due to such globalization effects, the need to strengthen food quality monitoring systems in almost all countries is becoming more evident [**1,2**]. Moreover, food quality control is crucial for consumer safety as well as for the food industry, and therefore, proper monitoring of the food contaminants and quantification of food constituents is becoming more vital for food processing industries and for the consumers [**3**]. Regulatory agencies like Food Safety and Standards Authority of India (FSSAI), United States Food and Drug Administration (USFDA) and the European Food Safety Authority (EFSA) has laid out some strict guidelines which ascertains the maximum levels for certain contaminants in foodstuffs to maintain a decent level of human health and consumer protection [**4-6**]. Nowadays, analysis of food products is performed at the end of the production process using traditional techniques such as chromatography, mass spectrometry, ultraviolet detection, or fluorescence techniques either individually or in combination with other separation techniques [**7-10**]. These conventional approaches have certain limitations such as these are time consuming, laborious and expensive. These processing also require large sample volumes and highly trained personnel [**11,12**].

Also, traditional quality determination methods including sorting and grading by visual screening, measurement of firmness, sensory evaluation and adulterant content measurement have been implemented by several food processing industries. On the other hand, efforts have been made to explore rapid, robust and point-of-care approaches to assess food quality. In recent times, EIS measurements have gained immense popularity for detecting adulterants in food products [**13-21**]. Also, this technique has been applied to several biological systems [**20,22-28**[. With the sustained development of devices and analytical methods in this domain, several novel measurements have been developed which can be implemented for rapid, low-cost, precise and point-of-care diagnosis of adulterated foods.

As of today, the applications of EIS technique on food quality monitoring have not yet been studied up to the required level and the current dissertation demonstrates EIS measurements on a variety of foods and their corresponding adulterants and proposes new ideas for their detection and quantification.

## 1.3 Types of adulteration

Adulteration in foods and beverages can be broadly classified into two main types: intentional or deliberate adulteration and unintentional or unknown adulteration. Generally, the food products face adulteration in diverse circumstances when there is high demand from the consumer's point of view and also when their supply is low. Therefore, it is important to identify the nature of adulteration.

### 1.3.1. Intentional/deliberate adulteration

Intentional or deliberate adulteration is that kind of adulteration where dishonest traders or vendors purposefully adulterate different food products in order to promote the level of their essential nutrients after reducing the necessary amount in order to increase their profit margin. In general, chemicals such as urea, melamine, ergot, starch, flour, roasted barley powder, vegetable oils, water, skim milk, sand, cane sugar, molasses, stone, brick powder, chalk powder, chicory grinded papaya seeds, are used as adulterants in marketed foodstuffs [**29**]. Compared to unknown adulteration, this type of adulteration is most detrimental since the entire process of mixing and reducing is being controlled by the traders without having any scientific background or any idea of consequent side effects.

### 1.3.2 Incidental/unknown adulteration

Incidental or unknown adulteration demonstrates the lack of proper hygienic conditions of food products and bio-consumables right from the start of production until when they are consumed. Possible places of adulteration involve food production, handling, processing, storing, transporting and when these are being sold.

## 1.4 Objective of the research

The objective of this research is to investigate the electrical impedance, capacitance, resistance and other electrical properties of different pure and adulterated food samples and to propose alternative approaches for their measurement. Primarily, a better understanding of the electrical properties of food samples in their aqueous solutions has to be developed.

Secondly, this research will briefly present the theoretical background of the dielectric properties of the pure and adulterated samples considered in this work. This study will also focus on the frequency dependence of the dielectric medium of the pure and adulterated samples.

Thirdly, this research will establish a rapid response method for food adulteration monitoring by using electrical impedance spectroscopic analyses in conventional conductivity cell prototypes as well as in the miniaturized on-wafer devices.

## 1.5 Challenges

Adulteration is an act of adding extraneous substances (adulterants) into the food items or reducing their necessary nutrients for monetary profit. The carelessness and lack of appropriate hygienic conditions during processing, storing, packaging, transportation and marketing of food stuffs can also be categorized as adulteration. This ultimately results in the fact that consumer becomes the victim of different unexpected fatal diseases. Therefore, it is important for the consumer to have an idea about the common adulterants used and their ill effects on health and hygiene which may prevent a long term catastrophe.

As a result, study of the factors affecting safety of food products have emerged as a thrust area within the food supply chain and have attracted a lot of attention of researchers from various fields, sectors, governments and regulatory bodies [**4-6**]. Foods are often subjected to adulteration (intentional or unintentional) while supply chains usually deal with untreated foodstuffs that could be toxic and harmful to consumers if they are not processed properly [**1-3,6**]. Different vulnerabilities of food supply chains, that explains the food safety and security issues imposed on foods and beverages are generally a direct result of these supply chains being highly interconnected.

## 1.6 Overcoming the challenges

From the consumer's point of view, food industries should provide good quality foods and other bio-consumables [**30**]. In this context, efforts have been made to enhance their quality using genetic engineering. Also, researchers have been using quality evaluation techniques for several decades which started from a single parameter measurement such as weight, size and firmness [**31**]. More complex biochemical and biophysical analysis were applied after this to observe internal quality changes, such as titratable acidity measurement, nutritional analysis and enzyme degrade analysis. With the progress of information technology, quality assessment of foods and beverages has also evolved such as computer aided imaging techniques, which can identify precise internal quality deterioration and even display the deterioration location. Human visual evaluation has been the primary empirical method for the grading and sorting of pure foods from the adulterated ones. Another option is to use Near-Infrared (NIR) spectroscopy to analyze colour variations [**32**]. Firmness is also a very good indicator which has been used as a measure of maturity and ripeness of fruits [**33**]. Over the past two decades, resonant frequencies, acoustic response, and ultrasonic approaches have been employed to investigate the firmness of food materials [**34-36**]. Another detection parameter, termed sweetness, is normally referred to as the total soluble solid content and is measured in °Brix [**37**]. Additionally, infrared (IR) methods also have the capabilities to determine total sugar content directly from fruits [**37-39**]. Acidity parameter is generally determined by titration approach. Titratable acidity measures total number of hydrogen ions inside a solution. In chromatographic techniques, Gas Chromatography (GC) is considered to be one of the most popular methods currently used to conduct measurement of sample odour properties [**40-42**]. The concentration of ethylene can be detected using this method. Apart from the above mentioned approaches, there are several other new techniques which have emerged as suitable candidates for the applications of food quality assessment. These include nuclear magnetic resonance (NMR) spectroscopy which can be used to determine the internal bruising and ripeness of fruit and vegetables [**43,44**]. Alternatively, optical spectroscopy techniques such as FT-IR, NIR, UV-Vis and electrical spectroscopy techniques such as EIS, electrochemical impedance spectroscopy and electrical impedance tomography techniques can be employed to detect toxic and carcinogenic adulterants in foods [**13-21,45-53**].

## 1.7 Characterization tools

### 1.7.1 FTIR Spectrophotometer

Fourier Transform Infrared Spectroscopy, also known as FTIR Analysis or FTIR Spectroscopy, is an analytical technique used to identify organic, polymeric, and in some cases, inorganic materials. The FTIR analysis method uses infrared light to scan test samples and observe chemical properties. FTIR relies on the fact that most of the molecules absorb light in the infra-red region of electromagnetic spectrum. This absorption corresponds specifically to the bonds present in that particular molecule. The optical measurements in this research work were performed by using Perkin Elmer, Spectrum Two spectrometer at room temperature by performing a total of sixty-four scans recorded for each spectrum in the spectral region ranging between 4000 cm$^{-1}$ and 400 cm$^{-1}$ with a nominal resolution of 4 cm$^{-1}$. The background emission spectrum of the IR source is first recorded, followed by the emission spectrum of the IR source with the sample in place. The ratio of the sample spectrum to the background spectrum is directly related to the sample's absorption spectrum. For the measurement scheme, 100 mg of IR grade Potassium Bromide (KBr) pellets were prepared by using a pestle and mortar, and the KBr matrix was cold-pressed into a transparent disk. Pure KBr pellet was used to obtain the reference background spectrum. The resultant absorption spectrum from the natural vibration frequencies of a bond indicates the presence of various chemical bonds and functional groups present in the sample. FTIR is particularly useful for identification of organic molecular groups and compounds due to the range of functional groups, side chains and cross-links involved, all of which will have characteristic vibrational frequencies in the infra-red range. **Fig. 1.1** shows the FTIR spectrometer (Perkin Elmer, Spectrum Two) used in our study to perform all necessary measurements.

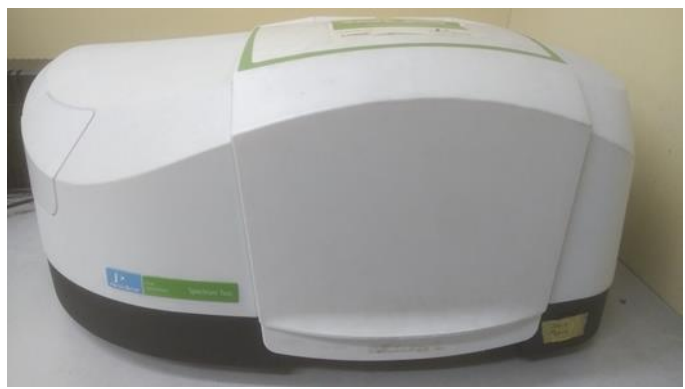

**Fig. 1.1**: Picture of the FTIR spectrometer used for the current study.

### 1.7.2 UV-Vis Spectrophotometer

The UV-VIS-NIR spectrometer system (Perkin Elmer Lambda 1050) measures the intensity of light ($I$) passing through a sample of thickness $d$ and compares it with the initial intensity of the incident light ($Io$). The ratio of $I/ Io$ is called the transmittance '$T$' and is generally expressed in arbitrary unit (a.u.). The UV-VIS-NIR spectrophotometer system has been employed for investigating the nature of variation in absorbance intensity for pure and adulterated food samples in the current research work.

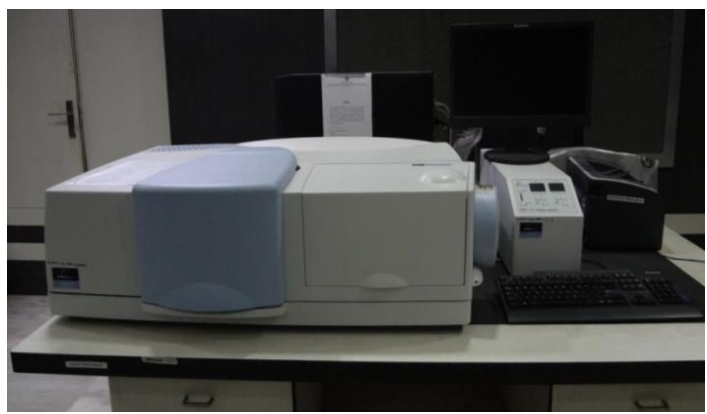

**Fig. 1.2**: Picture of the UV-Vis/NIR spectrophotometer used for the current study.

The light sources used inside the spectrophotometer includes a tungsten (W) lamp of wavelength range 300 nm – 2500 nm, a Deuterium arc lamp for UV region (190 nm – 400 nm) or a Xenon arc lamp for the entire UV to near infra-red (NIR) region of 160 nm to 2000 nm. A monochromator is employed to fix the wavelength of light and is synchronized with the readout circuit or computer. The light passed through the sample is collected by a detector circuit and is fed to the computer. Each time the wavelength of the light is varied and the measured absorbance A is recorded with wavelength. **Fig. 1.2** shows the UV-Vis/NIR spectrophotometer (Perkin Elmer LAMBDA 1050) employing which all relevant measurements have been performed.

### 1.7.3 Probe Station connected with the LCR Meter and Source Measuring Unit (SMU)

All electrical characterizations of the pure and adulterated samples considered in this research work have been performed by using ECOPIA Thermal Control probe station (Model No. ETCP-2000) connected with the TEGAM LCR meter and KEITHLEY Source Measuring Unit (SMU 2611B). Hence, both D.C. and A.C. measurements can be performed using it. The measurements were controlled by using LAB-VIEW software for rapid turnout of the experimental data. Also, EIS measurements for all the samples has been performed within the frequency range of 100 Hz to 5 MHz and the voltage amplitude is maintained to be 100 mV peak-to-peak, respectively. Both the LCR and SMU used in this work possess internal noise-reduction techniques and have been standardized with open-short compensation technique before the measurements.

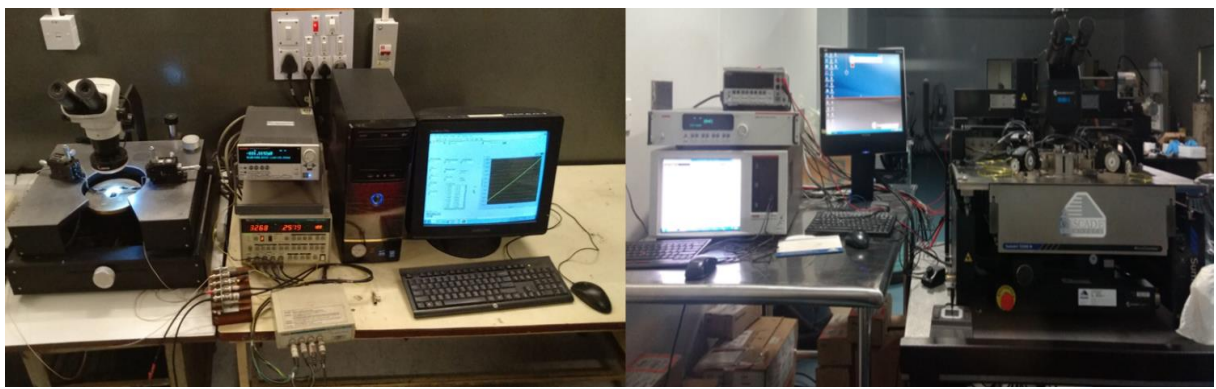

**Fig. 1.3**: Picture of two probe stations connected with LCR and Source Measuring Unit (SMU) for performing electrical measurements.

The fabricated device is placed inside a probe station which has four micro-manipulators. Each manipulator has gold (Au) coated tungsten (W) microprobe of tip-diameter ~ 20 μm. Specially, the junction capacitance-voltage measurement can be performed by varying the A.C. frequency from 1 kHz to 10 MHz. In addition, many other features such as series and parallel mode measurements, capacitance-time, capacitance-frequency, and impedance spectroscopy measurements make such a system a versatile tool for characterizing bio-sensing devices. **Fig. 1.3** shows two probe stations (ECOPIA ETCP-2000 and Semiconductor Characterization System SCS 4200) connected with LCR meter and SMU which have been used in this work for electrical measurements.

### 1.7.4 Spectroscopic Ellipsometer

A spectroscopic ellipsometer allows the measurement of refractive index and thickness of thin films. The instrument focuses on the reflection at a dielectric interface which depends on the polarization of incident light while the transmitted light through a transparent layer alters phase of the incoming wave, depending upon the refractive index of the material. An ellipsometer can be used to measure film thicknesses in the range of ~1 nm up to several micrometer. **Fig. 1.4** shows the spectroscopic ellipsometer (SE 850, Sentech Instruments GmbH) used in the current work.

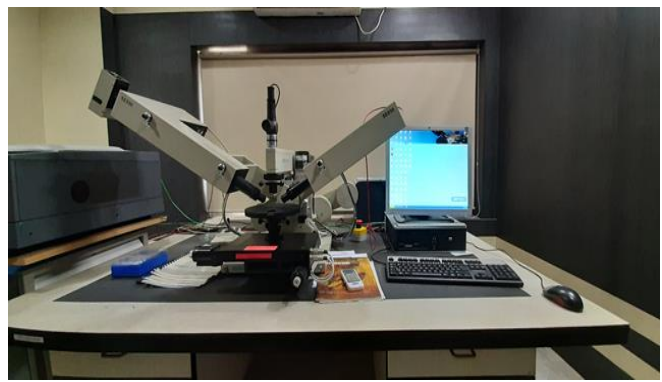

**Fig. 1.4**: Picture of the spectroscopic ellipsometer used in this work.

## 1.8 Organization of the thesis

This dissertation aims to design and fabricate on-wafer devices for the detection and quantification of different adulterants present in food samples. The detection processes involve EIS technique supported by optical spectroscopic techniques such as NIR and FT-IR using conventional conductivity cell prototypes and finally on-wafer devices. The fabrication of such devices is aimed to deliver superior performance in terms of sensitivity, speed, and cost-effectiveness.

In this context, a detailed background study of the materials considered along with their different adulterants has been performed and provided in Chapter 1. This chapter also includes the introduction of some of the characterization tools used for characterization and measurements.

Prior to commencing the experimental studies, an extensive chronological literature review has been performed and is described in detail in Chapter 2. Moreover, discussions on the successive progresses in this area of research have also been elaborated.

The impact of carbohydrate adulteration and its detection and quantification by EIS technique by using conductivity cells has been performed in Chapter 3. This chapter also presents a brief theoretical background on the measuring system including the dielectric properties of the materials used and the formulated electrical equivalent circuit of the experimental setup.

Chapter 4 emphasizes on the processes of quantitative estimation of sucrose syrup in several floral varieties of honey by employing EIS technique supported by FT-MIR technique to address the real time issues of tampering quality of honey. Different electrical measurements are performed on such pure and adulterated honey samples to observe the nature of changes in their respective electrical parameters.

Chapter 5 highlights the process of precise estimation of metanil yellow as an adulterant in turmeric powder samples using EIS, FT-MIR and UV-Vis spectroscopy techniques. This study also focusses on the chemical ion transport phenomenon that occurs inside the system due to the interactions among metanil yellow, de-ionized (DI) water and turmeric molecules. This phenomenon is responsible for the variation of measured electrical and optical parameters which varies with percent adulterant content and has been described in detail in the present chapter.

In contrast to the conventional measuring techniques employed in the previous studies, this work mainly focusses on the design and fabrication of LOC devices for the process of detecting a variety of organic and inorganic adulterants in milk. Experimental steps related to device fabrication and post-device fabrication sensing have been performed to observe precision and accuracy of the measurements performed and is discussed in detail in Chapter 6.

Chapter 7 concludes the present research work with its possible future scopes of further progresses and applications.

# Chapter 2

# Literature Review: Chronological development of food adulterant sensors

## 2.1 Origin of food adulterant sensors

In the past few decades, the impact of food fraud has been an alarming problem at global scale involving the social, environmental, health as well as economic aspects. The ever-growing number of food and bio-consumable adulterants and their consequent adverse effects on different foods, agriculture and biological health must be prevented by any means since the entire globalized food supply chain may be in question.

The idea of sensing or identifying food adulteration first originated in the year 1793 by Frederick Accum, who probably, was the first person to unmask the real nature of food adulteration [**54**]. He observed the indiscriminate practice of food adulteration that was going on in the contemporary British society. In 1820, his book on "*A Treatise on Adulterations of Food and Culinary Poisons*" has created remarkable turbulence in the society. In course of time, he mixed different adulterants in food products and raised public interest in this matter [**54**]. Since then, there has been a massive increase of using toxic and carcinogenic food adulterants and also their detection techniques have been developed accordingly.

In reality, serious adulteration paves the way for many diseases ranging from mild to life-threatening conditions like vision problem, liver problem, skin diseases, and several stomach disorders such as diarrhea [**55**]. Widely and commonly observed phenomena such as asthma, skin diseases and cancer are caused due to the intake of fish, fruits, meat or milk adulterated with chemical like formalin [**55,56**]. Human health is highly sensitive to food adulteration and sometimes shows immediate side effects like diarrhea, dysentery and vomiting [**55**]. In addition to immediate effects there may be many long term adverse effects of adulterated food. Long term effects like colon, peptic ulcers, liver diseases like cirrhosis and liver failure, heart diseases, blood disorder, bone marrow abnormality and kidney damage have been observed due to adulterants like colouring dyes, calcium carbide, urea, burnt engine oil and sometimes even due to excess amount of permitted preservatives [**55**].

In this context, several rapid, robust, point-of-care detection techniques in the domain of food safety and bio-sensors are being employed globally to restrict the usage of toxic adulterants in food and bio-consumables through qualitative as well as quantitative detection processes.

## 2.2 Emerging technologies in food and bio-science: Their advantages and disadvantages

Substitution of any component in food sample is performed either in the form of total substitution or partial substitution. Detection of partial substitution is difficult as compared to total substitution since, prior to the investigation of the adulterant; proper identity of adulterants should be known. Moreover, in partial substitution, the adulterant mimics the colour as well as taste of the pure sample which requires a highly precise recognition system to detect and quantify. Several methods, based on morphological and anatomical characterization, characteristic markers (odor, color, texture) and chemical testing, have been developed to authenticate marketed food samples and to check for any sorts of toxic adulterants [**57**]. The first strategy in the process of detecting adulterants in foods and beverages is through proper demonstration of the presence of foreign substances in them [**58**]. Such adulterants can be detected by various techniques based on the type of adulterant to be detected. These systems include physical, chemical, analytical as well as electrical techniques some of which are described in the following:

**Physical techniques**- Various physical methods for detection of adulteration including microscopic and macroscopic visual structural analysis as well as analysis of food by analyzing the physical parameters like morphology, texture, solubility, bulk density etc. have been designed but these methods do not guarantee qualitative adulterant detection. Visual structure analysis utilizing macroscopic and microscopic features is very useful in case of microbial detection particularly in case of fungi [**59-61**]. Also, microscopic inspection of certain spices such as chillies, cumin, coriander and cloves lead to easy detection of starch [**55,62**]. Through optical microscopy, adulteration of honey with cane sugar and sugar based products are detectable [**63**]. With the advent of electron microscopy it is now possible to detect the botanical origin of honey efficiently by analyzing the surface pattern of pollen from honey. Scanning electron microscopy (SEM) has also been used to study pollen from apple varieties [**64-67**]. However, electronic microscopic analysis is not a cost-effective routine technique and requires a meticulous sample preparation, although it offers advanced technical advancements in the detection system.

**Chemical/biochemical techniques**- Various chemical and biochemical methods for detection of adulterants can be categorized mainly into four different techniques such as chromatography, spectroscopy, immunology and electrophoresis [**55**]. Although these methods are comparatively more accurate and sensitive than physical techniques, there industrial applicability is hampered by cost and need of specialist training. The basic analytical approach involves various steps such as: a) extraction with a suitable solvent; b) cleanup for the removal of interfering matrix components; c) chromatographic separation and; d) selective detection. Among the analytical techniques, HPLC (High-performance liquid chromatography) is the most widely used technique. HPLC can be used as a quality control tool as it can separate various chemical constituents from mixtures; it is also used for characterizing food products or to detect adulteration. The adulteration of quince jams with apple [**68,69**], olive oil with hazelnut oil [**70-72**], citrus juices with flavones glycosides [**73-76**], phenolic compounds in black tea liquors [**77**], proline isomers and amino acids in wines [**78,79**] are some of the examples where adulterant has been successfully detected by using HPLC. Another technique namely GC is being used for separating volatile organic compounds. GC along with mass spectrometry (MS) and FTIR has been widely used for adulterant detection as these are nondestructive techniques. Gas chromatography is generally used to discriminate among different varieties of the same samples, adulteration detection, and organic compound authentication and identification.

Among the spectroscopic techniques, NIR helps in rapid detection of adulterants in the raw material, however, it is unable to identify the contaminant/adulterant. NIR spectroscopy can also be used as a tool to detect fraud and adulteration of soya based products used as animal feed [**80,81**]. Alternatively, NMR spectroscopy detects an adulterant and also provides its relevant structural identification [**55**]. FTIR spectroscopy, through detailed spectral investigation can differentiate adulterated sample from unadulterated samples and can identify the adulterant by extracting the specific constituent atomic bonds [**14,82**]. FTIR spectroscopy has shown excellent potential for the detection of milk adulterants and can be used in dairy industries [**83-87**]. Also, Atomic Absorption Spectrometry can be utilized as a validation method for the analysis of lead contamination in all foods except oils, fats and extremely fatty products [**88**]. Spectrometric techniques have also been utilized to determine gamma oryzanol content (%) in oils from spectrophotometer absorption measurements [**89**].

A combination of chromatographic and spectroscopic techniques has also shown high potential in adulterant detection techniques, for example, GC-MS has shown potential to detect honey adulteration with commercial syrups [**90,91**]. Solid phase micro-extraction-gas chromatography-mass spectrometry (SPME-GC-MS) has also been successfully employed for the detection of adulteration of ground roasted coffee with roasted barley [**92**]. Another combination of spectroscopy and chromatography, liquid chromatography-mass spectrometry (LC-MS) has been successfully applied to determine adulterants in herbal remedies [**93,94**]. Recently, the combination of FTIR spectroscopy and chemometric methods have been labelled as a rapid method for adulteration detection [**82,95,96**]. Electrophoresis technique has also been utilized in detecting food frauds, for instance electrophoretic analysis had potential to detect and quantify additional whey in milk and dairy beverages [**97**]. Among them, capillary electrophoresis method has been utilized to determine the adulteration of cow milk in goat milk products and adulteration in basmati rice [**98,99**].

Among the immunological techniques, enzyme-linked immunosorbent assay (ELISA) is the most widely used form of immunoassays in adulterant detection and has distinct advantages such as high sensitivity, easy accessible, reliable, low cost and high throughput over other techniques [**100**]. ELISA can be successfully used to determine adulteration of milk samples [**100,101**] but the entire detection process is very costly. Moreover, it can also be produced in the formats that are compatible with the industrial food processing environment.

Although certain physical and chemical techniques are easy and more convenient for routine adulterant detection in food products but they may not provide exact quantitative as well as qualitative results. Structural evaluation, i.e., detection of adulterant on the basis of its microscopic and macroscopic features of plant parts in grounded form requires high expertise. Similarly, chemical profiling is very useful and has ability to detect adulterants like synthetic drugs or phytochemicals [**55,57**], although, it is normally avoided by researchers as it involves the use of rare and expensive chemicals [**55**]. DNA-based methods have the potential to replicate these approaches [**102,103**] for which the food processing industries are employing several DNA based detection techniques, however, very few methods among them have proved robust enough to be used.

**DNA based Methods /Molecular Techniques**

Among the various techniques for detecting adulterants, the use of DNA based molecular tools could be more ideal for the marketed samples of plant origin, especially, when the adulterants are biological materials. Discrimination of adulterants from original food item can easily be done by molecular markers if both adulterant and the original food show physical resemblance. Also, certain omics techniques which involve analysis and manipulation of DNA, RNA, protein or lipid have become an integral part of molecular biology, genetics and biochemistry. Mainly three strategies are being followed in the arena of adulterant detection which incorporates DNA based methods such as PCR, sequencing and the process of hybridization. Primary steps of sequencing based detection involves variation in species specific region inside the genome due to insertion, deletion or transversion which acts as the key to differentiate and detect biological adulterants from original food samples [**55**]. Sequencing by hybridization method is generally utilized for the detection of adulterant on the basis of minute changes in nucleotide strand relative to known DNA sequence and the detection can be done from a variety of species, all at the same time [**104**]. However, these techniques have certain disadvantages such as these processes are time consuming; require a large amount of sample, skilled technicians and well-maintained experimental conditions. On the contrary, PCR based methods are simple, sensitive, specific and low-cost, thus offers a high potential in adulterant detection and authentication of commodities [**105,106**]. Real-time PCR method is a very fast detection method and has been used in the detection of adulterants in several food products [**105-110**]. It is now been frequently used as an analytical tool for adulterant detection in food industry, mainly due to its speed and specificity in analysis of food and its ability to amplify DNA sequences from highly fragmented DNA found in processed food. Another version of PCR, i.e., Duplex PCR has also been utilized for quantitative detection of poultry meat and milk adulteration [**111,112**]. Another unique technique namely Randomly Amplified Polymorphic DNA (RAPD) has also been used to detect adulterants in commercial food items due to its low operating cost and ability to discriminate different botanical species such as chilli powder [**113**], although, it lacks reproducibility. However, development of SCAR markers *i.e.* sequenced RAPD markers enable easy, sensitive and specific screening of commercial samples for adulterants and eliminates the problem of reproducibility of RAPD marker. Various SCAR markers have been developed for adulterant detection [**114-116**]. Another molecular method based detection system is becoming increasingly popular in rapid authentication of various food commodities titled loop mediated isothermal amplification (LAMP) technique [**117-119**]. Various species identification and detection of contaminants including antibiotics, pesticides, residues etc. are also possible through the molecular techniques.

## 2.3 Current status of adulterant detection techniques

In the past few decades, the increasing number of food adulterants or contaminants has raised alarms about the food safety and has led to remarkable improvements in the relevant analytical methodologies. Nowadays, food processing industries employ modern analytical techniques which help to meet the global demands on food safety and quality, leading to the development of more conclusive analytical methodologies for easy and cost-effective detection [**7**]. From the consumer's point of view, there is a need to complement the old and conventional techniques with the more sensitive detection ones including optical and electrical spectroscopy. Among such spectroscopies, the infrared based techniques such as NMR, NIR and FTIR are mostly preferred. In recent years, combination of both spectroscopic and separation techniques such as LC-MS, GC-MS and CE-MS are being used in the industries. In the arena of biological contaminant detection in food samples, the molecular techniques have progressed significantly where living organisms such as enzymes, antibodies, their DNA strands are used. Recent innovative molecular methods and DNA-based techniques allow fast and authentic detection of microbial contaminants in food. Novel biosensors have also been developed to detect microbial contaminants and selective hormones in food thereby providing high degree of specificity and sensitivity [**120-122**]. A similar novel molecular approach includes the use of peptide nucleic acid (PNA)-based technologies for food analysis and food authentication [**123,124**]. Also, adulterant detection in various foods and bio-consumables by employing novel electrical detection systems are emerging and proving to be less time consuming, cost-effective and superior for point-of-care applications in comparison to those conventional analytical, optical and chromatographic detection systems. For example, EIS is a powerful technique that can be used in a wide range of applications, such as microbiological analysis [**125,126**], food products screening [**13-21**], corrosion monitoring [**127-129**], quality control of coatings [**130-133**] and cement paste [**134,135**], characterization of solid electrolytes [**136,137**] and human body analysis [**138-**

**141**]. **Fig. 2.1** summarizes the different characterization techniques that are generally used to detect adulterant content in foods and bio-consumables.

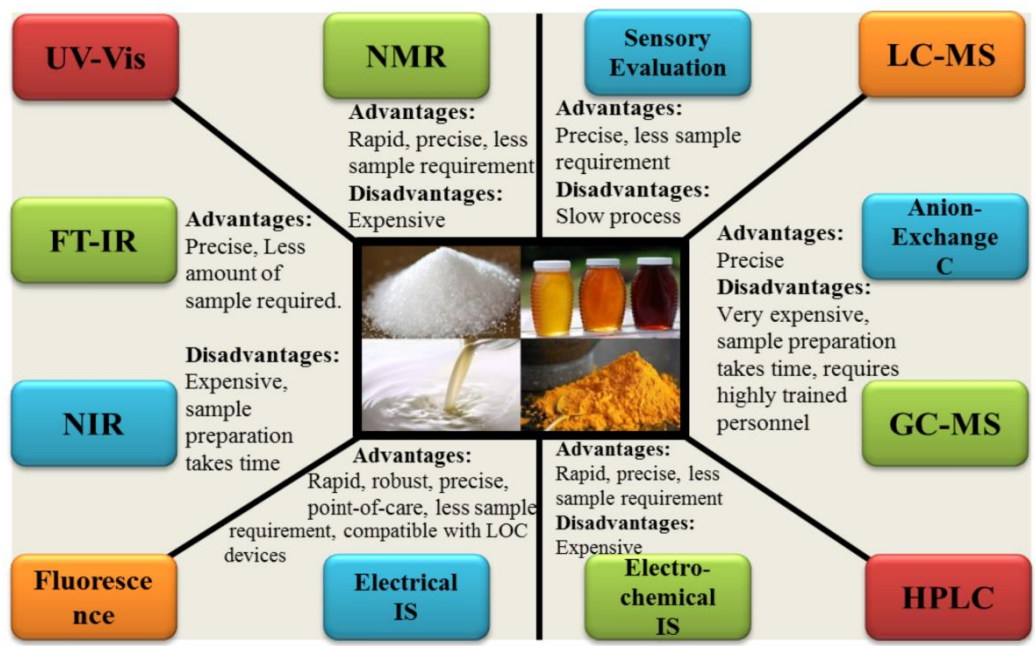

**Fig. 2.1:** Schematic representation of different characterization techniques employed to detect adulterant content in foods and bio-consumables.

## 2.4 Role of Electrical Impedance Spectroscopy in food safety

The most effective detection techniques are expected to be rapid, low-cost and precise in order to reduce the need for expensive and time-consuming laboratory analyses. Due to low adulterant content in materials under consideration, the techniques should also be capable of dealing with small amounts of the chemical compounds of interest in the investigated samples. Electrical measurements at low frequencies fulfill the above requirements and hence they may offer valuable information in an effective way. Due to the measurement specifics, these techniques have a potential to become especially useful in investigations of foods of low conductivity, e.g., honey [**142,143**], as well as milk and dairy products such as kefir and yogurt [**144.**]

EIS dates back to 1894 when Walther Nernst measured the dielectric constant of aqueous electrolytes and other organic fluids [**145**]. However, it was only in the mid-1980s that the interest in EIS really grew substantially owing to computer-controlled digital instruments which allow rapid and precise measurements for complex data processing and analysis. EIS is an experimental technique based on analyzing changes of the electrical parameters such as impedance, capacitance and conductance in a system, which appear due to the frequency variations of an applied AC voltage [**13-28,142,143**]. It is successfully used in various and diversified fields of science such as electrochemistry and materials science [**146-149**], pharmacology [**150,151**] and geophysics [**152,153**]. In case of a sufficiently wide frequency range, the EIS measured impedance spectrum incorporates relevant information about the electric properties of the system, comprising the sensor and material under test [**13-17,154**]. In real-time scenario, selecting an appropriate range of frequencies for investigations requires initial determination of the considered material characteristics such as capacitance and conductivity [**13-17,23**]. EIS is a method which analyzes the electrical properties of different materials and systems by incorporating alternating electrical signals at different frequencies and measuring the output signals [**13-17**]. The most straightforward way of interpreting EIS result is to construct an electrical equivalent circuit (EEC) for similar impedance spectrum analogous to the experiment [**13-17,23**].

Applications of EIS can be divided into three aspects, that is, electrical impedance tomography in medical imaging [**155,156**], quality and safety assessment in food industry and phytophysiology in agronomy [**157-162**]. Research objects and targets of EIS applied in food are abundant and extensive, including dry matter content in fruits [**163-167**], ripening of banana and mango [**168-171**], changes in potato and spinach tissues during or after heating [**172,173**], moisture content of carrot slices during drying [**164,166,**], quality evaluation of meat [**174-178**], salt and moisture content of fish [**159,179**] and real-time detection of bovine milk adulteration [**19,180**]. It is also used in the determination of additive contents in the natural juices [**21,159**] and quality assessment of cooking oil [**51,181,182**].

Therefore, to summarize, the role of EIS extends to a large number of interdisciplinary areas of research such as:

- Food and beverage industry,
- Agro-based industry,
- Blood analysis,
- Cell counting and sorting,
- Biomarker detection,

- Protein engineering,

- Particle size detection,

- Bacterial detection,

Further, the worldwide progress of in fabrication technologies have opened up new opportunities to integrate the EIS technique with novel microfluidic LOC devices as the measurement platform thereby improving the standards of detection quality with distinct advantages as discussed in the following section.

## 2.5 Alternative approach to rapid, low-cost and point-of-care sensing: Advent of LOC technology

LOC devices offer a concrete platform for miniaturization, parallelization, integration and automation of a variety of bio-chemical, bio-electrical and physico-chemical processes [**183**]. The primary advantage of LOC device is that it is a low-cost system with the use of small volume (in micron or nano level) of samples/reagents with high speed of operation and parallel implementation of test circuits to control fluidic transport from the source/reservoir to the sensing area. Silicon, as the base material has been considered by several microelectronic laboratories all over the world for LOC fabrication. Since the digital microfluidics technology requires necessary advances in micro- and nano-fabrication, therefore, several design and fabrication techniques were developed. Initially, the process of photolithography is employed to implement different subsequent steps such as resist coating on substrates, baking, UV exposure and lastly developing. Other than silicon, glass is also considered to be an alternative substrate while using this technique. Although, silicon and glass offer excellent mechanical and chemical properties, the chip fabrication on these substrates require high manufacturing expenses and high processing complexity. In recent years, a different approach in device fabrication is seen which involves the use of PDMS (polydimethylsiloxane), a polymer material which has been found to be reasonably cheaper than the conventional chemicals used in lithographic techniques and also offer high impact real-time applications. Recently, the implementation of novel mask less techniques such as inkjet printing, 3D printing and laser micromachining for the process of pattern transfer on substrates are getting more and more research attention. Several materials which have been considered as substrates for LOC or bio-chip fabrication are summarized below.

**Silicon LOC**: The idea of industrial fabrication of LOC devices was first started on the silicon substrate. Since, almost all of the recent micro-technologies are based on silicon micro-machining, therefore it became the automatic choice for device fabrication due to its excellent mechanical and chemical properties. Silicon micromachining offers high precision and it has the ability to integrate a lot of complex microelectrodes on a single chip. The concept of parallelization has been successfully implemented on silicon LOCs. Moreover, silicon based bio-chips can accommodate hundreds of unit cells (electrodes), all, on the same chip in a cost effective manner. However, the production cost of silicon LOCs are much higher as compared with polymer or paper devices. Also, silicon is not optically transparent (only IR) which limits its role in several real-time applications. Furthermore, silicon, as the substrate, has high electrical conductivity in comparison with glass and paper which restricts its usage in high voltage LOC applications. **Fig. 2.2** shows a silicon substrate based biosensor device fabricated in our laboratory.

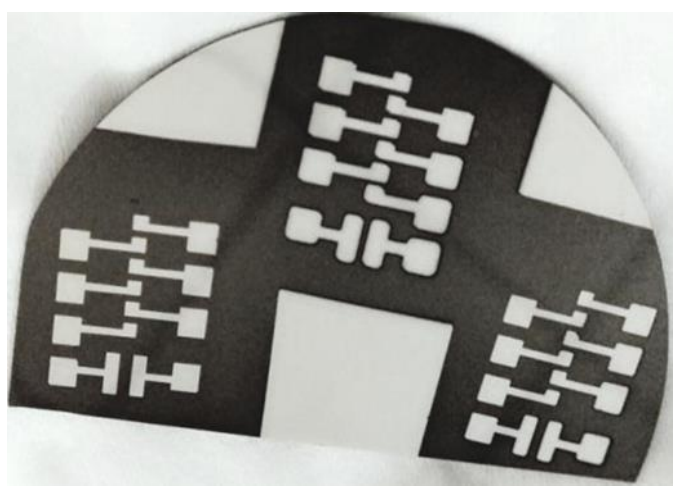

**Fig. 2.2**: Picture of a silicon substrate based biosensor device.

**Glass LOC:** In the recent years, glass based LOCs and bio-chips provide a potential low-cost alternative to expensive silicon DMF devices. Glass substrates offer optical transparency, compatibility with micron level machining and inertness to almost all available chemicals. Also, complex reproducible electrode design integrations can be easily implemented on glass substrate which makes it a good candidate for digital microfluidic (DMF) LOC devices. Another striking feature of glass is its bio-compatibility which extends its range of high impact real time bio-sensing applications. Although, the cost of glass substrate is negligible as compared to its silicon counterpart, still, the cost of other fabrication schemes involving it is similar to that of silicon fabrication technology. Therefore, glass based LOCs are still not used in research laboratories. **Fig. 2.3** shows a glass substrate based biosensor device fabricated in our laboratory.

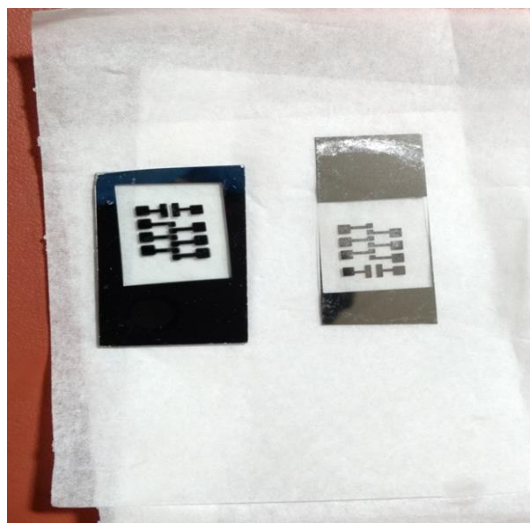

**Fig. 2.3**: Picture of a glass substrate based biosensor device.

**PDMS LOC:** PDMS is often used by various research laboratories for LOC prototype fabrication. In general, soft lithography technique is commonly used to fabricate PDMS structures. Being a transparent and flexible elastomer, PDMS is widely used to fabricate LOC devices in a very cost-effective manner. Moreover, PDMS structures provide great resolution and can contain sub-0.1 µm features. Also, another advantage of these PDMS LOCs and bio-chips is their quick production as compared to silicon or glass based microfluidic devices. In spite of these excellent features, PDMS shows severe limitations for industrial production due to certain drawbacks. They have extreme sensitivity towards specific chemical exposure, ageing problem, ultra high bio-molecular or drug absorption tendency etc. Therefore, since PDMS has ageing problems and it also absorbs hydrophobic molecules, it is difficult to fabricate electrodes on a PDMS chip. **Fig. 2.4** shows a PDMS substrate based LOC as developed by microfluidic Chipshop.

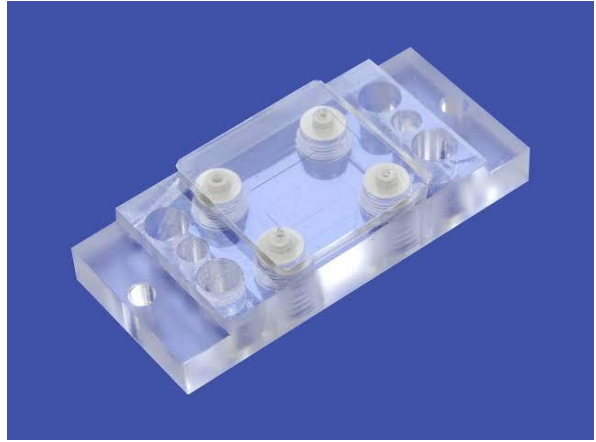

**Fig. 2.4**: Picture of a PDMS substrate based LOC.

**Plastic LOC**: The latest advancement in DMF LOC fabrication technology is the incorporation of plastic as the substrate material. Plastic is very easy to manufacture, cost-effective, lightweight and is readily available in the market. Such substrates can easily integrate several test circuits on it for electrical signal processing, which can be further used in high impact applications involving micro-electromechanical system (MEMS) devices. They also provide the option of printing patterned electrode designs on it by using inkjet printers. Electrically conductive inks can be used to print or replicate any user-defined designs on a plastic substrate. Also, they offer novel economic strategies in several bio-medical and healthcare applications. The production cost of a single plastic LOC is very much less as compared to silicon or glass or polymer LOC devices. Such devices are manufactured for one time use and can be easily disposed. Therefore, with more developmental research in this field, plastic based LOCs could open up the important fields of bio-sensing, bio-medical, food safety and healthcare systems while making it very much accessible to limited-resource populations. **Fig. 2.5** shows a plastic substrate based biosensor device fabricated in our laboratory.

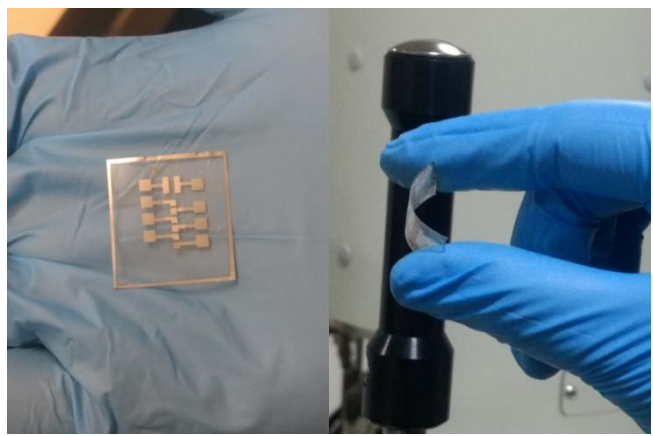

**Fig. 2.5**: Picture of a plastic substrate based biosensor device.

## 2.6 Advantages of LOCs over conventional devices

As per the literature, DMF LOC devices provide innumerable significant advantages when compared with conventional detection platforms **[184]** and the most significant advantages are:

**(i) Low cost:** Fabricated LOC devices are less expensive than other conventional sensing devices. Significant advancement in microelectronic technologies has enabled the lowering of cost for sample transport and analysis, faster processing and test with less sample consumption.

**(ii) High parallelization:** Owing to its unique micro-electrode integration capability, LOC system can accommodate hundreds of such circuits for test and analysis to be executed simultaneously on the same platform. This will help to detect specific problems and provide target oriented solutions for each process at the same time.

**(iii) Ease of use and compactness:** Although, the overall structure of LOC devices extend up to a few centimeters, still, it allows the incorporation of a large number of real-time biological and biochemical processes within that small volume. Therefore, a device with an area of just a few square centimeters is capable of performing complex tasks and analyses which requires full scale analytical laboratories to execute.

**(v) Faster response time and diagnosis:** LOC devices offer rapid transport, mixing and separation of sample droplets within a few milliseconds. The entire process of sample extraction from reservoirs and transportation to the sensing chamber via electrode paths followed by analysis and diagnosis takes negligible time as compared to all these processes performed in a clinical laboratory.

(vi) **Low sample requirement:** LOC technology requires negligible amounts of liquid sample for diagnosis and chemical processing as compared to the requirement of bulk volumes of samples in conventional laboratories. This decreases the overall cost of analysis.

**(iv) Human error reduction capability:** LOC devices do not require human intervention while processing which largely reduces the risk of process failure due to human error as can be observed in analytical laboratories.

## 2.7 Summary

The process of detection and quantification of adulterants in food and bio-consumables has been going through a chronological development in the past few decades which includes sustained development in the analytical and characterization techniques. Modernization of these techniques have led to the emergence of novel systematic, strategic, precise and point-of-care detection technologies for a wide range of applications in the fields of bio-electronics, bio-chemistry and bio-medical engineering. Out of these, the chemical methods have been successfully used to detect adulteration which is based on the chemical reaction between contaminants and the reagent or sample. Also, chromatography due to its high separation efficiency tool has become an important separation technology in the food compositional analysis, especially for trace contents such as pesticide residues, food additives, etc. Molecular methods have been successfully used as a tool for enhancing sensitivity and specificity of authenticity testing and microbial contaminant identification. However, all these techniques are highly expensive and require skilled personnels to facilitate the processes. Also, the reagents or chemicals used in these processes are much costly hence the overall detection cost increases. On the contrary, EIS has been proved to be a promising detection technology with advantages of being fast, nondestructive, inexpensive, easy to implement and shows potential to replace the traditional methods in order to save time as well as cost.

The current literature review illustrates that EIS technique has the potential for high end applications in assessing the quality of food and beverages. This study also suggests the ideal combination of EIS technique and on-wafer sensor devices to achieve enormous progress and developments in the field of food adulterant detection and bio-sensing.

# Chapter 3

# Impact of carbohydrate adulteration:

# Detection and quantification

## 3.1 Introduction: Need for a rapid and low-cost detection technique

For the past few decades, there has been an enormous increase of consumption of sugar and other carbohydrates among the global community in the form of fruit juices and beverages [185]. Sucrose, in its "sugar" form is used as one of the fundamental ingredients for sweetening of several daily consumables. It is also used as the most common preservative for jams, processed fruits and condensed milk. Different categories of carbohydrates find wide applications in the areas of cellulose, plastic, ink manufacturing and transparent soap making industries. With the sustained progress of food processing industries, adulteration has accordingly been a very routine practice for profit making. This poses serious threat to the public health security and therefore it requires to be addressed in every possible means. For instance, the mixing of chalk powder ($CaCO_3$) in normal sugar is a very common practice in India [186]. Also, the addition of sodium carbonate ($Na_2CO_3$) in sugar has been a severe threat to the global consumers. Such materials are miscible with the actual products and physically impossible to distinguish. As a matter of concern, with the emergence of innumerable adulterant materials and different manipulative refining techniques, it is getting very difficult to identify the presence of such adulterants in the daily consumables. Therefore, rapid detection of such adulterants is of immense importance.

Globally, several techniques are available to quantify the presence of different adulterants with varying wt% in the daily consumable products. Generally, traditional spectroscopy techniques such as FTIR, UV-VIS, Raman, NMR and chromatography techniques such as HPLC and GC have been widely used to detect the presence of adulterants in foods and beverages. Some of these techniques are rapid [187,188] while others are time consuming [90,189-192] in detecting adulterants [193-197]. However, the EIS and CS techniques can be very much effective in comparison to all these traditional methods for the detection of such

*Most of the work of Chapter 3 are published in the papers*
  1. *"Quantitative estimation of soda ash as an adulterant in aqueous sucrose solution by employing electrical impedance and capacitance spectroscopy" by C Das et al. (2019), Measurement, 148, 106937.*
  2. *"Analytical modelling of electrical impedance based adulterant sensor for aqueous sucrose solutions" by S Chakraborty, C Das et al. (2017), Journal of Electroanalytical Chemistry, 784, 133-139.*

adulterants due to its cost-effective, robust and simple approach. Also, it opens up the opportunity for integrating the existing integrated circuit (IC) technology which can lead to the development of system-on-chip (SOC) devices.

EIS technique generally uses an equivalent circuit to analyze and characterize the observed experimental results in response to the application of a small AC signal. Relative permittivity/dielectric constant is one of the fundamental properties of any material and EIS measures the electrical properties on the basis of this parameter [198,199]. Such technique actually characterizes the physicochemical properties of the bulk system in the form of a solution [200] and also has been primarily used to detect real-time cellular and subcellular growth [201]. This has also been widely used for investigating the electrical properties of plant tissues [202], DNA detection system [203], human mesenchymal stem cell growth [204] and detection of antibodies [205,206]. It is also being exploited to investigate the different structures of organic and inorganic materials over a wide and continuous range of frequencies [207]. EIS is also a useful tool for detecting artificial chemical additives/adulterants in beverages [21], fruit juices [208], dairy products [209-211], in determining the floral origin and electrical properties of honey [142,143] in examining specific biological tissues [22,173] and detecting pathogens [126,212,213].

Therefore, EIS is a potential alternative means for detecting small fractions of adulterants present in different consumable products. In this context, developing an appropriate theoretical model can also be useful for the overall understanding on the working principle of EIS [16,17]. In the current study, a detailed investigation has been conducted on the impedance spectroscopy of sucrose solutions by determining its impedance, capacitance and conductance. $Na_2CO_3$ has been incorporated as an adulterant in such solutions in a controlled manner. The variation of impedance, capacitance and conductance of the adulterated solution has been studied in detail and the relevant sensitivity of the method is determined in terms of the measured parameters.

## 3.2 Theoretical Background

The present work deals with a bulk aqueous sucrose solution, in which sodium carbonate is added in a controlled manner as the adulterant. The bulk sucrose solution contains water and sucrose, both of which are polar in nature and therefore makes it a purely polar system.

However, the addition of ionic sodium carbonate (Na₂CO₃) compound into it converts the solution to be an effective mixture of both polar and ionic compounds. In such a mixture, sodium carbonate dissociates to form sodium and carbonate ions as shown in Eqn. 1.1,

$$Na_2CO_3(aq) \rightleftharpoons 2Na^+(aq) + CO_3^{2-}(aq) \tag{1.1}$$

Therefore, the final solution contains polar molecules along with dissociated sodium and carbonate ions and their interactions will govern the alterations of the output electrical parameters considered here. The principal theory related to the interactions between such polar and ionic compounds has been discussed in the following section.

### 3.2.1 Langevin-Debye polarization model

The mixed system is assumed to be in thermal equilibrium. If an external field $\boldsymbol{E}$ is applied on it, the potential energy of a constituent molecule of dipole moment $\boldsymbol{p}$ can be expressed as,

$$W = -\boldsymbol{p}.\boldsymbol{E} = pE <cos\alpha> \tag{1.2}$$

where $\alpha$ is the angle between $\boldsymbol{p}$ and the electric field $\boldsymbol{E}.$

The $<cos\alpha>$ needs to be calculated by weighting $cos\alpha$ over all possible orientations with the Boltzmann's factor, and hence gives the Langevin-Debye expression for polarization as:

$$\vec{P} = np[coth\gamma - \frac{1}{\gamma}] \tag{1.3}$$

where $\gamma = pE/kT$, k and T being the Boltzmann's constant and absolute temperature respectively, and n is the number of dipoles per unit volume.

The electric displacement vector, $\vec{D} = \varepsilon\vec{E}$ and according to Gauss's theorem for dielectrics, there exists a certain amount of polarisation charge in the system along with its total free charge [**214**] and further it yields the electric displacement vector: $\vec{D} = \varepsilon_0\vec{E} + \vec{P}$; where $\varepsilon_0$ is the permittivity of free space. Thus, the polarization is related to the applied electric field as,

$$\vec{P} = (\varepsilon - \varepsilon_0)\vec{E} \tag{1.4}$$

Comparison of Eqns. (2) and (3) yields,

$$\frac{\varepsilon}{\varepsilon_0} = \frac{np^2}{3kT\varepsilon_0} + 1 \tag{1.5}$$

## 3.2.2 Frequency dependency of dielectric medium

For relatively lower ionic density, the effective dipole moment of the aqueous sucrose dipoles will govern overall dielectric nature of the system. Therefore, the effective dielectric constant of the solution can be written as [16],

$$\frac{\varepsilon}{\varepsilon_0} = \frac{1}{3kT\varepsilon_0}\left[n_d p_d^2 + n_i\left\{p_i\left(1 + \frac{n_{i1}}{n_{i2}}\right) - p_w\right\}^2\right] + 1, \quad [n_i \leq n_d] \tag{1.6}$$

where, $\varepsilon$ and $\varepsilon_0$ denote permittivity of the system and free space respectively, $k$ is the Boltzmann constant and $T$, the temperature of the system. $n_i$ and $n_d$ denote the quantity of ionic and sucrose dipoles per unit volume, whereas $p_i$ and $p_d$ depict their respective dipole moments. Here, $p_w$ is considered to be the dipole moment of a pure water molecule. Also, $n_{i1}$ and $n_{i2}$ are two consecutive numbers of ionic dipoles per unit volume considered here due to the increase of ionic density inside the system.

Thus, for the ionic dipole dominated system, the effective dielectric constant can be written as,

$$\frac{\varepsilon}{\varepsilon_0} = \frac{1}{3kT\varepsilon_0}\left[n_d p_d^2 + n_i\left\{p_i\left(1 + \frac{n_{i1}}{n_{i2}}\right) - p_w\right\}^2 + n_i'\left\{p_i\left(1 + \frac{n_{i1}'}{n_{i2}'}\right)\right\}^2\right] + 1, \quad [n_i > n_d] \tag{1.7}$$

where, $n_i' = n_i\left(1 + \frac{n_w}{n_i}\right)$, $n_w$ being the total water dipoles per unit volume. Therefore, the effective capacitance of the system can be calculated by using the formula:

$$C_{SOL} = \frac{|\varepsilon^*|A}{d} \tag{1.8}$$

where, d is the distance between the electrodes and A denotes the electrode area. The ε* used in Eqn. (7) depicts the frequency dependent complex permittivity of the system, where, $\varepsilon^* = \varepsilon - \frac{j\sigma}{2\pi f}$, σ and f depict the overall conductivity of the system and frequency of the applied electric field, respectively.

## 3.3 Materials and methods

### 3.3.1 Sample preparation

For preparing the sample solutions, sucrose powder (Merck) is mixed in electronic grade DI water (Millipore $^{TM}$) for five different concentrations of sucrose ranging from 13.72 $^{O}$Bx to 68.6 $^{O}$Bx. One degree Brix ($^{O}$Bx) represents 1 gram of sucrose in 100 grams of net solution, i.e., in the current experiment 1 gram of sucrose is dissolved in 99 grams of DI-water to produce 100 grams of the solution. It signifies the solution strength in terms of percentage by mass. Initially, 80 ml of DI-water and 13.72 $^{O}$Bx (i.e. 0.1 volume fraction) of sucrose are mixed inside a glass cylinder and subsequently adding sodium carbonate (Sigma-Aldrich) as adulterant with varying weight percentages (wt%) ranging from 0.0075% to 5%. This process is repeated for four other sucrose concentrations of 27.44, 41.15, 54.87 and 68.6 $^{O}$Bx, designated respectively by the volume fractions of 0.2, 0.3, 0.4 and 0.5 with respect to DI water in the solution.

### 3.3.2 Scheme of the experimental setup

In this study, all electrical measurements were performed by using a computer-interfaced LCR meter (TEGAM, Model 3550), as shown in **Fig. 3.1(a)**. A custom-made parallel-plate conductivity cell with unity cell constant is shown in **Fig. 3.1(b)**. The electrodes used in this work are uniformly coated with platinum of 99.9 % purity. The equivalent circuit of the measurement setup is shown in **Fig. 3.1(c)**.

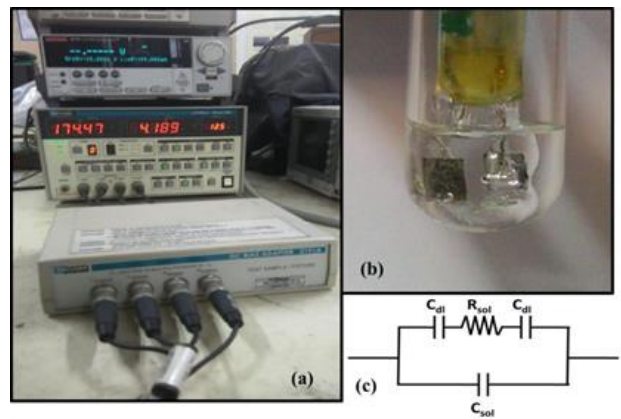

**Fig 3.1**: (a) The experimental setup for EIS measurements, (b) Real-time picture of a conductivity cell inside the test tube containing sample solution and, (c) equivalent circuit of the system.

## 3.4 Experimental procedures

The bias voltages of different frequency range are applied to a pair of metal electrodes, directly immersed into the medium under investigation (i.e. the DI-water/sucrose solution). The input signal along with the resultant signal is internally processed to produce a frequency-dependent transfer function relevant to net impedance (Z) of the system. The real part of impedance provides the sample resistance whereas the imaginary part deals with the capacitive/inductive reactance. The spectrum is achieved by measuring frequency dependent impedance within the frequency range of 50 Hz to 4 MHz which primarily depends on the experimental setup and chemical composition of the samples. However, the capacitance values depend only upon the intrinsic properties of the sample under investigation.

### 3.4.1 Equivalent circuit design

The equivalent circuit in **Fig. 3.1(c)** shows double-layer capacitors $C_{dl}$ with a resistor $R_{sol}$ in series and another capacitor $C_{sol}$ in parallel to the series combination. The equivalent circuit also explains the formation of $C_{dl}$ which appears due to the generation of ionic layers surrounding the platinum electrodes at the interface of the sample and the electrodes. Also, $C_{sol}$ and $R_{sol}$ depict the solution (pure or adulterated) capacitance and resistance, which are directly related to its dielectric nature. Furthermore, $C_{stray}$ and $L_{stray}$ signify the parasitic capacitance and inductance which arise due to the presence of certain coaxial cables and possible stray components inside the system **[39]**. Thus, the system impedance, which comprises of both $C_{sol}$ and $R_{sol}$, is calculated as the following:

$$|Z| = [(Z_1 \| Z_2) \| Z_{St}] + Z_L \tag{1.9}$$

$Z_1 = R_{sol} - j/\pi f C_{dl}$ ; $Z_2 = -j/2\pi f C_{sol}$. 'f' is the measurement frequency.

$Z_{St} = -\dfrac{j}{\pi f C_{stray}}$ ; $Z_L = 2j\pi f L_{stray}$

Thus, |Z| or the system impedance is the function of both $R_{sol}$ and $C_{sol}$. At very low frequencies, the measured net impedance is primarily dominated by the electrode polarization effect. In the mid-frequency range the resistive part dominates, and at the relatively higher frequency range, the dielectric nature of the solution becomes significant.

### 3.4.2 Data acquisition and process flow

Initially, the LCR meter is interfaced with LabVIEW software (National Instruments Inc.) by a Keithley GPIB cable and the instrument address is set to 35 followed by the optimization of the A.C. voltage and the data averaging time as schematically shown in **Fig. 3.2**.

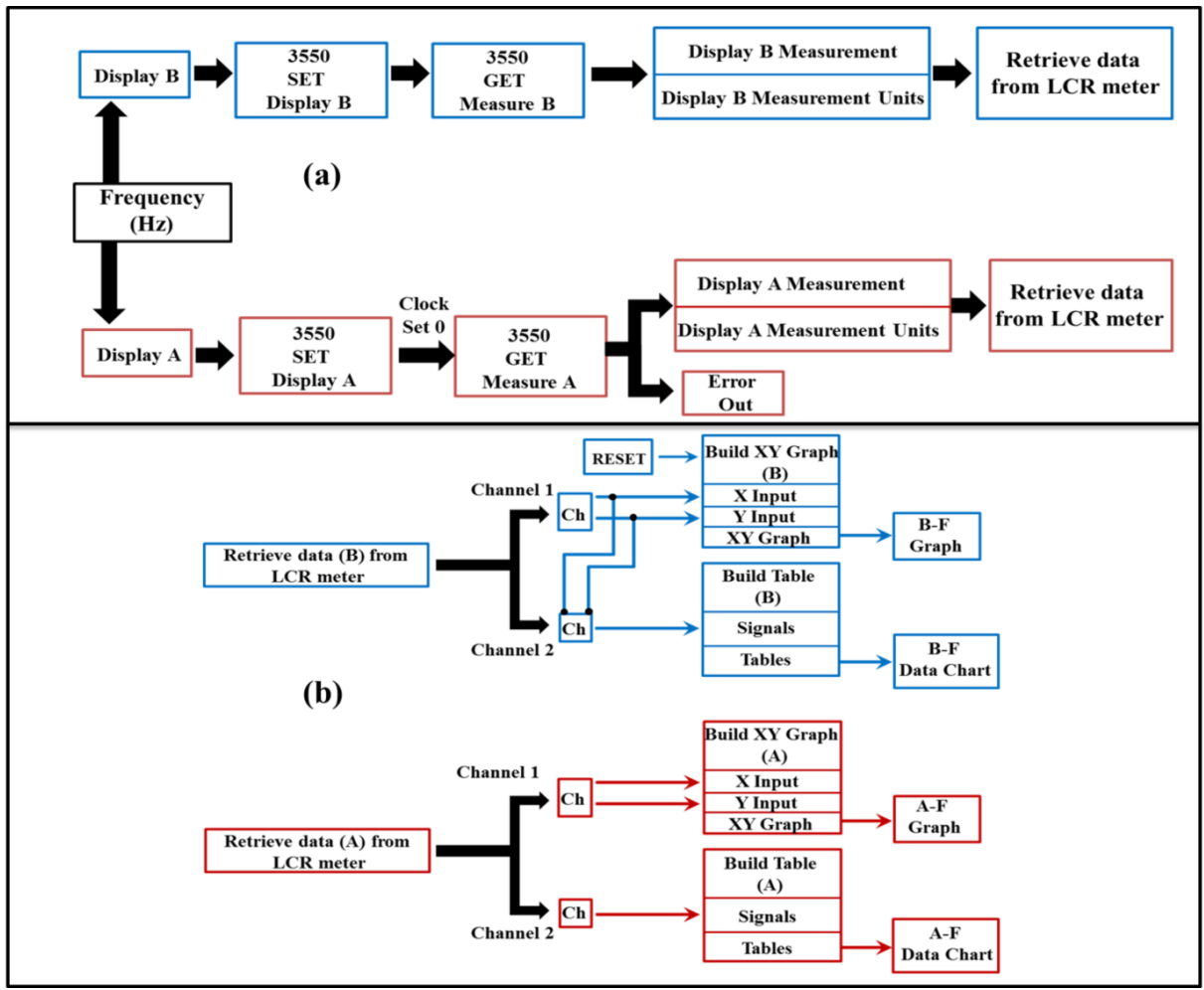

**Fig 3.2**: Block diagram of back panel of the LabVIEW measurement system for LCR Parameter Analyser design showing: (a) the measurement section and (b) the display section.

The flowchart, depicted in **Fig. 3.3**, illustrates the working principle of the LCR parameter analyzer design from the initial data input process to the final stages of parametric display.

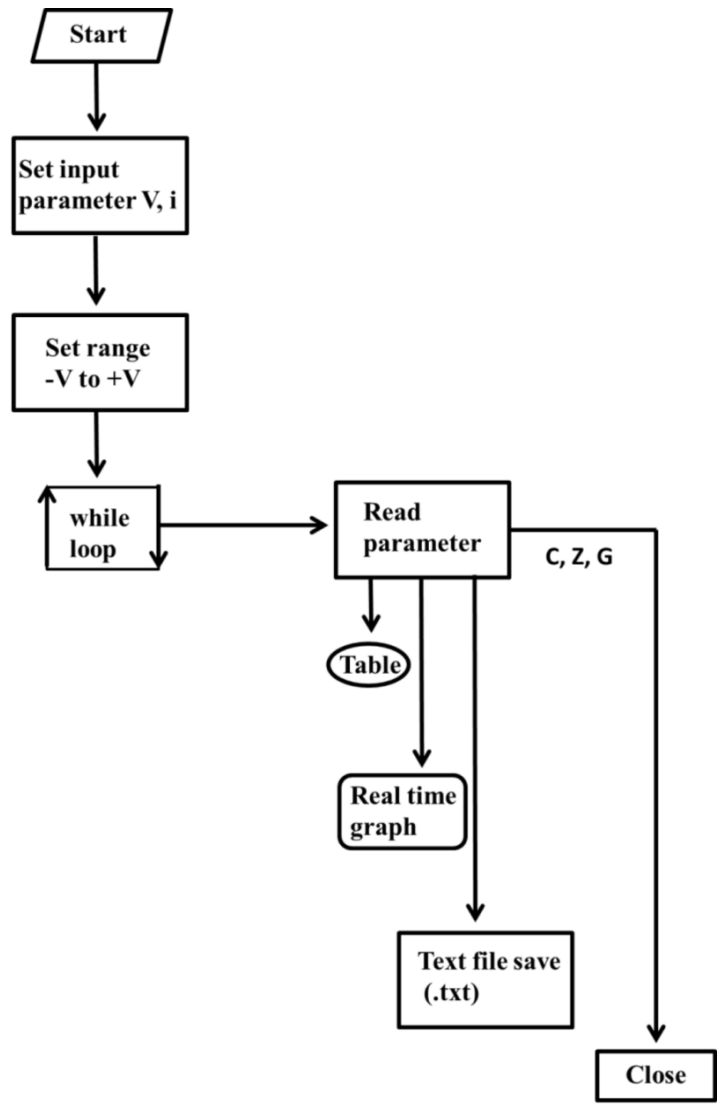

**Fig. 3.3**: Flow-chart illustrating the LCR Parameter Analyser design

For the frequency variation, a while loop is created in which the frequency is set to vary in the logarithmic scale. For each frequency value, inside the "while" loop, the data for inductance (L), capacitance (C), resistance (R) and other parameters are extracted sequentially and recorded parallely in three different modes. For real time data observation, two sets of tables and graphs are created in the front panel which shows two user-defined parameters such as A = L, C, R etc. and B = conductance (G), reactance (X), phase angle ($\psi$) etc. at any random point of time as depicted in **Fig. 3.4**. After the end of 'while' loop, the instrument automatically stops further acquisitions and the data is recorded in a text file where the file name can be saved each time prior to the measurements.

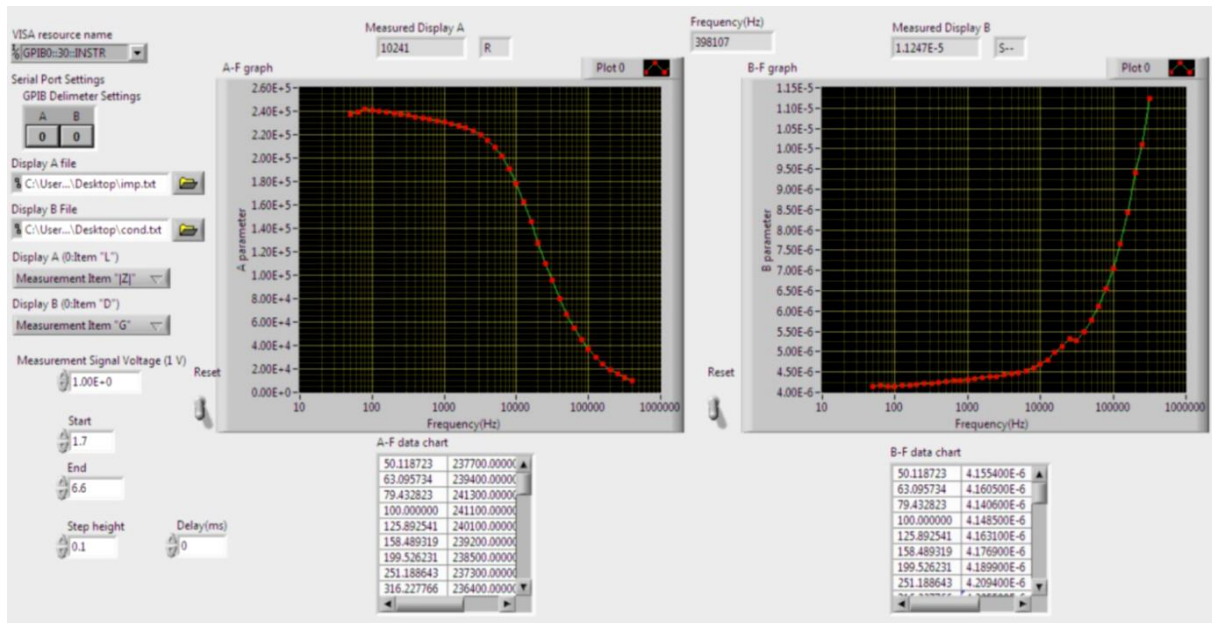

**Fig 3.4**: Front panel (user interface) of the LabVIEW measurement system showing the frequency spectrum of the electrical parameters.

A 1 V peak-to-peak AC voltage in the frequency range of 50 Hz to 4 MHz is applied to the electrodes for all the measurements and the experiments are repeated for three different times to verify reproducibility of the results. The TEGAM 3550 LCR meter is fully equipped with noise-reduction techniques and is calibrated by open-short compensation technique prior to the final measurements.

## 3.5 Results and Discussion

### 3.5.1 Electrical properties of sucrose-DI water mixture

The variation of theoretically and experimentally obtained impedance, capacitance and conductance with different volume fractions of sucrose in DI water-sucrose solution for different adulterant concentrations is shown in **Fig. 3.5(a), (b)** and **(c)**, respectively. The adulterant concentrations vary in the range of 0% to 5% in reference to the sucrose content of the solution. The measurements are performed for DI water/sucrose solution and DI water/sucrose/soda ash mixture solution for the sucrose volume fractions of 0.1, 0.2, 0.3, 0.4 and 0.5 respectively and the impedance values for pure DI water/sucrose solution measured for these volume fractions are obtained to be 158 kΩ, 141.53 kΩ, 162.1 kΩ, 230 kΩ and 160 kΩ, as shown in **Fig. 3.5(a)**. Therefore, for pure DI water/sucrose solution (i.e. 0% adulterant concentration) the nature of variation of impedance, capacitance and conductance is observed to be non-linear with the volume fraction of sucrose in the solution. Similarly, the experimentally obtained capacitance and conductance values are obtained to be 41.16 pF, 40.53 pF, 43.91 pF, 43.14 pF, 43.06 pF and 6.03 µS, 6.73 µS, 5.88 µS, 4.14 µS, 5.95 µS for the considered sucrose volume fractions, respectively, as shown in **Fig. 3.5(b)** and **(c)**. It is also apparent from **Fig. 3.5** that the theoretically obtained values estimated on the basis of Eqns. 1.6 and 1.7 exhibit a good agreement with the experimentally acquired data.

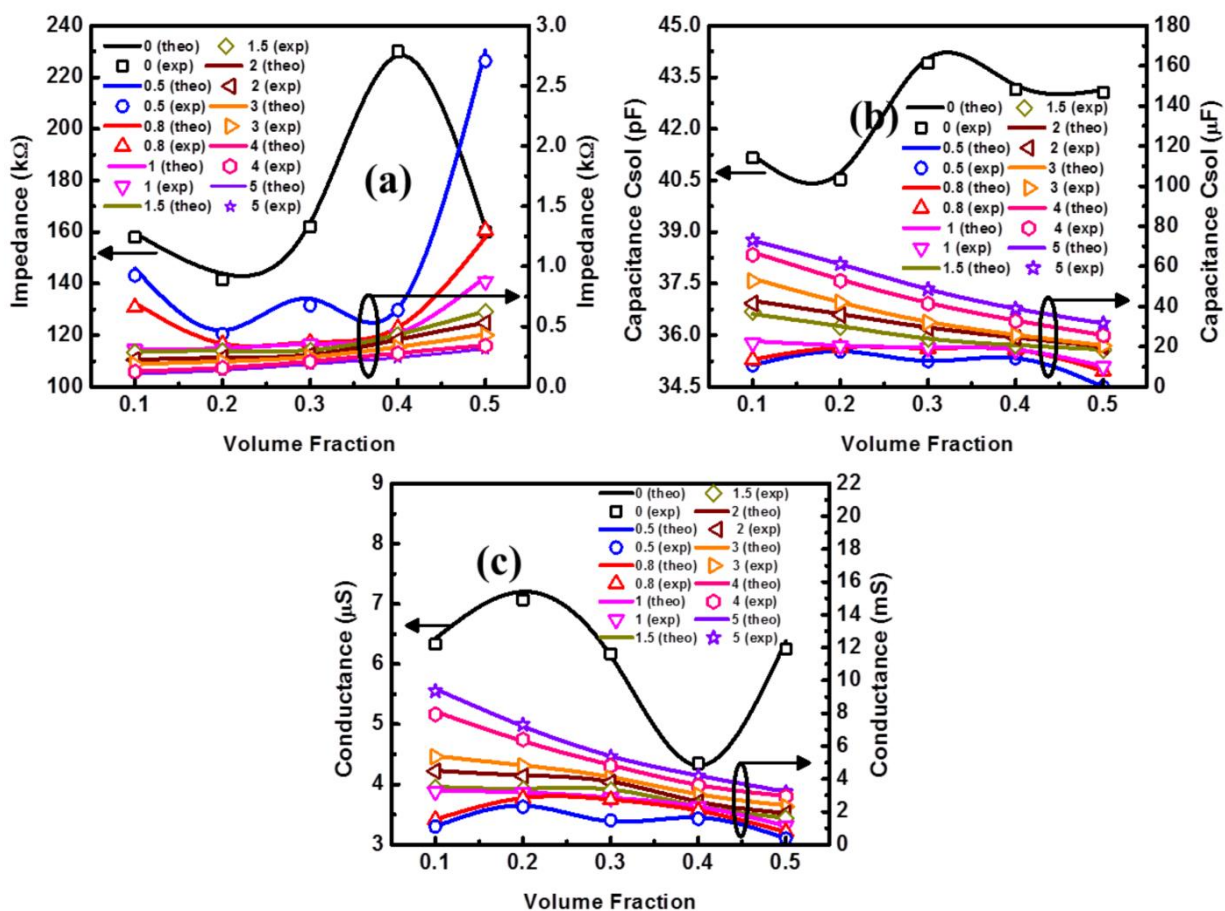

**Fig. 3.5**: Comparative plots of both theoretical and experimental variation of: (a) impedance, (b) capacitance and, (c) conductance with volume fraction of sucrose in sucrose-DI water mixture for eight different adulterant concentrations.

Interestingly, with gradual addition of controlled weights of $Na_2CO_3$ in sucrose-DI water solution, the non-linearity or quasi-oscillatory nature of the electrical parameters diminishes and a linear trend is observed. This linearity increases with increase of $Na_2CO_3$ content inside the solution. All the measurements are repeated for three different sets and **Figs. 3.6(a), (b)** and **(c)** show the variation of impedance, capacitance and conductance respectively, for all such sets with relevant error bar. The plots indicate the reproducibility of results within a variation range of ±5%. For verifying the reliability of the measurements, the results for three adulterant concentrations such as 0%, 1% and 5% in five different volume fractions of sucrose are shown.

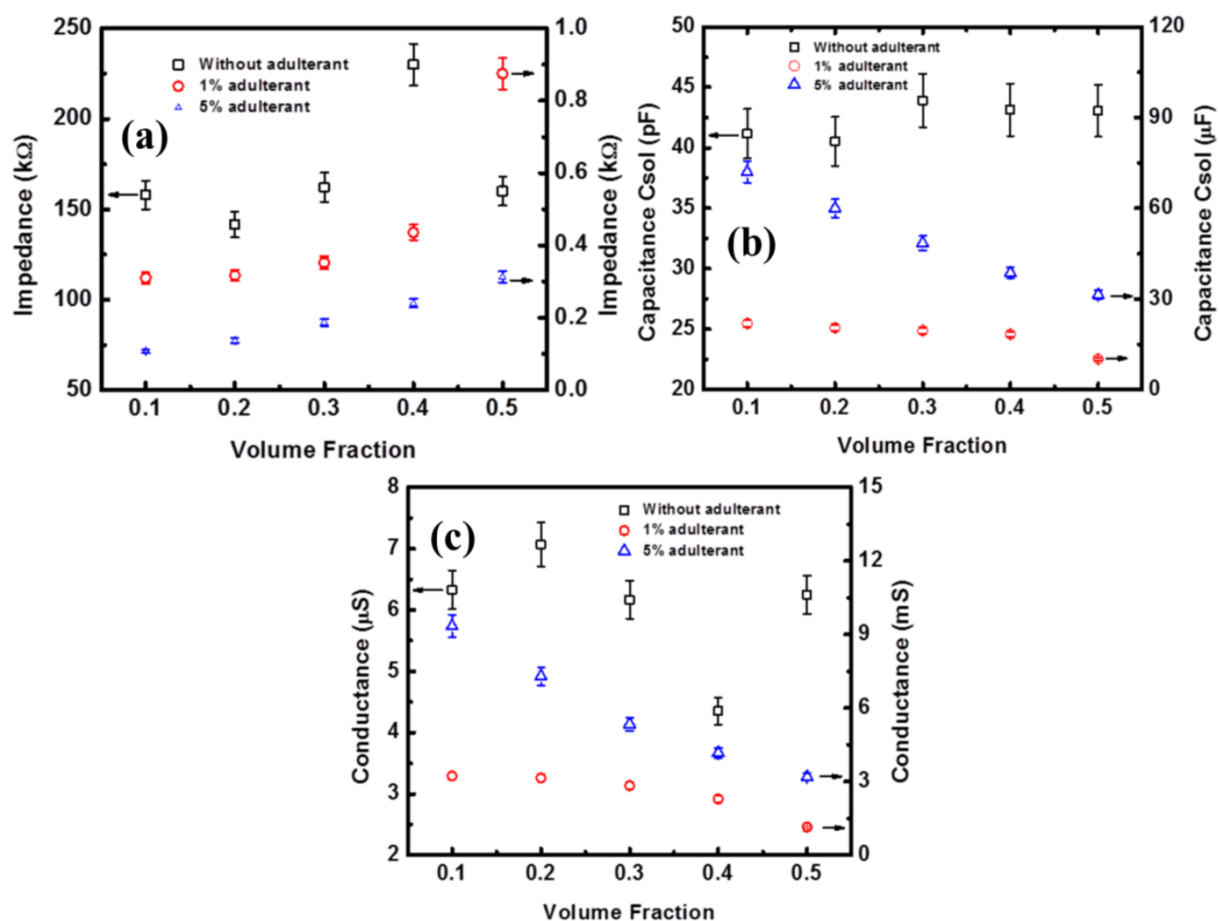

**Fig. 3.6**: Plots of: (a) impedance, (b) capacitance and, (c) conductance versus volume fraction with ±5% error bars for pure sucrose-DI water along with three different adulterant concentrations.

### 3.5.2 Electrical properties of sucrose-DI water-soda ash mixture

The impedance values within 1% of adulterant content are measured to be within the range of 157.15 kΩ to 310 Ω for all of the sucrose volume fractions. Above 1% adulterant content, similar variation is observed to be linear within the range of 287 Ω to 107 Ω. For the capacitance and conductance, such variations are measured to be in the range of 42.5 pF to 2.18 μF and 4.14 μS to 3.22 mS within 1%, and 3.73 μF to 72 μF and 3.48 mS to 8.9 mS above 1% respectively. The measured electrical parameters exhibit a linear variation with the adulterant content, as apparent from **Fig. 3.7(b)**, **(d)** and **(f)** respectively.

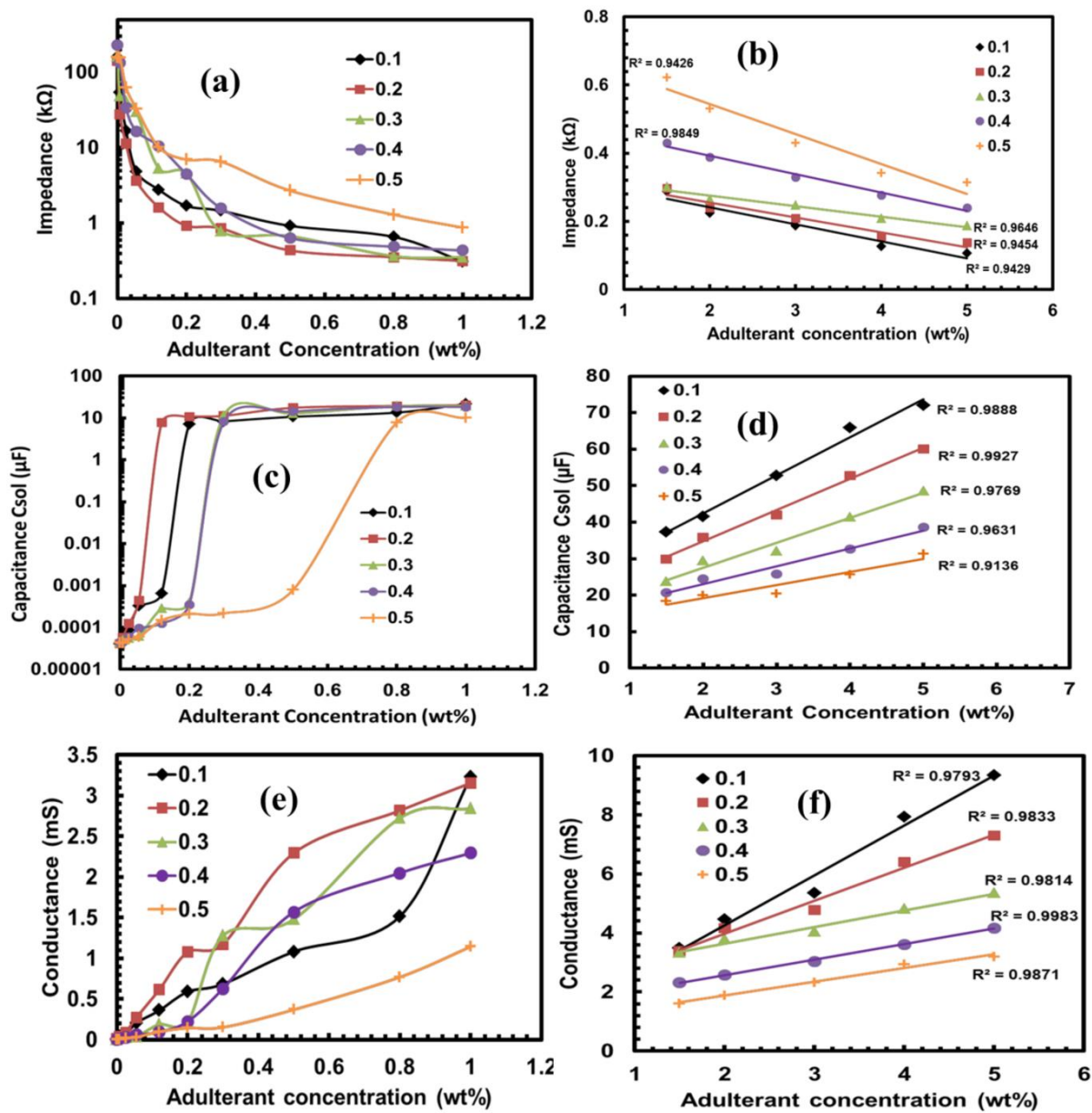

**Fig. 3.7**: Plots of impedance with adulterant concentrations ranging from: (a) 0% to 1% and (b) from 1.5% to 5% with correlation analysis, (c) plots of capacitance with adulterant concentrations ranging from 0% to 1% and, (d) from 1.5% to 5% with correlation analysis, (e) plots of conductance with adulterant concentrations ranging from 0% to 1% and, (f) from 1.5% to 5% with correlation analysis for five different volume fractions of sucrose.

It is also apparent from the above figures that for less than 1% (wt% with respect to sucrose) of adulterant concentration, the nature of variation of the electrical parameters are non-linear and repetitive. However, for the adulterant concentration $\geq 1\%$, the impedance increases steadily with the increase of sucrose volume fraction. Similarly, the capacitance decreases with increasing sucrose concentration for $>1\%$ of the adulterant content. The nature of variation of the conductance is observed to be similar to that of the capacitance. Such a change from quasi-oscillatory variation to linear is attributed to the dominance of sodium and carbonate ions over the pure polar sucrose dipoles in the solution. For a fixed volume fraction of sucrose, net impedance of the solution decreases with the increase of adulterant concentrations, while an opposite trend is observed for capacitance. The effective conductance of the system increases with increasing adulterant concentration for a particular volume fraction of sucrose. **Table 3.1** summarizes all the three sets of measurements performed for each volume fraction of sucrose consisting of their respective linear fitting equations (y = mx + c) and uncertainty values of both m and c for all the sets considered here.

**Table 3.1:** Summary of the linear fitting equations, their respective $R^2$ values and uncertainty values of both m and c considered for three sets of measurement performed.

| Volume Fraction | Set I | | Set II | | Set III | | Uncertainty of m* | Uncertainty of c* |
|---|---|---|---|---|---|---|---|---|
| | y = mx + c | $R^2$ | y = mx + c | $R^2$ | y = mx + c | $R^2$ | | |
| 0.1 | 1.69x + 0.86 | 0.9793 | 1.78x + 0.88 | 0.9777 | 1.77x + 0.89 | 0.9796 | (1.69,1.80) | (0.86,0.89) |
| 0.2 | 1.19x + 1.74 | 0.9833 | 1.14x + 1.77 | 0.9823 | 1.13x + 1.88 | 0.979 | (1.11,1.19) | (1.71,1.88) |
| 0.3 | 0.56x + 2.52 | 0.9814 | 0.57x + 2.59 | 0.9825 | 0.57x + 2.64 | 0.9813 | (0.56,0.57) | (2.51,2.65) |
| 0.4 | 0.52x + 1.5 | 0.9983 | 0.54x + 1.60 | 0.9969 | 0.55x + 1.57 | 0.9964 | (0.52,0.55) | (1.50,1.61) |
| 0.5 | 0.47x + 0.94 | 0.9871 | 0.48x + 0.93 | 0.9806 | 0.5x + 0.92 | 0.9943 | (0.46,0.5) | (0.92,0.94) |

*Results given as 95% confidence intervals on slope m and intercept c for k=2 and p=0.95

### 3.5.3 Variation of the EIS parameters

The nature of variation of the electrical parameters including impedance, capacitance and conductance is observed to be dominated by polar molecules in the sucrose solution with lower adulterant concentrations. However, the gradual increase of adulterant concentration leads to the net increase of ionic density of the system and for a critical adulterant concentration, when the number of ionic dipoles exceeds the number of polar dipoles, the nature of system dipole moment changes abruptly due to the dominance of ionic dipoles. Therefore, the gradual increase of adulterant concentration leads to an increase of net dielectric constant of the system.

### 3.5.4 Frequency dependent co-efficient of sensitivity analysis

For the adulterant concentration of ≥ 1%, such measuring system can be used as a capacitive sensor and the relevant coefficient of sensitivity can be defined as:

$$\beta = \frac{\sum(w - \langle w \rangle)(C - \langle C \rangle)}{\sum(w - \langle w \rangle)^2} \tag{1.10}$$

where, w and C denote the adulterant concentration (in wt%) and capacitance of the system respectively for a given sucrose volume fraction; <w> and <C> represent the average values of adulterant concentration and relevant capacitance at the given sucrose volume fraction. **Fig. 3.8** shows the plot of sensitivity versus volume fraction of sucrose in DI-water solution at four individual frequencies such as 1 kHz, 10 kHz, 100 kHz and 1 MHz respectively.

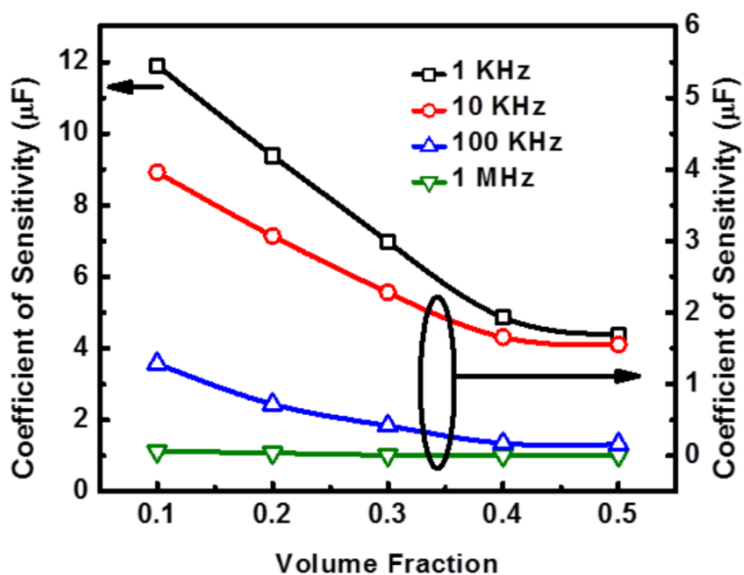

**Fig. 3.8**: Plot of sensitivity with volume fraction of sucrose at four different frequencies - 1 kHz, 10 kHz, 100 kHz and 1 MHz.

It is apparent from such plots that the sensitivity of the measuring system decreases with sucrose volume fraction, which is logical as the increment in sucrose volume fraction increases the relative dominance of pure polar dipoles, dictating the dielectric nature to be non-linear and repetitive. The sensitivity plots are also observed to be frequency dependent and show distinct variation in capacitance values when measured at four different frequencies such as 1 KHz, 10 KHz, 100 KHz and 1 MHz.

## 3.6 Summary

Several electrical parameters such as impedance, capacitance and conductance of DI-water sucrose solution with varying adulterant concentrations have been measured by employing electrical impedance spectroscopy and capacitance spectroscopy techniques. The variation of such electrical parameters have been studied both theoretically and experimentally with sucrose concentration and such variation has been observed to be quasi-oscillatory in nature whereas upon adding controlled sodium carbonate as an adulterant, the quasi-oscillatory nature is found to change gradually to be linear. At 1% adulterant concentration, the quasi-oscillatory nature of the system changes completely and exhibits a linear variation which remains similar for the relevant measurements. The lowest measured capacitance values for adulteration percentages <1%, =1% and >1% are observed to be 42.5 pF, 10.1 µF and 3.73 µF, respectively. Linear curve-fitting technique was incorporated to check linearity of the plots above 1% adulteration and the standard error of regression (S) with 95% prediction interval is calculated to be ±0.005. Therefore, electrical impedance and capacitance spectroscopy can be exclusively used to sense a minimum amount of 1% $Na_2CO_3$ as an adulterant in sucrose. Also, this technique can be used for quantitative estimation of different toxic and carcinogenic adulterants in foods and other bio-consumables.

# Chapter 4

# Applications of EIS for assessing the quality of honey

## 4.1 Introduction

In the previous chapter, the impact of adulteration of carbohydrate and its possible detection and quantification by employing EIS technique has been discussed in detail. This chapter emphasizes on the process of quantitative estimation of adulterant in several floral varieties of honey by employing EIS technique supported by FT-MIR spectroscopy.

Natural honey is a sweet and viscous material produced by different species of the apidae family including Apis mellifera, Apis dorsata and Apis cerana bees from the nectar of different flowers obtained from a vast section of the rich floral diversities of the world. The chemical composition of honey depends mainly on the floral sources, geographical locations, climatic conditions, methods of production and different storage conditions [**215-218**]. Generally, pure honey is a slightly viscous and sticky solution containing 80–85% saccharides (mainly glucose, fructose, maltose and sucrose), 15–20% water, 0.1–5% protein, 0.1–2% ash and negligible amount of amino acids, enzymes and vitamins, phenolic antioxidants and some trace minerals [**216,218,219**]. Each of these major and minor ingredients is known to possess unique dietary, therapeutic and remedial properties. The major ingredients are almost similar for all types of honey samples throughout the world and only the minor constituents differ due to the variation of floral sources and temperature conditions which affects the overall physicochemical properties of honey [**217**].

To maintain the international standards, each and every honey sample must undergo certain processing stages starting with the process of collection from the hives followed by refining, processing, storing and finally delivering to the consumers. Natural honey samples must contain specific ingredients as dictated by the IS 4941: 1994 regulations [**220**] and should not be adulterated by any low-cost chemicals or sweeteners. Globally, cheaper sweeteners such as Sucrose Syrup (SS), Corn Syrup (CS), jaggery, beet sugar, High Fructose Corn Syrup (HFCS), Maltose Syrup (MS) etc. are generally used to adulterate almost all categories of honey for monetary profit [**195,221-224**]. Besides, sugar or sucrose is a fundamental

*Most of the work of Chapter 4 are published in the paper "FT-MIR supported Electrical Impedance Spectroscopy based study of sugar adulterated honeys from different floral origin" by C Das et al. (2017), Talanta, 171, 327-334.*

ingredient of honey and is being used in the process of adulterating pure honey since several decades [**194,195,221,225**]. In this context, authentication of pure honey is one of the most significant issues for its quality regulation and food safety. Therefore, to address such issues, a rapid detection of adulteration in honey is very important.

Several techniques can quantify the presence of adulterants with varying wt% in pure honey products. In general, traditional fast spectroscopy techniques such as NIR, FT-MIR spectroscopy, Raman Spectroscopy, NMR [**82,188,194,195,221-224,226-232**] and comparatively slower chromatography techniques such as HPLC, Liquid Chromatography (LC) and GC [**90,196,233-235**] are widely used to assess the quality of honey. Another innovative method called isotope ratio mass spectrometry is also being employed to address this issue [**236**]. However, these analytical methods are expensive and complex with slow turnout rate and therefore not suitable for rapid, real time, point-of-care applications. In this context, the practical implementation of EIS technique in the domain of honey floral origin and its safety monitoring has been a topic of intensive research [**142,143,198,225**] and thus makes it very much effective in comparison to all these conventional methods due to its cost-effective, robust and simple approach for detecting toxic adulterants in honey. Also, FT-MIR spectroscopy is a rapid, robust and accurate technique and therefore employed by various research groups to assess food quality [**82,195,222,224,229-231**]. Both EIS and FT-MIR spectroscopy techniques find potential applications in bio-electronic engineering comprising the domain of devices and sensors [**143,237,238**], due to its use in detecting artificial chemical additives and toxins in fruits and beverages [**21,239**]. Also, EIS technique has been used to detect adulteration in grape juice [**208**], dairy products [**19**] and the presence of pathogens [**126,213**].

However, in comparison to the rapid progress of other detection technologies, very few researches are being dedicated to develop EIS as a rapid, cost-effective and portable technique. In this perspective, a thorough investigation has been conducted by using FT-MIR supported EIS technique on the pure honey samples along with their controlled adulterated solutions. Also, variation of electrical parameters such as impedance, capacitance, conductance, I-V and optical parameters including absorbance of the pure and adulterated samples has been studied in detail and the sensitivity of the method is determined in terms of system conductance. The primary objective of this work is to implement EIS for sensing and quantifying SS as an adulterant in honey from various floral species. The results obtained from these measurements are compared with those obtained from FTIR study to testify EIS as a rapid, cost-effective and precise technique which can be further used for estimating and sensing the presence of different adulterants in foods.

## 4.2 Materials and methods

### 4.2.1 Sample procurement and preparation

Honey samples used in the current work are collected from botanists and apiarists from several diversified locations in the state of West Bengal, India. All of the honey samples were normally stored at 5-7 °C and kept at room temperature (25±1°C) for 4-5 hours prior to the measurements. Each sample was homogenized with a stirrer before being measured.

### 4.2.2 Summary of the honey samples with their floral sources

**Table 4.1** summarizes the family and floral sources of the honey samples considered in the current work.

**Table 4.1**

Summary of the types of honey along with their respective floral sources used in the present work.

| Honey Sample (Common Name) | Floral Sources | Family |
|---|---|---|
| **Acacia** | Robinia pseudoacacia | Fabaceae |
| **Albizia** | Albizia lebbeck | Mimosaceae |
| **Brassica** | Brassica juncea | Brassicaceae |
| **Mangrove** | Avicennia officinalis | Acanthaceae |
| **Karanj** | Pongamia pinnata | Fabaceae |
| **Neem** | Azadirachta indica | Meliaceae |
| **Tamarind** | Tamarindus indica | Fabaceae |

### 4.2.3 Preparation of sucrose syrup (SS) as an adulterant

The sucrose-water mixture, consisted of 100 g molecular grade sucrose (Merck, India purity >99%) and 100 g de-ionized water (Millipore [TM]) was added in a 500 ml flask. The masses of sucrose and water were measured in an electronic balance (Mettler Toledo AL 204) having a precision of 0.01 mg. The flask was placed on a magnetic-stirred hot plate and for constant stirring, a medium-sized magnetic bead was inserted into the solution and the stirring speed was maintained to 500 R.P.M. During heating, temperature of sucrose–water mixture was kept between 80-100 $^O$C, and the mixture was continuously stirred to make sucrose melt in water evenly. Such process of heating and stirring was performed for 10 minutes and then, the solution was weighed. The procedure continued until the sucrose concentration in the solution reached 80% and the weight of the solution was measured to be 125 g which contains 100 g sucrose and 25 g DI water. The final solution was left to cool at room temperature before being used for measurement. No noticeable crystal structures have been observed inside the solution at the time of measurements.

### 4.2.4 Measurement of the honey-SS solution

For the adulteration process, 0.1 ml of pure SS (80% conc.) containing 120 mg of pure sucrose was added to 1.2 g each of pure acacia, albizia, brassica, karanj, neem, mangrove and tamarind honey to make adulterated honey–SS with different sucrose concentrations ranging from 0% (pure honey) to 70% at 10% interval based on w/w ratio at room temperature. Since, accuracy of the measured electrical property of the solution highly depends on the presence of air bubbles between the probes and solution, therefore, effort was made to completely remove them during the experiment. All measurements were conducted at least three times at room temperature (25±1°C). **Table 4.2** summarizes the net content of adulterant, percentage of adulteration and their distinctive levels as adulterant in the samples and it is worthy to mention that the accuracy of the weight balance is ± 0.1 mg.

**Table 4.2.**

Summary of the concentration of adulterated honey solutions with their levels of adulteration

| Weight of pure honey (g) | Content of Adulterant (g) | Percentage of adulteration (w.r.t pure honey) | Level of Adulteration |
|---|---|---|---|
| | 0 | 0 | Nil |
| | 0.12 | 10 | Low |
| | 0.24 | 20 | Low |
| 1.2 | 0.36 | 30 | Medium |
| | 0.48 | 40 | Medium |
| | 0.6 | 50 | Medium |
| | 0.72 | 60 | High |
| | 0.84 | 70 | High |

## 4.3 Experimental procedures

### 4.3.1 Ash content measurement

As the electrical conductance of honey is governed mainly by its ash and mineral salt contents, therefore determination of ash content is a significant parameter that can be used as an identification marker of honey florals. The process step of extraction of total ash content inside the original honey is shown in **Fig. 4.1**.

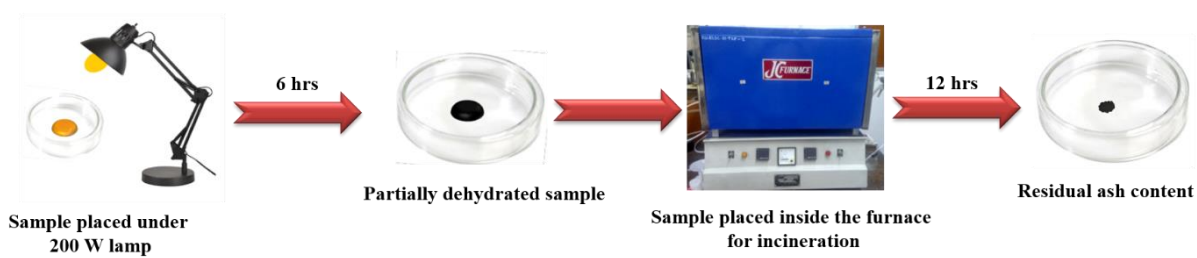

**Fig. 4.1**: Process flow for measuring ash content in honey.

All of the honey samples weighing 5 grams each was placed under a 200 W yellow lamp until their color turned black. Subsequently, these were incinerated at 600 °C in an annealing furnace overnight until the honey samples were completely dehydrated. Then, these were cooled and stored in desiccators before being weighed by the electronic balance.

### 4.3.2 Total soluble solids and moisture content measurement

The total soluble solid content and water content were determined by using a hand-held brix refractometer (RHB-90ATC, SINOTECH, Zhangzhou, China) at 20 °C following the IS 4941:1994 (Annex B) regulations [**220**]. **Fig 4.2(a)** shows picture of the refractometer used in this study and **Fig 4.2(b)** represents the internal scaling tables which are seen through the eye-piece.

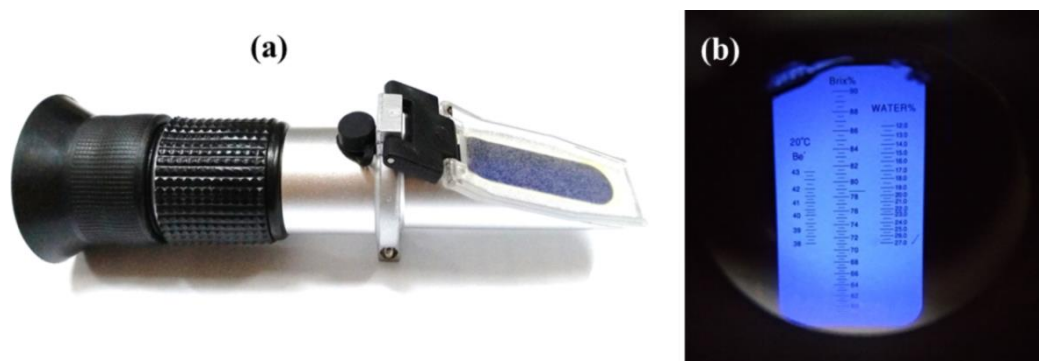

**Fig. 4.2**: (a) Hand-held brix refractometer used in this study and, (b) internal scaling tables as viewed through the eye-piece.

Sample measurements were performed by placing two to three drops of the raw honey sample upon the transparent base of the brix refractometer. Inside the transparent base, when seen through the eye-piece, different scaling tables are present utilizing which one can observe the water, total soluble solids and Be content of any sample. Percentages of moisture content in the honey samples was also verified from their respective refractive indices. **Table 4.3** summarizes the total soluble solid content, ash content and moisture content of all the honey samples used in this work.

**Table 4.3**

Summary of the physico-chemical properties of seven types of honey samples considered for the current work.

| Honey Sample | Total soluble solids (%) | Ash Content (%) | Moisture Content (%) |
|:---:|:---:|:---:|:---:|
| Acacia | 82.1±0.77 | 0.05±0.01 | 16.2±0.4 |
| Albizia | 81.5±0.42 | 0.1±0.02 | 16.8±1.0 |
| Brassica | 81±0.61 | 0.12±0.01 | 17.5±0.3 |
| Mangrove | 80.5±0.48 | 0.15±0.03 | 18±0.7 |
| Karanj | 79.2±0.56 | 0.27±0.02 | 18.8±0.5 |
| Neem | 79.8±0.82 | 0.21±0.01 | 18.5±1.0 |
| Tamarind | 81.8±0.75 | 0.08±0.01 | 16.5±0.7 |

### 4.3.3 EIS measurement

Impedance measurements were performed on each honey sample and the data are recorded simultaneously. Measurements for all the samples have been repeated five times in order to check reliability of the results. The EIS measurements have been executed by a precision LCR Meter (TEGAM 3550) in the frequency range of 100 Hz–5 MHz, with voltage amplitude of 100 mV peak-to-peak. The LCR Meter was interfaced with a personal computer by LabVIEW software for superior parametric visualization. The LCR meter is fully equipped with noise-reduction techniques and is calibrated by open-short compensation technique prior to the final measurements.

### 4.3.4 Current-voltage (I-V) measurement

The current-voltage (I-V) measurements of pure and adulterated honey samples were performed by using Keithley Source Measuring Unit (SMU 2611B) within a voltage range of -10 V to +10 V with a minimum resolution of 0.1 V and the output current values were measured accordingly.

### 4.3.5 FT-MIR measurement

The optical measurements were performed by using JASCO FTIR-6300 Spectrophotometer at room temperature by performing a total of sixty-four scans recorded for each spectrum in the spectral region ranging between 4000 cm$^{-1}$ and 400 cm$^{-1}$ with a nominal resolution of 4 cm$^{-1}$. Individual honey samples were liquefied at 40 °C prior to their measurements in order to remove any unnecessary solidified content. 100 mg of IR grade Potassium Bromide (KBr) pellets were prepared by using a pestle and mortar, and the KBr matrix was cold-pressed into a transparent disk. Before performing any measurement, pure KBr pellet was used to obtain the reference background spectrum. 5 µl drops of each sample were applied onto the surface of the pellets and the resultant spectra were measured. Each of the spectral measurements has been repeated three times to ensure the precision of the instrument used.

## 4.4 Results and Discussions

### 4.4.1 Electrical characterization and analysis

In this study, the influence of SS adulteration on individual honey samples and nature of variation of their respective electrical properties at 1 KHz frequency at room temperature were observed. For each sample, total eight measurements including pure honey (0% adulterated) and maximum adulterated honey (70% adulterated) are performed to calculate the values of relevant electrical parameters.

#### 4.4.1.1 EIS analysis of sucrose adulterated honeys

EIS analysis shows the variation of different electrical parameters such as impedance, capacitance and conductance in **Fig. 4.3(a)**, **(b)** and **(c)** respectively for the adulterated honey samples with different adulteration concentrations. It is apparent from such plots that the electrical parameters vary linearly with sucrose content (from 10% to 70%) for all of the pure honey samples. The capacitance and conductance values exhibit a linear decreasing trend while the system impedance demonstrates an increasing trend with the increase of sucrose content.

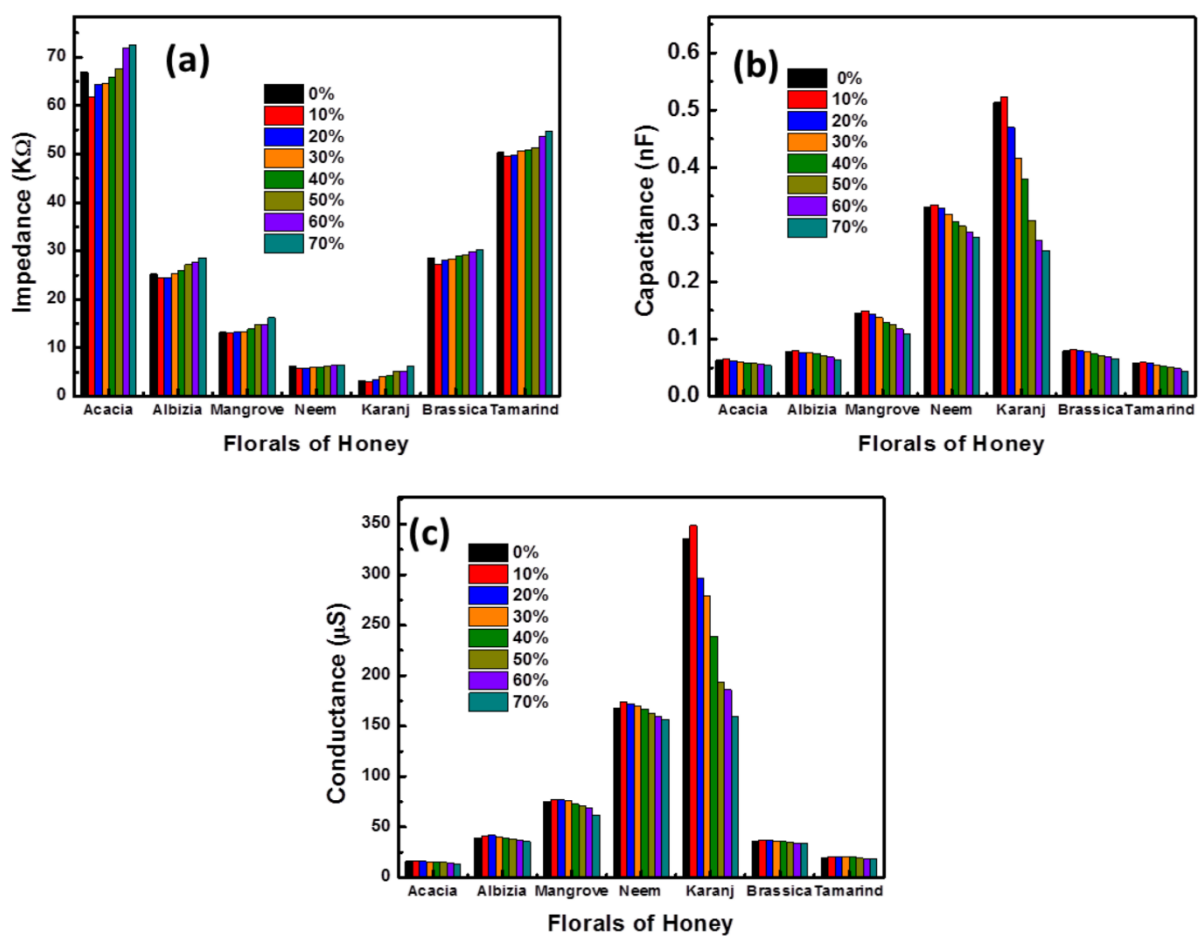

**Fig. 4.3**: Plots of (a) impedance, (b) capacitance and (c) conductance versus different florals of honey with eight different percentages of adulteration.

Since the primary constituents of honey (such as fructose, glucose, maltose and sucrose) are all polar in nature, therefore, honey can be assumed as a polar liquid. Also, since the adulterated SS contains molecular grade sucrose and DI water, both of which are also polar in nature, therefore, the addition or mixing of polar substance in a polar solution results in a quasi-oscillatory behavior of the electrical parameters for a specific range of adulterant concentration (0 to 10 %) [**16,17**]. It is evident from **Fig. 4.3(b)** that the capacitance values increase with adulterant content up to 10% concentration. However, for the adulterant concentration >10% the system capacitance exhibits a linear trend and decreases with added adulterant content. Similar nature is observed for the conductance values whereas it is reverse in case of impedance for all of the samples under investigation. Initial increment in capacitance values in lower adulterant concentration is attributed to the increasing polarization of the system for dipolar alignment relevant to polar-polar type of molecular interactions between the adulterant SS and constituent carbohydrates and water present in honey. Upon adding more adulterant in the solution such interaction sustains a threshold condition and the additional sucrose starts interacting with ionic components such as the ash components present in honey. Polar dipole moment is much smaller than ionic dipole moment, and hence, interaction of polar dipoles corresponding to sucrose molecules with the ionic dipoles effectively reduces the resultant dipole moment of the system, and thereby decreases the overall polarization and dielectric constant of the system. In terms of electrical conductivity, karanj and neem honey exhibited higher values while the lowest were shown by acacia and tamarind respectively. **Table 4.4** summarizes the minimum and maximum values of impedance, capacitance and conductance of pure and adulterated honey samples and thereby demonstrates the nature as well as their range of variation in terms of adulteration. It should be noted that the estimated error in the measurements was observed to be within ± 2%.

**Table 4.4**

Summary of the values of electrical performance parameters of the seven varieties of honey used in the present work.

| Honey Sample | Impedance(KΩ) | | Capacitance (pF) | | Conductance(µS) | |
|---|---|---|---|---|---|---|
| | Min (Pure) | Max (Adulterated) | Min (Adulterated) | Max (Pure) | Min (Adulterated) | Max (Pure) |
| Acacia | 66.94 | 72.51 | 54.14 | 63.53 | 13.61 | 16.02 |
| Albizia | 25.25 | 28.62 | 63.29 | 77.68 | 35.66 | 39.55 |
| Brassica | 28.57 | 30.25 | 65.79 | 79.28 | 33.45 | 35.58 |
| Mangrove | 13.25 | 16.237 | 110.87 | 146.11 | 61.78 | 75.19 |
| Karanj | 3.29 | 6.24 | 254.45 | 513.41 | 160.9 | 336.03 |
| Neem | 6.15 | 6.505 | 277.21 | 330.66 | 156.4 | 167.3 |
| Tamarind | 50.35 | 54.74 | 44.61 | 58.6 | 18.13 | 19.03 |

Note: Percentage of estimated error is ± 2%.

### 4.4.1.2 I-V analysis of sucrose adulterated honeys

**Fig. 4.4(a)** shows the current-voltage plots for all of the natural honey samples and **Fig. 4.4(b)** shows the variation of current for the percentage of adulteration present in honey for a particular voltage at +10 V. The maximum current values for pure acacia, albizia, brassica,

mangrove, karanj, neem and tamarind are measured to be 0.147 mA, 0.387 mA, 0.347 mA, 0.685 mA, 2.62 mA, 1.42 mA and 0.197 mA, respectively. Also, the minimum current values for those samples are measured to be 0.127 mA, 0.327 mA, 0.289 mA, 0.56 mA, 1.43 mA, 1.2 mA and 0.147 mA, respectively. Thus, a decreasing trend is observed with the addition of SS in honey. In general, pure SS is much less conductive than honey and from **Fig. 4.4(b)**, it is apparent that with its addition, current in the system effectively decreases which in turn increases the net impedance of the system. Therefore, the system can sense the inclusion of adulterant internally and vary the output electrical parameters accordingly.

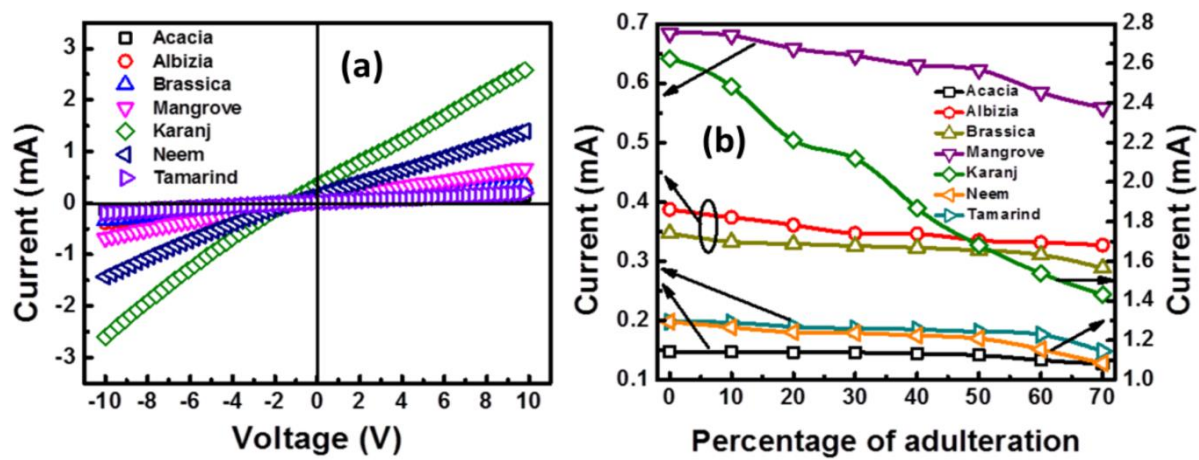

**Fig. 4.4:** (a) Plots of I-V characteristics of all of the honey samples considered and, (b) plots of the variation of current with different percentage of sucrose adulteration at +10 V.

### 4.4.2 Optical characterization and analysis

In this study, the influence of SS adulteration on individual honey samples and nature of variation of their optical properties in terms of absorbance were observed. For each sample, a total of eight measurements including pure honey (0% adulterated) and maximum adulterated honey (70% adulterated) are performed to calculate the necessary optical absorbance data.

### 4.4.2.1 FT-MIR analysis of sucrose adulterated honeys

**Fig. 4.5(a)** shows the FT-MIR spectra of seven different varieties of honey in the spectral region between 1800 cm$^{-1}$ to 700 cm$^{-1}$. This region is also termed "fingerprint region" as it depicts complicated series of absorptions resulting from all sorts of stretching, bending and vibrations of atoms occurring within a molecule. The significance of this region is largely due to the change of absorbance peaks and troughs with variation of compounds which easily helps to identify one material from the other. Therefore, for data analysis, this region is chosen to study the differences or modifications that occur in the spectra due to adulteration.

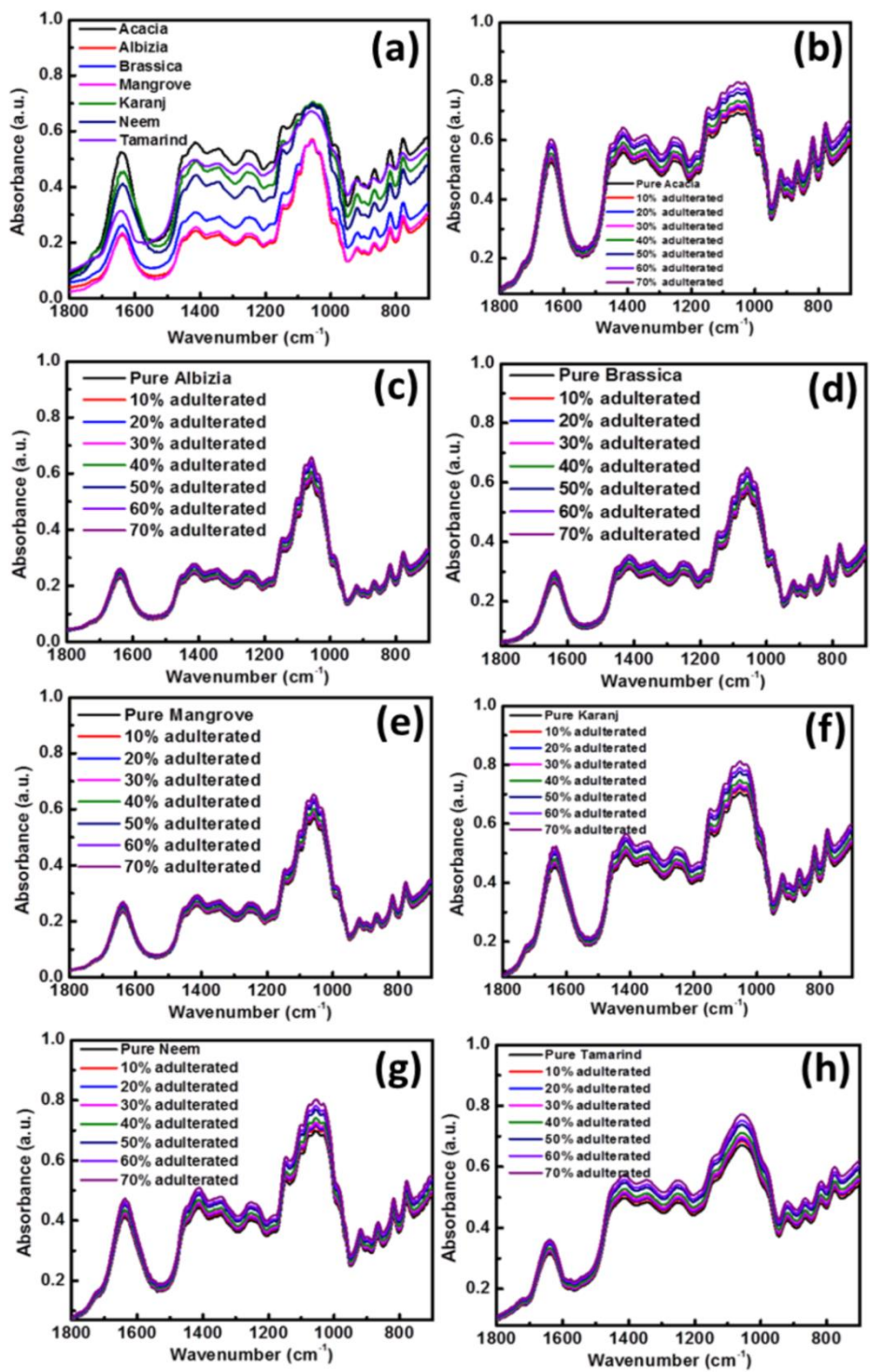

**Fig. 4.5**: (a). Absorbance plot for all of the pure honey samples in the spectral region 700–1800 $cm^{-1}$ and, (b)–(h) Absorbance plots of sucrose adulterated (from 0% to 70%) acacia, albizia, brassica, mangrove, karanj, neem and tamarind samples.

Each of the **Figs. 4.5(b)** to (**h**) depict the changes in absorbance values of acacia, albizia, brassica, mangrove, karanj, neem and tamarind samples with respect to the increase of adulterant. Since, honey contains almost 70-80% saccharides in different forms and, since, the adulterated sample is also pure SS, therefore, the MIR spectra will be mainly dominated by saccharide and water absorptions. Several modifications in the peaks and troughs were observed in the fingerprint region. Previous studies report the presence of absorption peaks originating primarily due to glucose, fructose and sucrose [**230**]. The presence of the lone peak in the spectral region between 1700 cm$^{-1}$ and 1600 cm$^{-1}$ is originating from the O-H bending of water, peaks between 1470 cm$^{-1}$ and 1150 cm$^{-1}$ represent the bending of C-C-H, C-O-H, and O-C-H groups and those within 1060 cm$^{-1}$ and 1020 cm$^{-1}$ are attributed to C-O, C-C and O-H vibrations from carbohydrates [**229**]. The region 950 cm$^{-1}$-700 cm$^{-1}$ is known as the "anomeric region of carbohydrates" consisting of two vibrations of C-O-C asymmetric stretching corresponding to α and β anomers near 920 cm$^{-1}$ [**240**]. Also, this region experiences strong C-H and C=C bendings due to the presence of carbohydrates in honey.

### 4.4.2.2 Variation of FWHM with adulterant content

In this context, FWHM of the peak at 1056 cm$^{-1}$ for all the pure and adulterated honey samples is estimated. Since FWHM of a material depicts structural distribution of the atoms inside a molecule, hence it will definitely change with the addition of similar compounds into the base material. **Fig. 4.6** shows the variation in FWHM values due to percentage change in sucrose adulteration for all the seven varieties of honey considered in the current work.

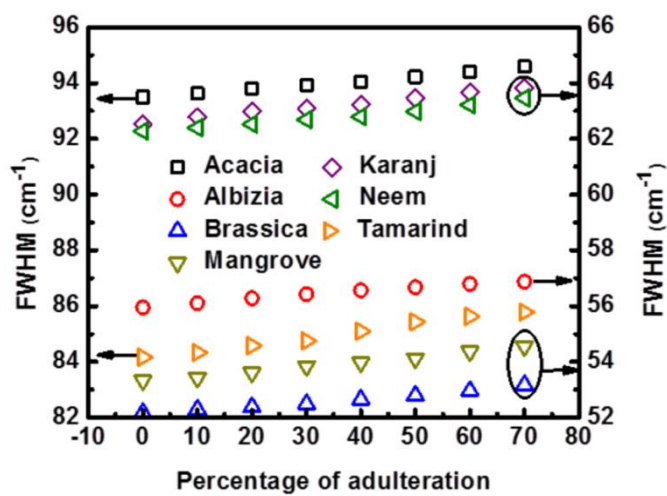

**Fig. 4.6**: Plots of the variation of FWHM values with percentages of sucrose adulteration for all the samples considered.

From the plots, it is apparent that FWHM of the peak at 1056 cm$^{-1}$ relevant to C-O, C-C and O-H stretching increases with varying adulterant concentrations. Similar nature is observed for each of the adulterated honey samples. FWHM signifies the width of a peak and its broadening is caused by a large distribution in the specific bond angles (θ). Distribution of such bonds can be investigated following the bond angle distribution function, given by [**241**],

$$\gamma(\theta) = 16\pi^2 \iint_0^R r_1^2 r_2^2 f(r_1) f(r_2) f(r_1, r_2, \theta) dr_1 dr_2 \qquad (2.1)$$

where, R is the maximum distance between two nearest neighboring atoms, $r_1$ and $r_2$ represent the function is apparently a radial average over the triplet correlation function *f(r$_1$,r$_2$,θ)*. Addition of sucrose molecules as adulterant lengthens the distance between the neighboring atoms from the central atom and thereby increases the bond angle distribution function [**241**]. Such increment consequently enhances the peak width which results to the broadening of FWHM.

As discussed earlier, effective polarization of the system under equilibrium alters with the varying adulterant content. Addition of SS as an adulterant perturbs the equilibrium condition by increasing the randomness of the system which consequently rises the entropy and hence the internal energy of the system. FTIR analysis also confirms the increment of system energy upon adding adulterant which corroborates the physical observation from the dielectric study of the system with/without any adulterant content.

## 4.5 Frequency dependent co-efficient of sensitivity analysis

For the adulterant concentration ≥ 10%, the present measuring system can be used as a sensor for assessing quality of honey and the relevant coefficient of sensitivity ($\beta$) can be determined by the following expression:

$$\beta = \frac{\sum(w - \langle w \rangle)(G - \langle G \rangle)}{\sum(w - \langle w \rangle)^2} \qquad (2.2)$$

where, w and $G$ denote the adulterant concentration (in wt %) and conductance of the system respectively for a given SS concentration; <$G$> and <w> represent the average values of conductance and adulterant concentration for a particular variety of honey. **Fig. 4.7** shows the value of co-efficient of sensitivity extracted for individual honey samples in terms of the system conductance.

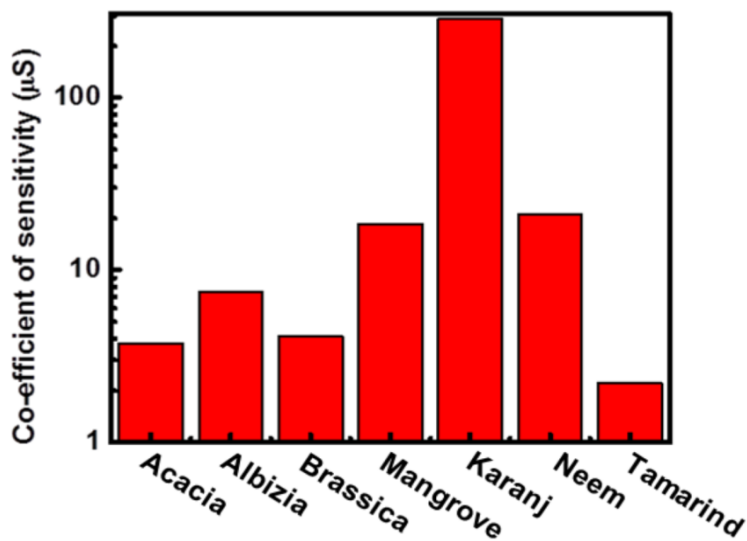

**Fig. 4.7**: Plots of the co-efficient of sensitivity for seven different florals of honey.

From **Fig. 4.7**, it is clearly observed that the value of co-efficient of sensitivity for individual varieties of honey differ from each other in terms of the measured conductance. The value of sensitivity for karaj honey is observed to be much superior in comparison to acacia, brassica and tamarind samples. This deviation is attributed to the change in measured conductance values due to the linear increment of sugar content in honey.

## 4.6 Physical corroboration of the EIS and optical observations

As discussed earlier, the effective polarization of the system under equilibrium alters with the varying adulterant content. Addition of SS as an adulterant perturbs the equilibrium condition by increasing randomness of the system which consequently rises the entropy and hence the internal energy of the system. FTIR analysis also confirms the increment of system energy upon adding adulterant which corroborates the physical observation from the dielectric study of the system with/without any adulterant content.

## 4.7 Summary

To summarize, EIS analysis and characterization of the adulterated honey samples from different floral species supported by FT-MIR study reveals that electrical and optical parameters investigated here significantly depend both on the varying floral species and varying adulterant content. EIS analysis has indicated certain variation in the measured electrical parameters for the different honey varieties. Also, from this study, it is observed that the nature of variation of impedance, capacitance and conductance with percentage change of the adulteration is linear. The results are also validated by analyzing their respective I-V characteristics which show linear decrease in net current values with the addition of SS and EIS and I-V results indicate that the system is enable to sense the presence of 10% SS content as threshold and successfully continues up to 70% adulteration. Simultaneously, FT-MIR study confirms the change in absorbance values for individual varieties of honey. Also, this study indicates the increase in absorbance values due to the increase in adulterant content. FWHM is calculated for the spectral peak at 1056 cm$^{-1}$ due to C-O, C-C and O-H stretching which tends to increase with the increase of adulterant content. Finally, electrical co-efficient of sensitivity is extracted for individual varieties of honey in terms of the measured EIS parameters.

The work conducted in this chapter provides a comprehensive understanding on the detection and quantification of exact sucrose content in honey by employing EIS and FTMIR techniques which confirms the variation of certain electrical and optical parameters due to the change in adulterant content. The current study also reveals that EIS can be used as an efficient tool for rapid, cost-effective and accurate quantification of adulteration in different other floral species of honey. Such technique can also be used for detecting and quantifying the presence of adulterants in other food products.

# Chapter 5

# Estimation of metanil yellow in turmeric by capacitive measurements

## 5.1 Introduction

The previous chapter deals with the sensing and detection of sucrose content in different floral species of honey by employing EIS and FT-MIR techniques. This chapter will highlight the process of precise estimation of metanil yellow as an adulterant in turmeric powder samples by using EIS, FT-MIR and UV-Vis spectroscopy techniques. This study also focusses on the chemical ion transport phenomenon that occurs inside the system due to the interactions among metanil yellow, DI water and turmeric molecules. The addition of metanil yellow in turmeric powder changes the measured electrical and optical parameters which vary with the percent adulterant content and has been described in detail in the present chapter.

Turmeric (Curcuma Longa L., family Zingiberaceae) is a commonly used bright yellow coloured spice extracted from the rhizomes of Curcuma longa. It exists in the form of a medicinal plant and is generally marketed as fresh turmeric powder, dry turmeric rhizomes, turmeric oleoresin and turmeric oil in the spice industries all over the world. The yellow colour of turmeric appears from its principal curcuminoid known as curcumin ((1E,6E)-1,7-Bis(4-hydroxy-3-methoxyphenyl)hepta-1,6-diene-3,5-dione) which acts as a marker for turmeric quality. Turmeric rhizomes contain an anti-inflammatory, anti-bacterial, anti-fungal, anti-viral and anti-cancerous property which is attributed mainly to the presence of curcumin [242-244]. Curcumin extract from turmeric has also been used to improve behavioral symptoms in patients suffering from alzheimer's disease [245].

The fresh and dried turmeric rhizomes are generally free from contamination, however, turmeric in powder form is often adulterated by various low-cost chemicals and dyes for monetary profit [246]. Few studies have reported about the issue of turmeric adulteration by curcuma zedoaria, a branch species of turmeric with similar properties of turmeric itself [246,247]. In this context, the most detrimental issue is its adulteration by non-permitted food colour MY ($C_{18}H_{14}N_3NaO_3S$, C.I. Acid Yellow 36 (13065)). MY is an acidic azo (N=N) dye comprising of a sodium salt of [m-(p-anilinophenyl) azo] benzene sulfonic acid and has been toxicologically classified under the category CII by the joint Food and Agriculture Organization of the United Nations/ World Health Organization (FAO/WHO) expert committee on food additives [**46**]. Several reports of study on rats demonstrate that effective consumption of MY for a longer period causes tumor growth [**248**], neurotoxicity [**249**], hepatocellular carcinoma [**250**], lymphocytic leukemia [**251**] and other similar chronic diseases. In this perspective, authentication of pure turmeric samples from the adulterated ones should be a major issue for its quality control. Therefore, to address such an issue, rapid and precise detection of percentage of adulteration in turmeric is necessary.

Globally, different techniques have already been used for quantitative estimation of various adulterants in turmeric powder. Sasikumar, Syamkumar, Remya, and John Zachariah, [**246**] reported the implementation of PCR based detection of adulteration in turmeric samples. Thin layer chromatography has been implemented by Sen, Gupta and Dastidar, [**247**] to detect the presence of curcuma zedoaria and curcuma aromatic in turmeric samples. Nallappan, Dash, Ray and Pesala, [**252**] reported about the identification of yellow chalk powder mixed in turmeric powder by implementing terahertz spectroscopy. Also, Sequence Characterized Amplified Region (SCAR) markers have been used by Dhanya, Syamkumar, Siju, and Sasikumar, [**253**] to detect the presence of Curcuma zedoaria and Curcuma malabarica in graded turmeric powder. Another approach to detect adulteration in turmeric by employing Laser Induced Breakdown Spectroscopy (LIBS) is performed by Tiwari, Agrawal, Pathak, Rai and Rai, [**254**]. Furthermore, conventional techniques such as FT-IR and FT-Raman spectroscopy have been used by Dhakal, Chao, Schmidt, Qin, Kim and Chan, [**255**] to evaluate the MY adulterated turmeric powder. However, most of such analytical methods use complex techniques, costly instruments and have prolong processing time which limits the requirement for rapid, low-cost and point-of-care applications. In this context, EIS is an emerging technique, ideal for the detection of adulterants in different foods and beverages in a much rapid, robust, simpler and cost-effective manner [**13-19,21,208**]. Apart from EIS, several optical spectroscopy techniques such as NIR, FT-IR and UV-Vis are being used to detect and quantify several toxic and carcinogenic adulterants in foods [**45,48,49**]. Therefore, the real-time implementation of EIS technique to detect the presence of various cheaper chemicals and coloured dyes in turmeric powder would be very beneficial. However, no such research or study, to the best of the author's knowledge, has been conducted on the detection and quantification of MY in turmeric powder by implementing EIS technique.

Therefore, in this context, a detailed study and analysis has been carried out by employing EIS technique, supported by optical spectroscopy techniques such as FT-MIR and UV-Vis spectroscopy on pure turmeric and its adulterated solutions. Variation of electrically measured parameters such as impedance, capacitance, current and absorbance (optical parameter) for all the measured samples has been thoroughly studied and analyzed in this work. Therefore, the primary focus of this research work is to implement EIS for systematic detection and quantification of percent weights of MY in turmeric powder. The acquired results from EIS study are corroborated with those obtained from the UV-Vis and FT-MIR studies to verify the accuracy and ability of EIS technique that can further be used for detection and quantification of various toxic and carcinogenic adulterants in foods, beverages and other bio-consumables.

## 5.2 Materials and methods

### 5.2.1 Sample procurement

Turmeric powder used in the present study is collected from dried turmeric rhizomes, powdered by using pestle and mortar. The obtained turmeric powder is then stored in a vacuum desiccator to avoid any moisture contact. MY (Sigma Aldrich, India) samples were also stored in a vacuum desiccator before being used for experimental purpose. DI water used in the present study is obtained from Milipore$^{TM}$ (Merck, India).

### 5.2.2 Preparation of metanil yellow-turmeric powder mixture

For the preparation of turmeric-metanil yellow powder mixture, the individual weights of turmeric powder, MY, percentage of adulteration and the level of adulteration are summarized in **Table 5.1**. The content of DI water is varied in accordance with the variety of electrical and optical measurements performed. The weights of turmeric powder and MY were measured with the help of an electronic balance (Mettler Toledo, AL 204) having a precision of ±0. 1 mg.

**Table 5.1**: Summary of the concentration of adulterated turmeric solutions with their level of adulteration.

| Weight of pure turmeric (mg) | Content of metanil yellow in solution (mg) | Percentage of adulteration (w.r.t pure turmeric) | Level of Adulteration |
|---|---|---|---|
| | 0 | 0 | Nil |
| | 1 | 5 | Low |
| | 2 | 10 | Low |
| 20 | 4 | 20 | Medium |
| | 6 | 30 | Medium |
| | 8 | 40 | High |
| | 10 | 50 | High |

## 5.3 Experimental details

### 5.3.1 Equivalent circuit of the measurement setup

**Fig. 5.1(a)** shows schematic representation of the conductivity cell dipped inside DI water-turmeric-MY solution, connected to an LCR meter or a Source meter at room temperature. The arrows represent constituent dipoles of the individual components present inside the system. **Fig. 5.1(b)** shows the equivalent circuit of the measurement setup used in the present work.

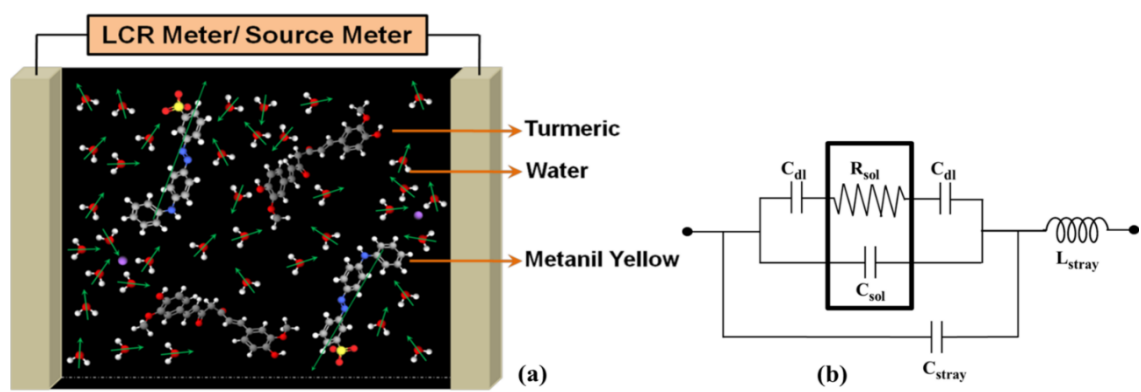

**Fig. 5.1**: (a) Schematic representation of the experimental system (**Red:** Oxygen**, Blue:** Nitrogen**, White:** Hydrogen**, Grey:** Carbon**, Yellow:** Sulfur**, Violet:** Sodium) and (b) electrical equivalent circuit of the measurement setup with dashed area depicting the inherent property of the sample solution.

### 5.3.2 UV-Vis spectroscopy measurement

The UV-Vis spectroscopic measurements of pure and adulterated turmeric solutions were performed in the spectral region between 250 nm to 550 nm with a scan speed of 480 nm/min and the data acquisition interval was limited to 1 nm. Primarily, 2.5 ml of DI water was taken in a quartz cuvette of path length 10 mm to record the reference spectrum. Afterwards, 20 mg of pure turmeric powder was dispersed in 250 ml DI water from which 2.5 ml was used for the measurement and spectral data were noted simultaneously. This process was repeated for all the varied weights of MY added in the turmeric-DI water solution.

### 5.3.3 FT-MIR measurement

FT-MIR measurements were performed in the spectral range of 1800 cm$^{-1}$ to 700 cm$^{-1}$ with a minimal resolution of 4 cm$^{-1}$. Complete sixty four scans were performed for each spectrum and the corresponding data were simultaneously recorded. Initially, Potassium Bromide (KBr, IR grade) powder (~100 mg) was grinded by a pestle and mortar before being hard-pressed into a thin transparent pellet by using a hydraulic pellet press. Prior to all of the measurements, the KBr pellet was used to acquire the background/reference spectrum. Then, 2 mg of pure turmeric powder was included into the mortar along with the KBr and cold-pressed again into a pellet and the resultant spectrum was measured. Afterwards, different weights of MY maintained in the same ratio (as in **Table 5.1**) for 5% to 50% adulteration were successively added into the mixture and the resultant spectra were noted.

### 5.3.4 EIS measurement

For EIS measurement, an electrical signal of variable frequency is applied to a pair of metal electrodes which are directly immersed into the solution under investigation. Both the input and resultant signals are internally processed to provide the frequency-dependent net impedance (Z) of the system [20]. The real part of impedance provides the sample resistance, whereas the imaginary part consists of capacitive/inductive reactance. The required spectrum is generally achieved by measuring the frequency dependent impedance, capacitance, and conductance. Such parameters primarily depend on the experimental setup as well as on the variation of sample compositions. In this work, EIS measurements for all the pure and adulterated turmeric samples have been performed within the frequency range of 100 Hz–5 MHz and the voltage amplitude is maintained to be 100 mV peak-to-peak, respectively. The TEGAM 3550 was interfaced with a desktop computer through Lab-View software for rapid turn out of the output data. Also, the LCR meter used in this work possesses internal noise reduction techniques and has been standardized with open-short compensation technique before performing the measurements.

### 5.3.5 I-V measurement

I–V measurements of pure and adulterated turmeric samples were performed within a voltage range of − 5 to + 5 V. To observe their relevant I–V characteristics, a time varying DC voltage sweep with a sweep rate of 0.1 V/s has been applied to the samples under test. Subsequently, the current values for all the samples are extracted and plotted for five different voltages of 1 V, 2 V, 3 V, 4 V, and 5 V, respectively.

## 5.4 Results and Discussions

### 5.4.1 Study of electrical properties of the metanil yellow (MY) adulterated turmeric

### 5.4.1.1 EIS analysis

In this study, the effect of MY on the electrical properties of pure turmeric at room temperature has been investigated. The frequency for electrical measurements is chosen to be 1 KHz and seven set of measurements from pure (0% adulterated) to 50% adulterated turmeric samples were performed for both EIS and I-V measurements.

### 5.4.1.2 Understanding the ion transfer scheme during the process of turmeric adulteration from EIS study

The systematic variation of impedance, capacitance and conductance with the change in MY content in pure turmeric solution for two frequencies of 1 kHz and 10 kHz is demonstrated in **Fig. 5.2(a)** and **(b)**, respectively.

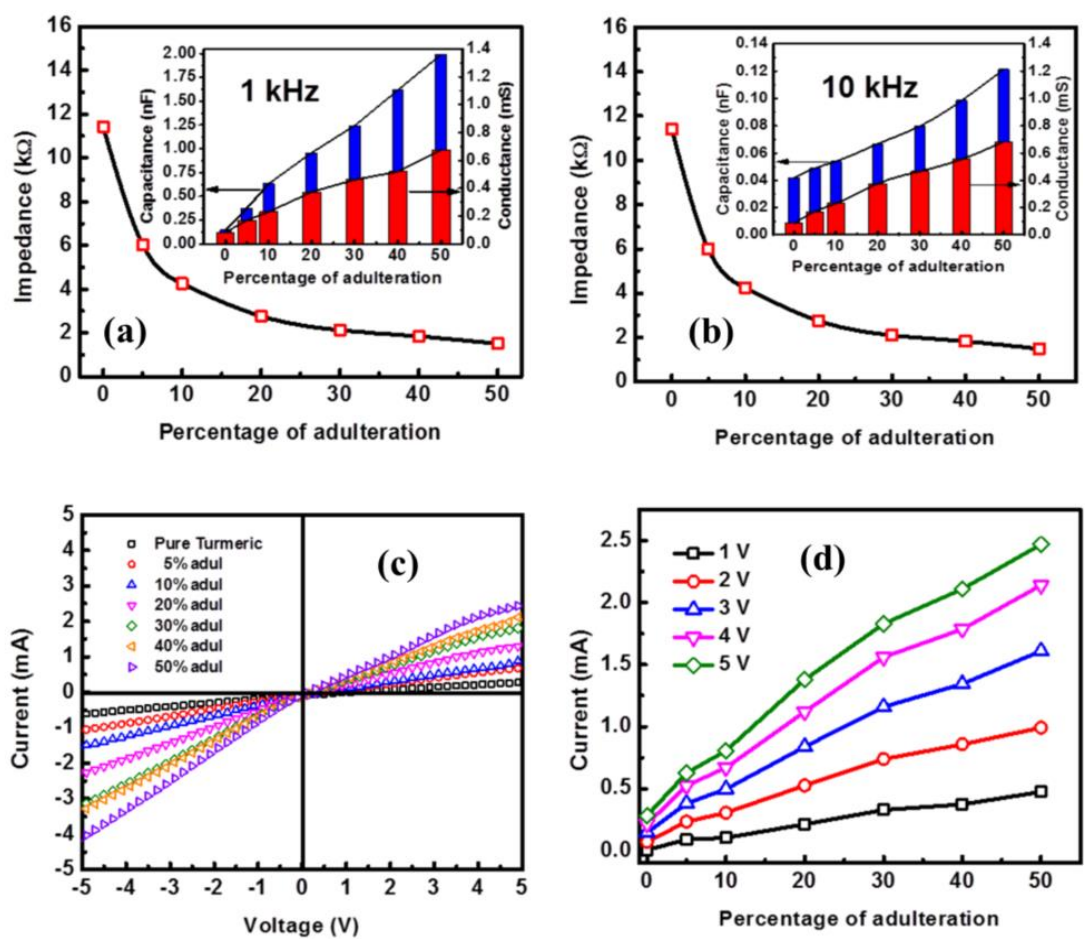

**Fig. 5.2**: Plots of (a) variation of impedance, capacitance and conductance (at inset) for different percentages of turmeric adulteration at 1 kHz and (b) at 10 kHz frequency, (c) current with applied voltage sweep (-5 V to +5 V) for different percentages of adulteration and, (d) current versus percentage of adulteration at 1 V, 2 V, 3 V, 4 V and 5 V, respectively.

It is apparent from the inset of **Fig. 5.2(a)** and **(b)** that the net capacitance and conductance increases with the increase of MY content in the adulterated turmeric solution. Also, the impedance plots suggest a decreasing trend with increasing percentage of adulteration.

Turmeric is partially soluble in DI water and therefore will result in maximum dispersion when added to it. Under this situation, very few ions are present inside the system which will contribute to the net system impedance or conductance directly. This can be realized from **Fig. 5.2(a)** and **(b)** that, at 0% adulteration (pure turmeric), the impedance is maximum and decreases with the addition of higher percentage weights of MY. Also, the plot suggests the conductance to be minimum for pure turmeric when compared with the other higher percentages of adulteration. MY, being a sodium salt of water soluble acidic azo dye, will get completely dissociated into anionic dye ions (X-SO₃⁻) and cations (Na⁺) in its aqueous solution [**256**] as shown in Eqn. 3.1. Here, X represents the portion of MY compound excluding the SO₃⁻ group. Thus, for every such increment in MY content, more number of Na⁺ and X-SO₃⁻ ions will get distributed inside the system. However, X-SO₃⁻ is a strong conjugate base of X-SO₃Na, and therefore, most of these ions will interact with the mobile hydrogen (H⁺) ions in the system and will form weak acid X-SO₃H as shown in Eqn. (3.3).

$$(X - SO_3Na) \rightarrow (X - SO_3^-) + Na^+ (MY\ dissociation) \tag{3.1}$$

$$H_2O \rightarrow H^+ + OH^- (water\ dissociation) \tag{3.2}$$

$$XSO_3^- + H^+ \rightarrow XSO_3H (weak\ acid) \tag{3.3}$$

At relatively lower MY concentrations, these Na+ ions will be surrounded by thick layer of hydration shells and may form water-separated ion pairs. However, with the increase of MY concentration, the newly appeared Na⁺ ions will interact with the outer layer hydration shells that were already surrounding a Na⁺ ion. This will result in the sharing of hydration shells among the total Na⁺ ions inside the system.

### 5.4.1.3 Gradual change of effective dipole moment with adulterant content

The steady increment of system capacitance due to the increase of MY content is attributed to the changes in effective dipole moment of the system due to continuous variation of their dipole orientations. DI water is pure polar in nature whereas turmeric is a non-polar compound having no permanent dipole moment. The interaction of non-polar molecules inside the turmeric leads to redistribute their electron density and thereby initiates the formation of instantaneous dipoles [**257,258**]. Thus, the dielectric behavior of a turmeric-DI water dispersed system will be primarily governed by such instantaneous dipoles. However, MY, like most other azo dyes, have a high permanent ionic dipole moment due to the arrangement of its constituent atoms, leading to an uneven distribution of electrons around the entire molecule [**259,260,261**]. Moreover, the ionic dipole moment is much higher than the polar and instantaneous dipole moments [**17**]. Therefore, when MY is added to the turmeric-dispersed DI water, then, the effective dielectric nature of the system will be dominated by it. Gradual addition of MY increases the effective dipole moment which in turn increases the effective polarization of the system and thus results to an increment in the effective permittivity and the capacitance of the system.

### 5.4.1.4 Current-voltage characteristics

The variation of current flowing through pure and adulterated turmeric solutions for a voltage range of -5 V to +5 V is shown in **Fig. 5.2(c)**. It is observed from the plot that, with the increase of voltage from -5 V to +5 V, the net current inside the solution increases. As described earlier, incorporation of MY in the turmeric-DI water solution results in the formation of $H^+$, $OH^-$, $X$-$SO_3^-$ and $Na^+$ ions. These ions contribute to the process of charge transfer, leading to current generation inside the solution. For a lower concentration of MY, total ionic concentration inside the solution will be low. With successive addition of MY in controlled wt%, the number of such ions in the system increases considerably, resulting in an overall increase of net current. This observation is confirmed by plotting the current with percentage of adulteration at five different voltages 1 V, 2 V, 3 V, 4 V and 5 V as depicted in **Fig. 5.2(d)**. The current is observed to increase with the increasing content of MY in the solution for all the voltages considered in this work which is attributed to the continuous generation of such ions inside the solution due to the gradual increase of MY content. For the validation of such EIS data, the plots of Re(Z) and |Im(Z)| for pure and adulterated turmeric solutions in the frequency range of 1 kHz to 5 MHz have been depicted in **Fig. 5.3(a)** and **(b)**.

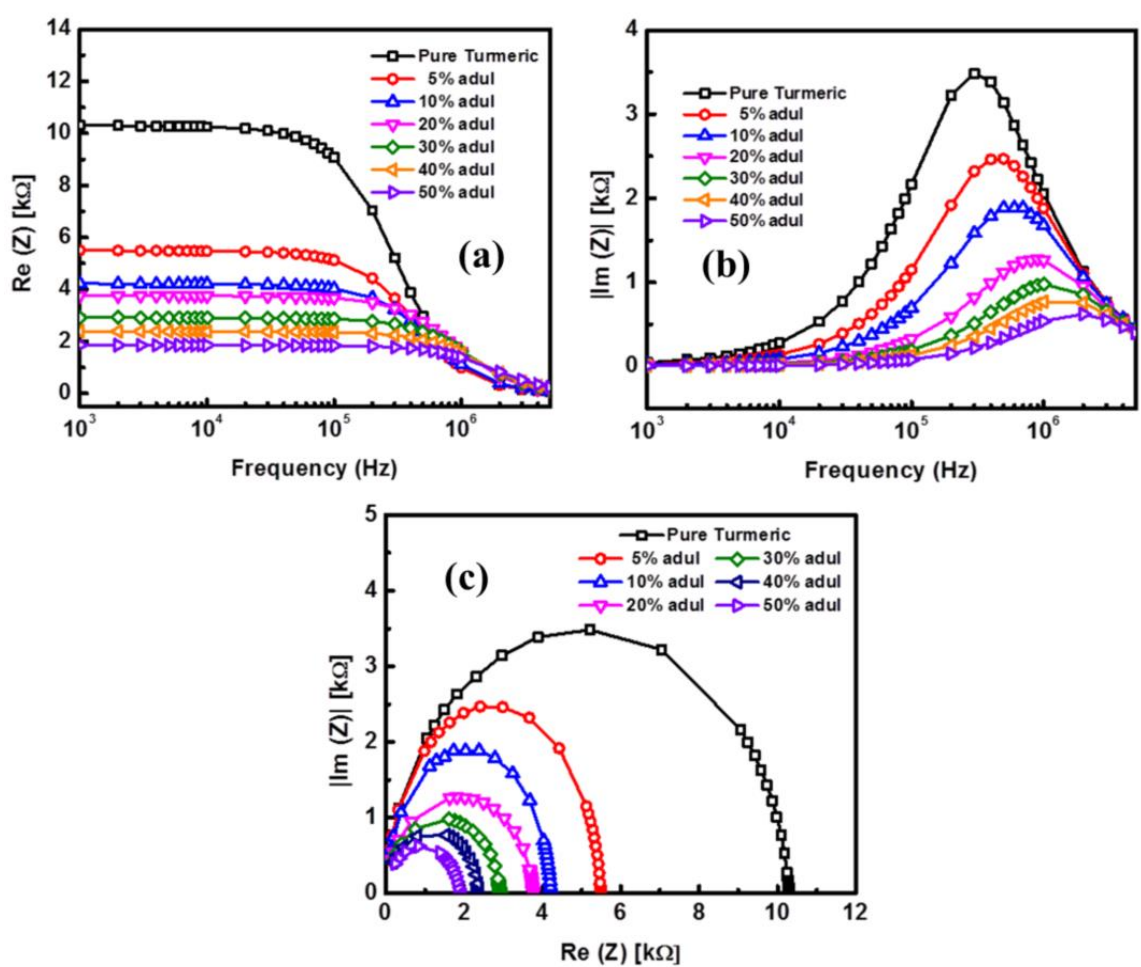

**Fig. 5.3**: Plots of: (a) Re(Z) and (b) |Im(Z)| versus frequency (1 kHz to 5 MHz) and, (c) |Im(Z)| versus Re(Z) for the pure and MY adulterated turmeric samples.

Both Re(Z) and |Im(Z)| are observed to decrease with the increase of adulterant content [**262,263**] in the solution. Also, a relevant nyquist diagram (|Im(Z)| versus Re(Z)) within the frequency range of 1 kHz to 5 MHz is plotted in **Fig. 5.3(c)** which clearly distinguishes the samples with different adulterant content.

### 5.4.2 Effect of metanil yellow on the optical properties of turmeric

In this work, the impact of incorporation of MY in aqueous turmeric solutions using both the UV-Vis and FT-MIR spectroscopy has been studied. The variation of absorbance for different adulterant concentration is measured for both the techniques and the results are analyzed in terms of the relative change of absorbance values.

### 5.4.2.1 UV-Vis absorbance study

The UV-Vis absorbance spectrum of pure and adulterated turmeric powder (0% to 50%) within the wavelength range of 250 nm–550 nm is shown in **Fig. 5.4(a)** and the comparative absorbance spectra of pure turmeric powder and MY in the wavelength range of 350 nm–550 nm are plotted in **Fig. 5.4(b)**, respectively. The variation of absorbance with increase in MY content inside the solution relevant to the absorbance peak at 442 nm is shown in **Fig. 5.4(c)** and **Fig. 5.4(d)** illustrates the FT-MIR spectra of pure and adulterated turmeric powder which shows a definite increase of absorbance with percentage addition of MY content.

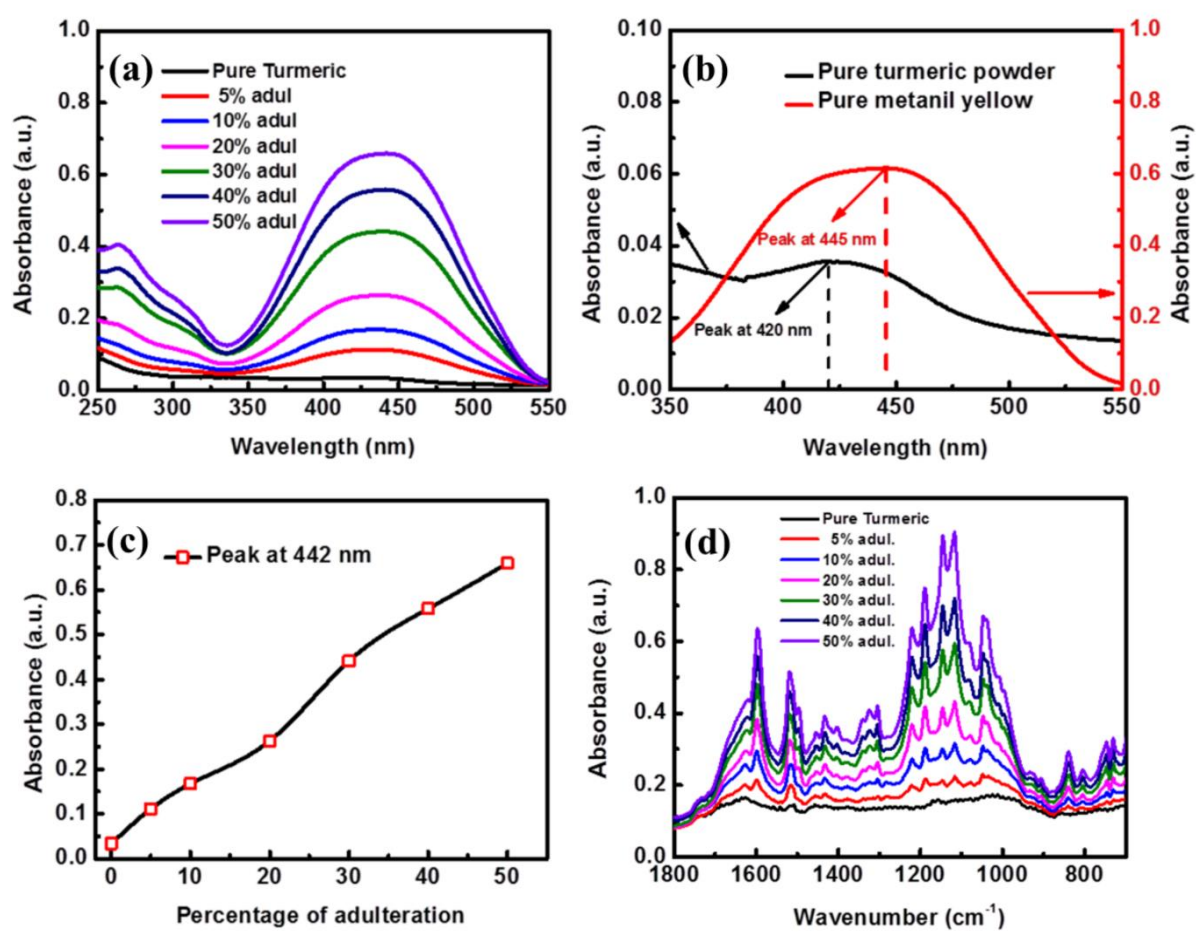

**Fig. 5.4**: (a) Plots of UV-Vis absorbance with wavelength for pure and adulterated turmeric solutions, (b) comparative absorbance plots for pure turmeric and pure MY solutions, (c) plot of the variation of absorbance with adulterant concentration at 442 nm and (d) FT-MIR absorbance plots for pure and adulterated turmeric samples in the spectral region 1800 cm$^{-1}$ to 700 cm$^{-1}$.

According to Beer-Lambert's law, more molecules are likely to interact with the incident light when concentration of a solution increases. With the increase of adulterant concentration, higher number of adulterant molecules will be present in the solution and therefore more light will be absorbed which will lead to the increase of overall system energy. It is also apparent from **Fig. 5.4(a)** that the absorbance of adulterated turmeric solution increases gradually with the increase of MY content in it. Therefore, these spectra justify the criteria for Beer-Lambert's law and highlight the adulteration sensing range [**264**]. **Fig. 5.4(b)** shows two definite absorbance peaks at 420 nm for pure turmeric [**265**] and at 445 nm [**47**] for pure metanil yellow. Therefore, with gradual increase of MY content in the turmeric dispersed DI-water solution, peak-shifting occurs from 420 nm (0% MY) to 442 nm (50% MY). The variation of absorbance with percentage of adulterant content in the solution relevant to the absorbance peak at 442 nm is plotted in **Fig. 5.4(c)** which shows an increasing trend. As discussed, the absorbance of a solution is directly proportional to the concentration of absorbing species inside the solution and the path length, as stated in Beer-Lambert's law. So, for a fixed path length, the absorbance will increase if the concentration of the absorbing species increases. In this context, the variation of absorbance also shows an incremental trend which is attributed to the gradual addition of MY in the solution.

### 5.4.2.2 Study of the FT-MIR spectral bands

**Fig. 5.4(d)** shows the FT-MIR spectra of pure and MY-adulterated turmeric powder in the spectral region between 1800 $cm^{-1}$ and 700 $cm^{-1}$. This spectral region is termed "fingerprint region" since it illustrates the complex chain of absorptions, which arises due to certain bond vibrations, bending and stretching of atoms in a compound. This fingerprint region helps to identify individual chemical compounds as it shows different absorbance peaks and troughs at distinct wavenumbers for different materials. Therefore, this region is closely examined to investigate the adulteration induced modifications in the spectra. The spectra of **Fig. 5.4(d)** suggest the increase of absorbance values with increasing adulteration percentage. **Table 5.2** summarises the different modes of stretching, bending and vibrations of pure and MY adulterated turmeric in the IR frequency range.

**Table 5.2**: Summary of the FT-IR spectral bands and their assignments for pure and adulterated turmeric powder.

| Pure turmeric | | MY adulterated turmeric | |
|---|---|---|---|
| IR (cm$^{-1}$) | Assignment | IR (cm$^{-1}$) | Assignment |
| 1739 | C=O stretching | 1597 | N=N stretching |
| 1683 | C=O stretching | | |
| 1626 | C=C stretching | | |
| 1508 | C–H bending | 1519 | C–C stretching |
| | | 1496 | C–C stretching |
| 1431 | C–H bending | 1455 | N=N stretching |
| 1376 | C–H bending in | 1403 | S=O stretching |
| | O=C–CH$_2$–C=O | | |
| 1287 | COH + CO(CH$_3$) stretching | 1304 | C–C stretching |
| 1207 | C–O–(CH$_3$) stretching | 1220 | N-H bending |
| 1181 | CH$_3$ deformation | 1188 | C–H bending |
| 1157 | C–H bending | 1145 | C–Nazo stretching, C–H bending |
| | | 1116 | C–H bending |
| 1130 | O-C-C stretching | 1048 | SO$_3$ symmetrical stretching |
| 1037 | C-O stretching | 1036 | SO$_3$ symmetrical stretching |
| 1023 | C-O stretching | | |
| | | 997 | Ring breathing |
| 920 | C–H bending | | |
| | | 906 | C–H wagging |
| 855 | C–H bending | 840 | C–H bending |
| 814 | C–H bending | 804 | C–H bending |

The FT-IR spectrum of pure turmeric powder reveals distinct confirmatory peaks of turmeric at 814 cm$^{-1}$, 855 cm$^{-1}$ and 920 cm$^{-1}$, 1181 cm$^{-1}$, 1376 cm$^{-1}$, 1626 cm$^{-1}$ 1683 cm$^{-1}$ and 1739 cm$^{-1}$ respectively. Absorbance peaks of turmeric below 950 cm$^{-1}$ reflects the C-H bending vibrational modes on aromatic sites which is different than that in MY. Similarly, IR peaks at 1181 cm$^{-1}$ and 1376 cm$^{-1}$ correspond to methyl deformation and C-H bending at O=C–CH$_2$–

C=O respectively which also confirms the presence of turmeric. Also, turmeric has two peaks for C=O vibrational modes at 1683 cm$^{-1}$ and 1739 cm$^{-1}$ respectively and a similar C=O peak at 1626 cm$^{-1}$ which is attributed to the presence of curcumin [**255**]. Now, with the incorporation of percentage weights of MY from 5% to 50% into pure turmeric-DI water solution, different absorbance peaks have been observed at 804 cm$^{-1}$, 840 cm$^{-1}$, 906 cm$^{-1}$, 997 cm$^{-1}$, 1048 cm$^{-1}$, 1145 cm$^{-1}$, 1188 cm$^{-1}$, 1220 cm$^{-1}$, 1403 cm$^{-1}$, 1455 cm$^{-1}$ and 1597 cm$^{-1}$ respectively [**255**]. The origin of such new distinct sharp characteristic peaks is attributed to the stretching, bending and vibrations of its different components present in the solution. With gradual increment of MY in the system, more such interactions will occur which will lead to the increase of absorbance peak intensity. The exact identification and confirmation of the presence of MY in the solution is indicated by distinct sharp vibrational modes at the N=N site (1597 cm$^{-1}$, 1145 cm$^{-1}$) and peaks occurring due to ring breathing (997 cm$^{-1}$) [**255**] and sulfate (1048 cm$^{-1}$) [**266**]. Alternatively, MY lacks several vibrational modes between 1740 cm$^{-1}$ and 1628 cm$^{-1}$ due to the absence of conjugated carbonyl groups which exists only in turmeric components [**255**].

## 5.5 Physical correlation of the electrical and optical properties

**Fig. 5.5(a)** and (**b**) shows the linear regression plots of the variation of capacitance at 1 kHz and 10 kHz (from EIS). **Fig. 5.5(c)** and (**d**) shows the same for current at 5 V (from I-V) and optical absorbance (from UV-Vis) values for pure as well as adulterated turmeric solutions.

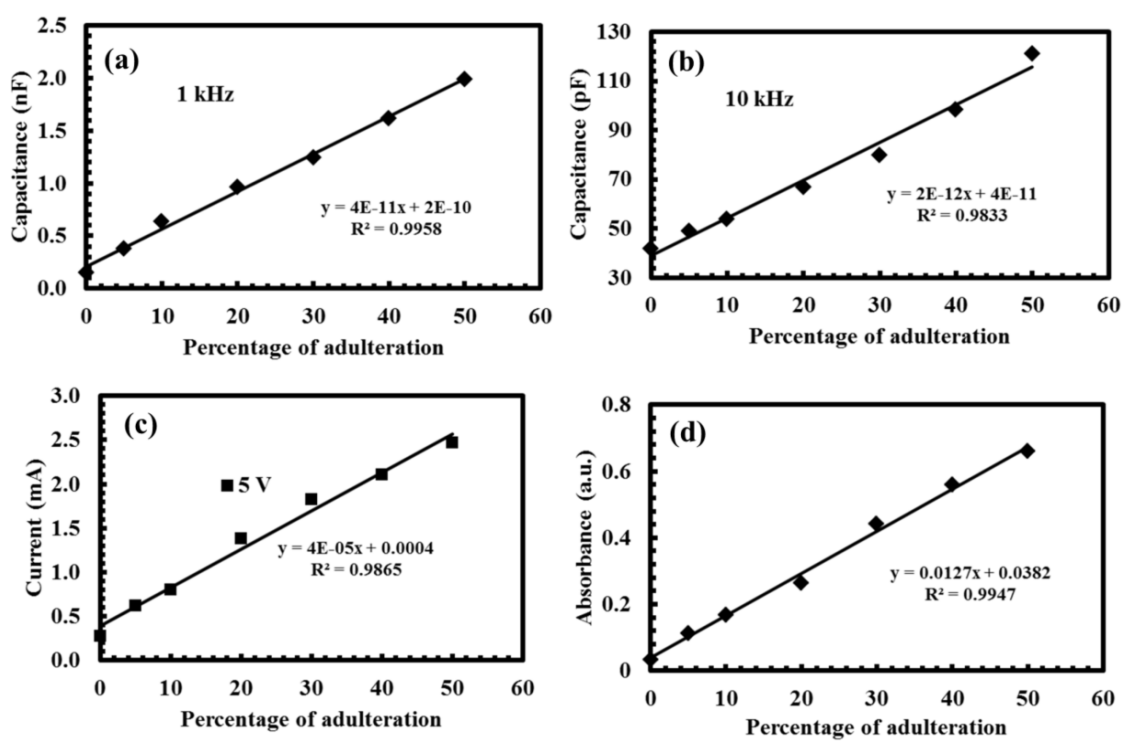

**Fig. 5.5**: Linear regression plots of (a) capacitance at 1 kHz (b) capacitance at 10 kHz, (c) current at DC 5 V and (d) absorbance at 442 nm for pure and different percentages of adulteration.

The values of coefficient of determination ($R^2$) are extracted to be > 0.9 for all of the electrical and optical measurements performed, which indicates superior sensitivity of the system to quantify the amount of adulteration. It is also worthy to mention that the repeatability errors for capacitance, current and absorbance in terms of percentage of adulteration are calculated to be 1.98% (~2%), 3.51% and 2.48% (~2.5%) respectively, for three sets of measurements of the parameters.

Now, previous discussion suggests that, inclusion of turmeric in DI water initiates the formation of instantaneous dipoles, which increases the randomness inside the system and thereby increases the effective polarization. These effects gradually get enhanced with the addition of percent weights of MY in the system. Accordingly, system entropy and therefore the internal energy of the system get increased. Alternatively, UV-Vis and FT-MIR spectral analyses also validate the increase of absorbance with compositional variation of MY and therefore indicate an increase of the total absorbed energy of the system which also supports the observations from EIS and I-V studies of the system.

## 5.6 Summary

Electrical impedance spectroscopic measurements have been performed on the adulterated turmeric samples which are also characterized by employing UV-Vis and FT-MIR spectroscopy techniques. All the measurements are observed to be very sensitive to detect and quantify the presence of MY in the solution. EIS study exhibited a steady variation of the system impedance, capacitance and conductance values with increasing adulterant percentage in the solution. I-V measurement showed the increase of current values with the gradual increase of MY in the solution. Analysis of both the EIS and I-V measurements indicate that the approach can be used to sense the presence of ≥5% MY in turmeric powder and it can be effectively applicable till the adulterant content is within 50%. Alternatively, UV-Vis and FT-MIR studies also confirm a systematic increase of absorbance values for gradual addition of MY in the solution and indicates an increment of the system energy. Such observation corroborates the results obtained from the EIS measurement. Linear curve fitting technique was incorporated to check the linearity of the plots which shows the $R^2$ (coefficient of determination) values to be well over 0.9. Moreover, the UV-Vis spectroscopy data indicate a significant characteristic peak-shift for absorbance from 420 nm for pure turmeric to 442 nm for the 50% MY-adulterated turmeric solution. Such a shift of peak with adulterant content can be used to confirm the presence of MY. Therefore, the present work establishes that EIS can be implemented as an effective tool for rapid, cost-effective and precise quantification of adulterants present in a turmeric sample. This technique can also be used for detecting and quantifying various toxic and carcinogenic adulterants in foods or other bio-consumables.

# Chapter 6

# Design and fabrication of on-wafer capacitive sensors for detecting different adulterants in milk by employing EIS technique

## 6.1 Introduction and background

Current developments in digital microfluidics have revealed the pathway for novel design and implementation of miniaturized devices (also called biochips or lab-on-chips) for different biochemical, biological and physico-chemical analyses [**267-270**]. These LOC or bio-chip devices are modernizing the entire processes of clinical diagnosis, on-chip immunoassays, detection of food adulteration, monitoring of environmental toxicity, bio-sensing, high-throughput DNA sequencing and other important real-time applications. The conventional idea of handling bulk volumes of liquid samples for performing laboratory analyses has been drastically changed with the progress of LOC systems. Instead of large volume of samples, now, typically micro-, nano- to pico-litres of samples are being used nowadays to perform repetitive automated laboratory processes. Thus, in comparison to all those traditional methods, these LOC devices or bio-chips can perform complicated sample preparation, variety of bioassay operations including sample analysis, mixing, separation and detection at a much faster rate and much lower cost margin [**267-273**]. Initially, microfluidic biochips started with the concept of continuous fluid flow in fabricated micro-channels. To maintain the required fluid flow in the microfluidic devices, certain mechanical micro-pumps, externally applied pressure sources and different electro-kinetic mechanisms were implemented. All these complex systems have exposed certain limitations in processing a variety of biological, biochemical and physico-chemical operations. However, the usage of syringe pumps, pressure controllers and micro-valves which are highly expensive and were often observed not suitable for complex, flexible microfluidic operations [**274**]. Alternatively, droplet-based digital microfluidic (DMF) LOCs or bio-chips offer several advantages over continuous fluid flow base microfluidics; hence this domain is rapidly progressing globally and as a result, attracting a lot of research attention. This DMF technology is based on the principle of electro-wetting-on-dielectric [**275-277**] which provides a relatively easier fluid-handling capability on an LOC as compared to analog microfluidics. The amount of reagents/samples which are difficult to obtain in considerable volumes can be manipulated using this technology by incorporating only micro- to nano-liters of the sample inside a single chip thereby reducing sample wastage and overall cost. Also, another great advantage of such technique is the fabrication and implementation of several test circuits on the same device which can be parallely performed thereby speeding up the entire process. The sample liquid in the form of droplets can be extracted from the source or reservoir applying specific voltage and can directly be fed to the parallel circuits instantaneously, all on the same chip. Unlike the continuous fluid flow microfluidic systems, DMF technology can alter the ability of a dynamically reconfigurable system which can drastically change its ongoing behavior, typically in response to any alteration within its environment. Nowadays, the advanced technological steps are able to transfer complicated pattern onto a chip which can be implemented to fabricate point-of-care LOC devices for target oriented applications. Similar to the fabrication processes of very large scale integration (VLSI) circuits and systems by incorporating innumerable transistors and other IC chips, DMF bio-chips too can be designed and fabricated in a similar way using unit cells or electrodes. Therefore, with all these striking features, DMF technology is expected to revolutionize the entire world of detection and sensing by implementing advanced LOCs and bio-chips for rapid, robust, cost-effective and point-of-care applications. However, it is worthy to mention that even though the DMF technology offers numerous advantages, still its role in bio-electrical, biomedical and food safety applications require further significant development.

## 6.2 Materials design

In this section, materials related to the design and fabrication issues in LOC devices were discussed thoroughly. Fabrication of a fully functional LOC device requires four different categories of materials such as substrate, electrode, insulating oxide and hydrophobic layer. Substrate selection is the initial step followed by metal electrode coating onto it. Next, a passivated insulating oxide layer must be incorporated to prevent the interaction between liquid sample and the metal electrodes. Finally, a hydrophobic layer must be developed at the liquid-insulator interface to prevent the wetting of the liquid samples during device operations.

### 6.2.1 Substrate selection

Selection of appropriate substrates for the fabrication of LOC devices is of utmost  importance. There are several factors upon which the selection of suitable substrate for specific LOC device depends. Among these, the initial and most important feature is the substrate's mechanical stability which should meet the device's overall design requirements. Also, electrical stability of the substrate must be checked thoroughly in order to avoid any leakage current in devices during real-time operations. The selected substrate should also provide good thermal properties for controlling heat dissipation during device processing. Additionally, the selected substrate must meet the necessary demands such as heat resistance, immune to chemicals, surface uniformity, bio-compatibility, flexibility etc. In this work, silicon, glass and plastic were chosen as substrates during the fabrication of LOC based sensor devices.

### 6.2.2 Electrode selection

For device fabrication, different types of metals including aluminum, copper, chromium, platinum and gold have been considered. Aluminum and copper were initially used as electrodes; however, these were rejected due to their chemical reactions with different biological samples under the application of an external bias voltage. Platinum and gold showed excellent immunity towards corrosions when used with bio- and food samples. Also, chromium, when coated with gold as a bilayer material exhibits high inertness to most of the chemicals and liquid samples used.

### 6.2.3 Electrode insulation

This section illustrates the method of incorporation and selection of several materials as dielectric layers in the fabrication of LOC devices for adulterant detection and quantification in food samples. Generally, an electrical insulator is defined as a material which contains electric dipoles and prevents the flow of current through it under the influence of any external dc electric field. The primary objective of this dielectric layer incorporation is to provide adequate insulation to metal electrodes from electrically conductive liquid samples during measurements. In general, a good electrical insulator material should possess the following characteristics:

- High dielectric constant,

- High dielectric breakdown strength,

- Good thermal stability,

- Good mechanical stability,

- Low leakage current and

- Immune to chemical abrasions.

Higher the dielectric constant of the material, higher will be its capacitive property whereas high dielectric breakdown strength of a material suggests that it can be used in high voltage applications. Also, properties such as low leakage current, good mechanical and thermal stability are required for the device reliability and process compatibility. Silicon dioxide ($SiO_2$) and aluminum oxide ($Al_2O_3$) are the two dielectric materials used in this study. Their overall dielectric, thermal, mechanical and bio-compatible properties are discussed below.

**(i) Silicon dioxide ($SiO_2$):** Silicon dioxide is a widely available dielectric material being extensively used in semiconductor microelectronic industries for insulation and protection of electronic devices. It offers reasonable value of dielectric constant, high dielectric breakdown strength and wide availability which make it an automatic choice for dielectric layer in semiconductor fabrication industries.

**(ii) Aluminum oxide ($Al_2O_3$):** Aluminum oxide or alumina is one of the oldest dielectric materials used in engineering industries as it offers excellent electrical insulation properties including solid hardness and brilliant abrasion resistance. Alumina provides excellent characteristic features such as thermal stability, with high melting point and is an electrically insulating material with high electrical resistivity. Also, alumina is highly corrosion resistant due to its high chemical stability. The dielectric, chemical, mechanical and thermal properties of $SiO_2$ and $Al_2O_3$ are summarized in **Table 6.1**.

**Table 6.1**

Summary of the different properties of $SiO_2$ and $Al_2O_3$.

| Material Properties | Silicon Dioxide ($SiO_2$) | Aluminum Oxide ($Al_2O_3$) |
|:---:|:---:|:---:|
| Atomic weight | 60.08 g/mol | 101.96 g/mol |
| Density | 2.65 gm/cc | 3.95 gm/cc |
| Dielectric Constant | 3.6-4.2 | 9.0-10.1 |
| Dielectric Strength | $10^7$ V/cm | $10^5 - 3.5 \times 10^5$ V/cm |
| Thermal Conductivity | $1.3 - 1.5$ W m$^{-1}$ K$^{-1}$ | 28 - 35 W m$^{-1}$ K$^{-1}$ |
| Resistivity | $10^{12} - 10^{16}$ ohm-cm | $10^{14}$ ohm-cm |
| Melting Point | 1700 $^o$C | 2072 $^o$C |

**6.2.4 Electrode hydrophobization**

Electrode hydrophobization is an important step in LOC fabrication technology. The dielectric layer may not provide an adequate surface to restrict electrowetting of the liquid sample. Thus, the surface of the dielectric layer needs to be modified with a hydrophobic coating to control the wetting property of the samples. In this work, we have used Teflon AF 2400 (1% conc.) as the hydrophobic material.

Teflon AF is an amorphous fluoropolymer which offers high chemical immunity, good temperature stability and lower surface energy. Generally, Teflon AF is easily soluble in several perfluorinated solvents at room temperature and as a result it can be spin-coated onto any substrate with varied thicknesses. In general, there are different grades of Teflon available in the market from which two commonly used grades are Teflon AF 1600 and Teflon AF 2400. The number 1600 and 2400 denotes the glass transition temperatures of Teflon at 160 $^o$C and 240 $^o$C, respectively. Moreover, the dielectric constant of Teflon AF is 1.9 which is the lowest among any known available polymers. Furthermore, it provides an excellent uniform hydrophobic coating with a high contact angle (~116$^o$).

## 6.3 Research scheme and process flow: A brief overview

The rate of milk consumption globally by human beings has been getting an incessant hike till date for its immense health benefits due to the presence of biochemical compounds with very high nutritional value, especially its vast content of protein, vital fatty acids, vitamins and minerals [**278,279**]. Therefore the malpractice of adulteration of milk and its byproducts is a very serious societal issue which is unfortunately targeted solely for monetary profit in a shadow economic mode [**280,281**]. In recent times, natural milk is adulterated with synthetic milk, tap water, starch, salt, pulverized soap, detergents, urea, melamine, allantoin, ammonium sulphate, sodium bicarbonate, cyanuric acid, benzoic acid and preservatives like formalin, hydrogen peroxide etc [**282-284**]. Inclusion of such adulterants decreases the percentage of valuable nutrients in milk and causes several human health-hazards such as acidity, ulcers, indigestion, asthma, malfunctioning of kidneys and even up to cancer and so on [**280,285**].

In the current work, adulterants in milk have been classified into polar and non-polar/ionic categories. In particular, diseases related to consumption of polar compounds such as urea, starch, melamine, allantoin and cyanuric acid constitutes blood sugar, diabetes, heart failure, renal failure, urinary blockage, tubular necrosis and, even cancer [**282-284,286**]. Alternatively, excess intake of non-polar/ionic adulterants can induce skin problems, eye damage and acute liver diseases in the human body [**283,285**]. Therefore, the issue of specific identification and quantification of polar and non-polar/ionic adulterants in milk need to be addressed for safe consumption of milk and its byproducts.

Numerous reports are available on the method of detection and quantification of the adulterants in pure milk products by implementing several techniques, such as, traditional spectroscopic techniques including NIR, FTIR, Raman Spectroscopy and NMR spectroscopy [**287-290**] as well as the chromatographic techniques such as HPLC, LC and GC [**291-293**]. However, most of such analytical schemes are associated with methodical complexities, costly measuring instruments and low throughput, which is less relevant in the context of current social demand for rapid, real time, point-of-care sensing applications. Further, with the advent of digital microfluidics based LOCs, the process of target-oriented bio-sensing is expected to observe considerable advancement in this domain including food adulterant detection [**18**]. However, all of the conventional spectroscopic and chromatographic techniques in their state-of-the-art are incapable to meet the portability requirements essential for chip based sensing platform. In this context, EIS can be successfully implemented in the LOC scheme for digital bio-sensing and food quality monitoring [**18,294**]. It is worthy to mention that EIS technique in general offers several applications in the domains of bio-electronic engineering [**295-299**], bio-medical study [**23,300-303**] and electronics in food safety [**13-17**].

In this perspective, the current work describes experimental detection and quantification of several polar and non-polar/ionic adulterants in milk by EIS scheme. Such technique is implemented on a bio-sensor chip fabricated in compatibility with digital microfluidics. The corresponding process flow for the fabrication of bio-chip has been illustrated systematically including the calibration of hydrophobic layer thickness to achieve appropriate contact angle of the microfluidic droplet of milk sample. The EIS study is performed to detect qualitatively as well as quantitatively the most commonly used milk adulterants including polar compounds, such as melamine, starch, urea, allantoin and cyanuric acid, as well as the non-polar/ionic ones, such as benzoic acid, ammonium sulphate, sodium bicarbonate as well as detergents and tap water. Prior to EIS detection, the presence of these compounds as chemical adulterants in milk is confirmed by FT-MIR study in detail. Finally, on the basis of measured EIS data, an impedimetric modulus-phase angle diagrammatic sensing method has been developed, which can precisely distinguish the presence of polar and non-polar/ionic adulterants in milk along with their relative quantity. Therefore, the general objective of the current research work is to explore a novel sensing method by employing on-chip EIS technique to address the issue of quantitative detection of adulterants in foods and other bio-consumables in a rapid, cost-effective and precise manner.

### 6.3.1 Sample procurement

Fresh cow milk samples were procured from different dairy farms and local markets. Various adulterants that were used in this work were procured from different sources and illustrated later. Adulterant such as tap water is collected from laboratory, detergent is procured from local market and different organic and inorganic adulterants such as Urea, Melamine, Starch, Sodium Bicarbonate, Allantoin, Ammonium Sulphate, Benzoic acid and Cyanuric acid were procured from Sigma Aldrich, India. Raw milk samples were stored in refrigerators at a constant temperature ($4^{o}$C) to avoid sample property degradation due to temperature fluctuations.

### 6.3.2 Preparation of milk-adulterant mixtures

Raw milk samples were kept at room temperature ($25\pm1^{o}$C) and homogenized by using a magnetic stirrer prior to all the measurements. The weight of all of the adulterated milk samples is maintained to 40 g for all percentages of adulteration. The weight percentages of soap in milk were considered to be 0.1%, 0.3%, 0.5%, 0.7% and 0.9% (w/w) respectively and is weighed by using an analytical weight balance (Mettler Toledo AL 204) having a precision of $\pm0.1$ mg. For on-chip measurement purpose, volumes of all the pure and adulterated milk samples were restricted to 5 µl using a micropipette (Tarsons, India). For conventional conductivity cell, the weight of milk-tap water solution is maintained to 40 g for all the percentages of adulteration and is summarized in **Table 6.2**. Similarly, the weight of milk-detergent solution is also restricted to 40 g and is depicted in **Table 6.3**. All the samples were homogenized in a magnetic stirrer before being measured and no noticeable crystals were found inside the solutions during the time of measurement.

**Table 6.2**

Summary of the weights of milk and water samples considered for different percentages of adulteration.

| Volume of milk (g) | Volume of water added (g) | Percentage of adulteration (%) |
|---|---|---|
| 40 | 0 | 0 |
| 36 | 4 | 10 |
| 32 | 8 | 20 |
| 28 | 12 | 30 |
| 24 | 16 | 40 |
| 20 | 20 | 50 |

**Table 6.3**

Summary of the weights of milk and detergent samples considered for different percentages of adulteration.

| Volume of milk (g) | Volume of water added (g) | Percentage of adulteration (%) |
|---|---|---|
| 40 | 0 | 0 |
| 39.96 | 0.04 | 0.1 |
| 39.88 | 0.12 | 0.3 |
| 39.80 | 0.20 | 0.5 |
| 39.72 | 0.28 | 0.7 |
| 39.64 | 0.36 | 0.9 |

### 6.3.3 EIS measurement

Electrical impedance measurements are performed by using a semiconductor characterization system and parameter analyzer (Keithley SCS 4200) at room temperature for all of the pure and adulterated milk samples considered in the current work. For all such measurements, the sensor chip with sample positioned on it is placed inside the probe station and connected to the SCS 4200 using a pair of tungsten probes and each of these measurements has been repeated for five times in order to verify the repeatability of measured values. It is worthy to mention that the SCS 4200 is fully equipped with noise-reduction techniques and has been calibrated in the current work prior to the measurements.

## 6.4 Fabrication of on-wafer biosensor device

This section illustrates the systematic process flow of on-wafer biosensor device fabrication, starting from silicon/glass cleaning to hydrophobic layer coating.

### 6.4.1 Substrate cleaning

Prior to fabrication, specific substrate cleaning techniques for Si have been performed as summarized in **Fig. 6.1**.

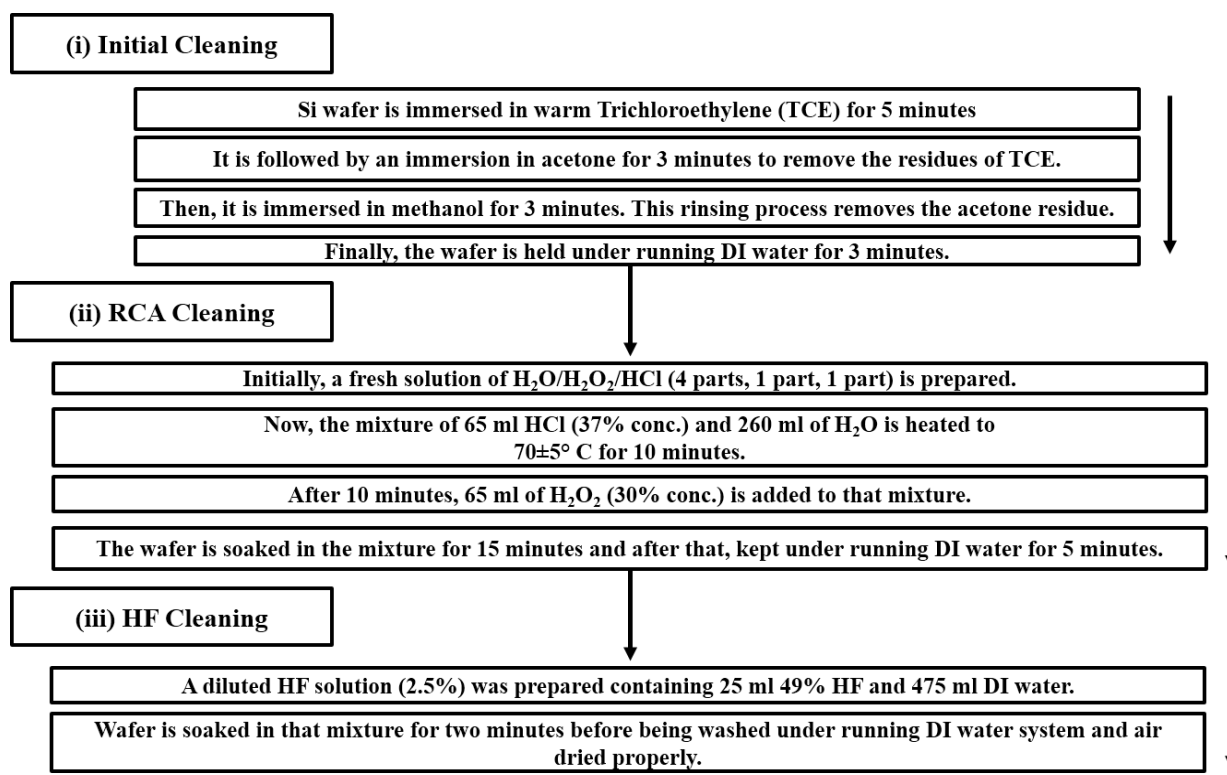

**Fig. 6.1**: Flowchart to illustrate the systematic cleaning process of Si wafers before device fabrication.

Finally, the Si wafer is cleaned and is ready to be used for the next stages of fabrication.

### 6.4.2 Sensor device fabrication: Materials, dimensions and techniques

This section illustrates different types of materials used and different techniques applied for fabricating the required sensor device. To electrically insulate the bulk Si from rest of the device, a layer of $SiO_2$ (~ 50 nm) was coated by using e-beam deposition technique. The fabricated $Si-SiO_2$ device was then annealed at 600°C for 1 hour inside an annealing furnace under constant argon (Ar) ambience for better adhesion. The next step involves a hydrophobic layer coating on top of the $SiO_2$ layer using a spin coater (SPIN NXG M1, APEX Instruments). For that, teflon AF2400 (1% conc.) is coated at six different spin speeds (in r.p.m). After coating the hydrophobic teflon layer, the devices were baked at three different temperatures such as 50 °C, 245 °C and 330 °C, separately. The teflon coating process is summarized in **Fig. 6.2**.

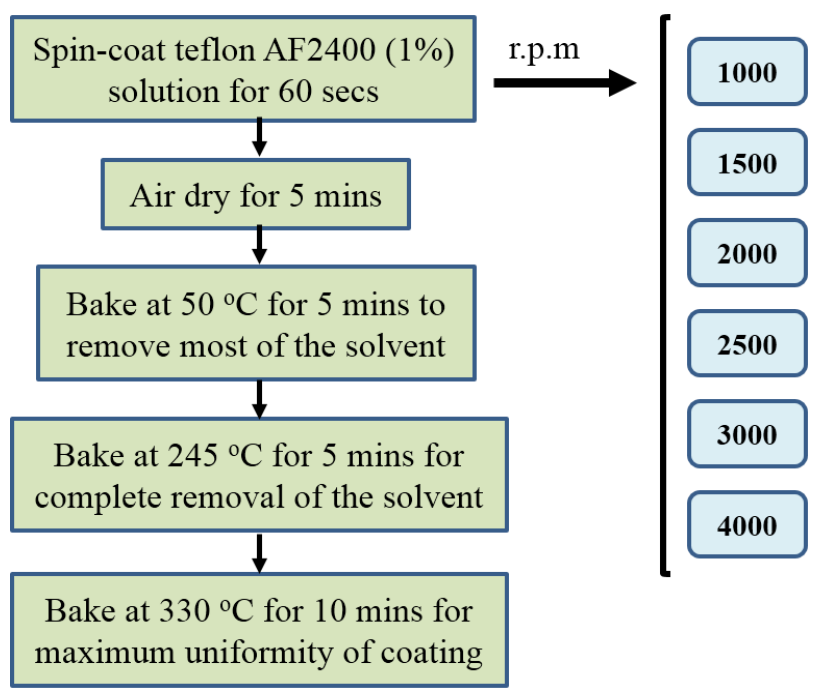

**Fig. 6.2:** Flowchart illustrating the teflon coating process.

Following the spin-coating, the teflon AF layer is air dried for 5 mins and then baked at 50 °C (122 °F) to remove most of the solvent. It is worthy to mention that the hydrophobicity is dependent on the thickness of teflon film [**304**] The time required to remove this solvent will increase with the thickness of the coated film and boiling point of the solvent. Also, since, teflon AF is a high molecular weight polymer, therefore, complete removal of solvent can take a long time unless it is heated above its glass transition temperature (Tg). To remove the last trace of solvent, the coated substrate is again heated to 5 °C (9 °F) above its Tg (i.e. 245 °C [473 °F] for teflon AF 2400) for 5 mins. Finally, for maximum uniformity of coating thickness and enhanced adhesion, the teflon layer is heated to 330 °C (626 °F) for 10 mins. Therefore, this three step baking process enables the polymer to spread uniformly over the substrate during the drying process. Also, it is observed that the liquid sample being placed between the electrodes typically attaches to the metal surface and spreads over the entire region thereby degrading the control on impedance measurement in the parallel plate capacitor model. Therefore, in order to restrict the liquid droplet from such spreading, it is necessary to reduce its wettability on the surface. Thus, teflon, which has a very strong hydrophobicity, is coated as a film on the platform for the liquid sample under test. To further enhance its hydrophobicity, the impact of teflon film thickness on wettability of milk is investigated in the current work.

Wettability of the microfluidic sample droplet placed on the hydrophobic layer is estimated by measuring the contact angle which is observed to depend significantly on the coated film thickness. The different thicknesses of teflon is obtained by varying the spin speed during coating in the range of 1000-4000 rpm and the corresponding contact angles are measured by using a contact angle meter (ACAM D2, APEX Instruments). The variation of such contact angles with different teflon thicknesses obtained due to varying spin speeds is depicted in **Fig. 6.3(a)** - **(f)**. Thickness values of the teflon layers are calculated by using the changes in amplitude component (Ψ) and phase difference (Δ) obtained from the spectroscopic ellipsometry (SE) measurements performed in the wavelength (λ) range 200–800 nm at 60° angle of incidence and 45° angle of polarization. It is observed that the highest contact angle is obtained for spin coating at 4000 rpm which leads to a teflon film thickness of ~70 nm. The experimental values of Ψ- Δ for this film are plotted with the wavelength of the incident light along with the simulated fitting curve for calculating thickness as shown in **Fig. 6.3(g)**.

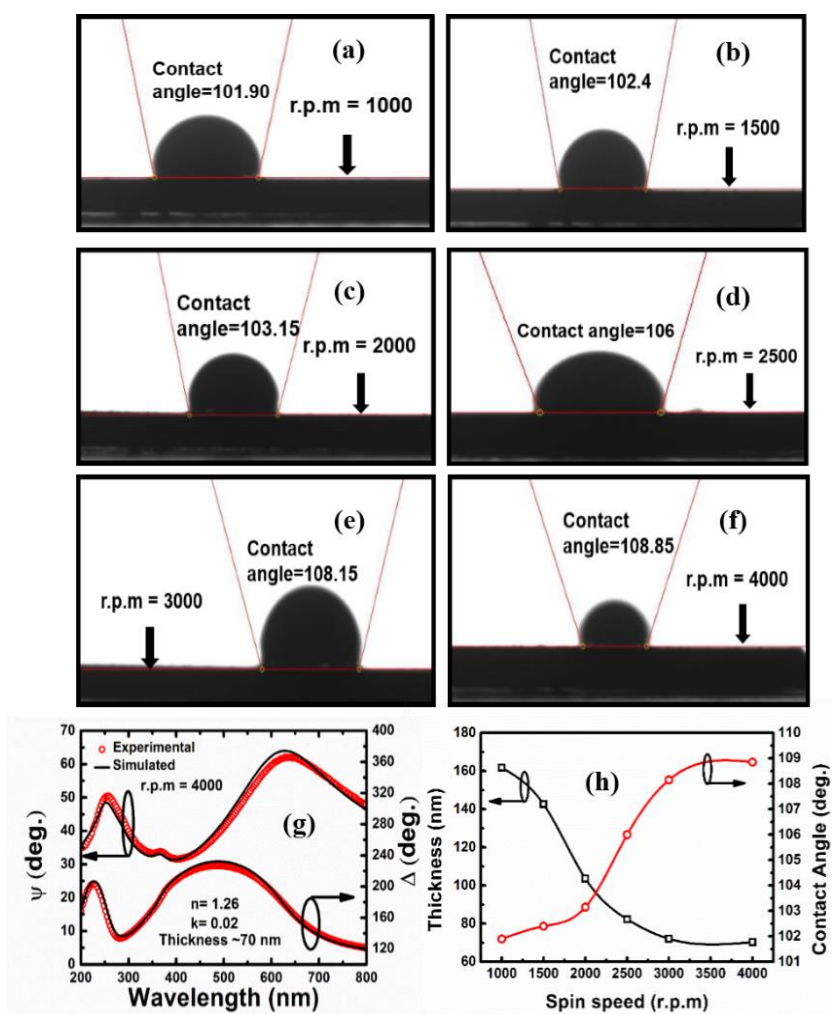

**Fig. 6.3:** (a) - (f) Plots showing the contact angle of milk sample for different coating spin speeds, (g) experimentally measured (symbols) and model fit (lines) spectroscopic ellipsometric parameters, $\Psi$ and $\Delta$, for the teflon film (4000 rpm spin speed) on $SiO_2/Si$ device in the range of 200–800 nm wavelength at 60° angle of incidence and, (h) figure showing the variation of teflon film thickness and contact angle for different spin speeds.

Such simulation is performed considering an air/teflon/$SiO_2$/Si-structure which is observed to fit with the experimentally obtained data up to an acceptable degree of accuracy. Also, the values of refractive index (n) and extinction coefficient (k) extracted from such fits are obtained to be 1.26 and 0.02 which shows good agreement with the previously reported results [**305**].

The thicknesses of all the teflon films for different spin speeds are obtained similarly and plotted in **Fig. 6.3(h)**. The variation corresponding contact angles for milk droplets placed on the films are also shown in the figure. It is apparent from **Fig. 3(h)** that the thickness of teflon film decreases (~160 nm to 70 nm) with spin speed (1000-4000 rpm) which is attributed to the increased centrifugal force with spin speed to spread out the teflon solution radially outward. With such decrease in thickness, the contact angle is observed to increase up to a saturation value of ~110° for the milk sample. Such improvement of hydrophobicity is originated from a corresponding increase in the surface tension due to an enhancement of surface-to-volume ratio with the reduction of film thickness. Measured film thicknesses of teflon AF2400 for the six spin speeds (in r.p.m.) are listed in **Table 6.4**.

**Table 6.4** Summary of the acceleration time, spinning time, spin speeds and corresponding film thicknesses of the teflon coated layers.

| Acceleration time (second) | Spinning time (second) | Spin speed (in r.p.m) | Achieved film thicknesses (nm) |
|---|---|---|---|
| 10 | 60 | 1000 | 161.79 |
| | | 1500 | 142.62 |
| | | 2000 | 103.66 |
| | | 2500 | 82.34 |
| | | 3000 | 72.08 |
| | | 4000 | 70.35 |

The fabrication of patterned electrodes was performed initially by using aluminum (Al) and copper (Cu). For this, Al and Cu sources were placed inside the thermal evaporation chamber and the chamber pressure was maintained at $1\times10^{-6}$ mbar by using a turbo molecular pump. Shadow mask was used to deposit metal electrodes onto the device surface. Thickness of the deposited metal films was controlled by a thickness monitor. Later on, the expensive metals such as gold (Au) and platinum (Pt) were used for metal electrodes. Both Au and Pt coatings were performed by using e-beam deposition as well as sputtering technique. Initially, 40 nm sputtered Au and Pt films were used as the control electrodes. Later, a thin chromium (Cr) adhesion layer (~ 20 nm) was introduced along with it to enhance the electrode stability by e-beam deposition technique. The fabricated sensor electrode is a parallel plated capacitor with dimensions 3 mm x 1 mm while the smaller square electrodes fabricated next to it having dimensions 1 mm x 1 mm respectively. The inter-electrode distance is maintained to 0.5 mm for all the fabricated electrodes. This entire process is initially performed on Si and then repeated on low-cost substrates such as glass and plastic. The process flow for sensor fabrication with a milk sample placed on the sensor electrodes and probe-connected to a semiconductor characterization system and parameter analyzer (Keithley SCS 4200) with probe-station connectivity are shown in **Fig. 6.4**.

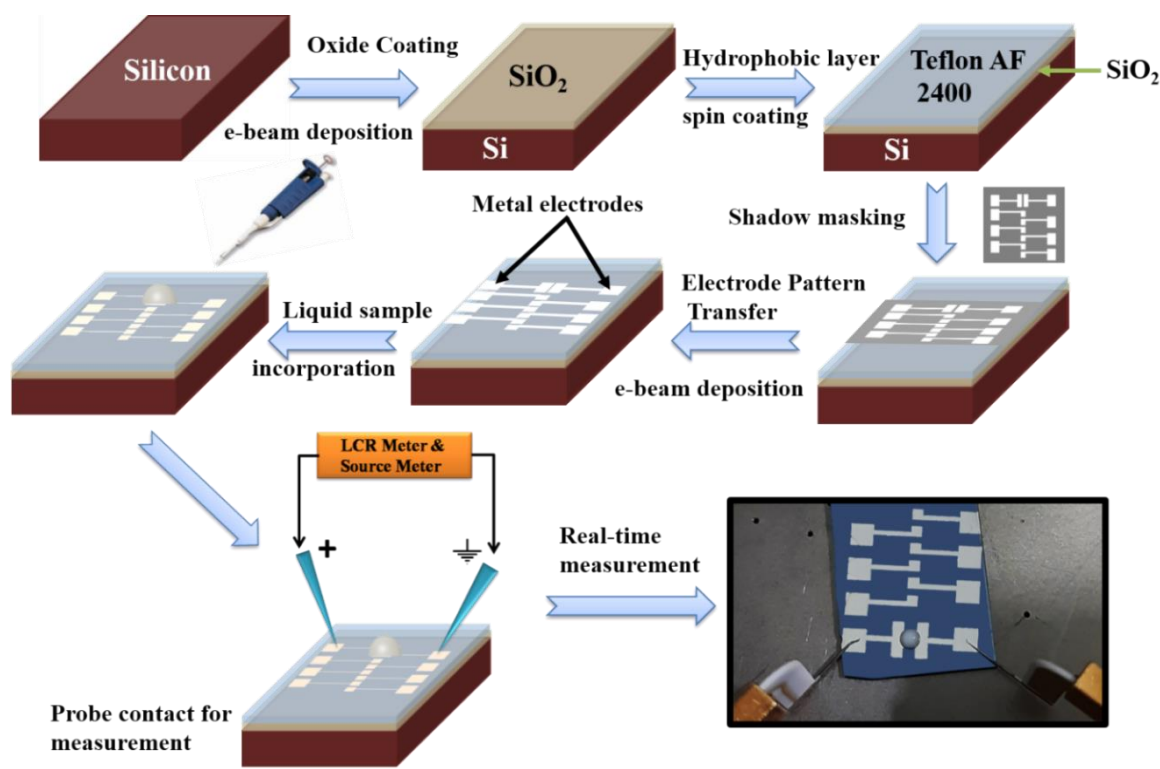

**Fig. 6.4**: Process flow of sensor device fabrication and image of a real-time fabricated device with liquid sample placed on it.

## 6.5 Use of the fabricated on-wafer device for milk adulterant sensing

Such on-wafer sensors are extremely useful for monitoring the food safety issues. Therefore, the on-wafer sensors were fabricated for detecting a variety of adulterants in several milk samples. The primary objective of this research was to develop a comprehensive understanding on the processes of detection and quantification of toxic and carcinogenic adulterants used in milk by using on-wafer sensor platform.

## 6.6 Results and Discussion

### 6.6.1 Study of the electrical properties of pure and adulterated milk: EIS analysis (on conductivity cell)

In this study, the influence of tap water adulteration on individual milk samples and nature of variation of their relevant electrical properties at a particular frequency of 1 KHz at room temperature were observed. For each sample, total six measurements starting with pure milk (0% adulterated) up to 50% adulterated milk are performed to calculate the necessary electrical parameters.

**Effect of tap water adulteration on the electrical properties of milk: EIS and I-V analysis**

The variation of impedance, capacitance and conductance for five different milk samples with selected weight percentages of tap water are shown in **Fig. 6.5(a)**, **(b)** and **(c)** respectively. It is apparent from such plots that the electrical parameters vary linearly with the variation of tap water content (from 10% to 50%) for all of the pure milk samples. The capacitance and conductance values show a linear decreasing trend while the net impedance curve depicts a linear increasing trend with the increase in tap water content.

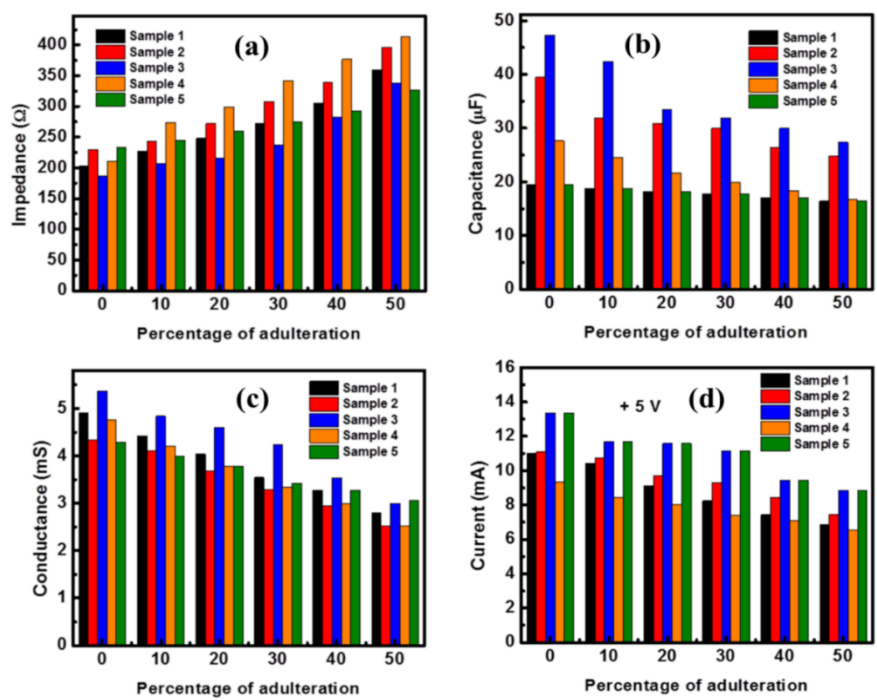

**Fig. 6.5**: Plots of (a) impedance, (b) capacitance, (c) conductance and (d) variation of current at +5 V for different percentages of tap water content in milk samples collected from five different sources.

Since, the primary constituents of milk are water (≥87%), lactose (≥4.9%), fat (≥3.4%), protein (≥3.3%), minerals (≥0.7%), therefore, it is an emulsion of fat in water. Water is a highly polar compound due to its O-H bonds and fats are generally non-polar due to the non-interaction of C-H bonds with any other polar molecules. Therefore, with high percentages of water and lactose (carbohydrate) present in milk, it is considered as a polar compound. However, tap water contains chlorine, fluorine compounds, hormones, nitrates, pesticides and other compounds. Also, the most important content of tap water is different salts of arsenic, aluminum, copper, lead, cadmium, barium, sodium, potassium, calcium and others. Therefore, upon adding tap water in the solution polar-polar interactions between the tap water molecules and resident polar groups in milk will occur. Consequently, after a threshold condition, the additional tap water molecules will start interacting with ionic components, such as soluble salt fractions, in the form of minerals present in milk. Since polar dipole moment is much smaller than the ionic dipole moment, the interaction of polar dipoles corresponding to tap water molecules with the ionic dipoles effectively reduces the resultant dipole moment of the system, and thereby decreases the overall polarization and dielectric constant of the system. In terms of electrical conductivity, milk is more conductive than tap water due to the presence of electrically charged compounds, like mineral salts. The electrical conductivity of milk is primarily determined by the sodium and chloride ions and the distribution of salt fractions between the soluble and colloidal forms. Therefore, by adding percent weights of tap water in the solution, net electrical conductance of the solution decreases linearly, for all the samples measured, as shown in **Fig. 6.5(c)**. Simultaneously, the net impedance depicts an incremental trend as shown in **Fig. 6.5(a)**.

**Fig. 6.5(d)** shows the variation of current for different percentages of milk adulteration in five different samples at a particular voltage + 5 V. As discussed earlier, tap water is much less conductive than milk and from **Fig. 6.5(d)**, it is apparent that, with its addition, the amount of charge flowing through the system decreases effectively which in turn decreases the overall current.

**Effect of detergent adulteration on the electrical properties of milk: EIS and I-V analysis**

This study illustrates the effect of addition of detergent as an adulterant on individual milk samples by observing the nature of variation of their relevant electrical properties at room temperature. For each sample, total six measurements starting with pure milk (0% adulterated) up to 0.9% adulterated milk are performed to calculate the necessary electrical parameters. The variation of capacitance, impedance and conductance for five different milk samples with selected weight percentages of detergent are shown in **Fig. 6.6(a)**, **(b)** and **(c)** respectively. It is apparent from such plots that the electrical parameters vary steadily with the variation of detergent content (from 0.1% to 0.9%) for all of the pure milk samples. The capacitance and conductance values show an increasing trend while the net impedance curve depicts a decreasing trend with the increase in detergent content.

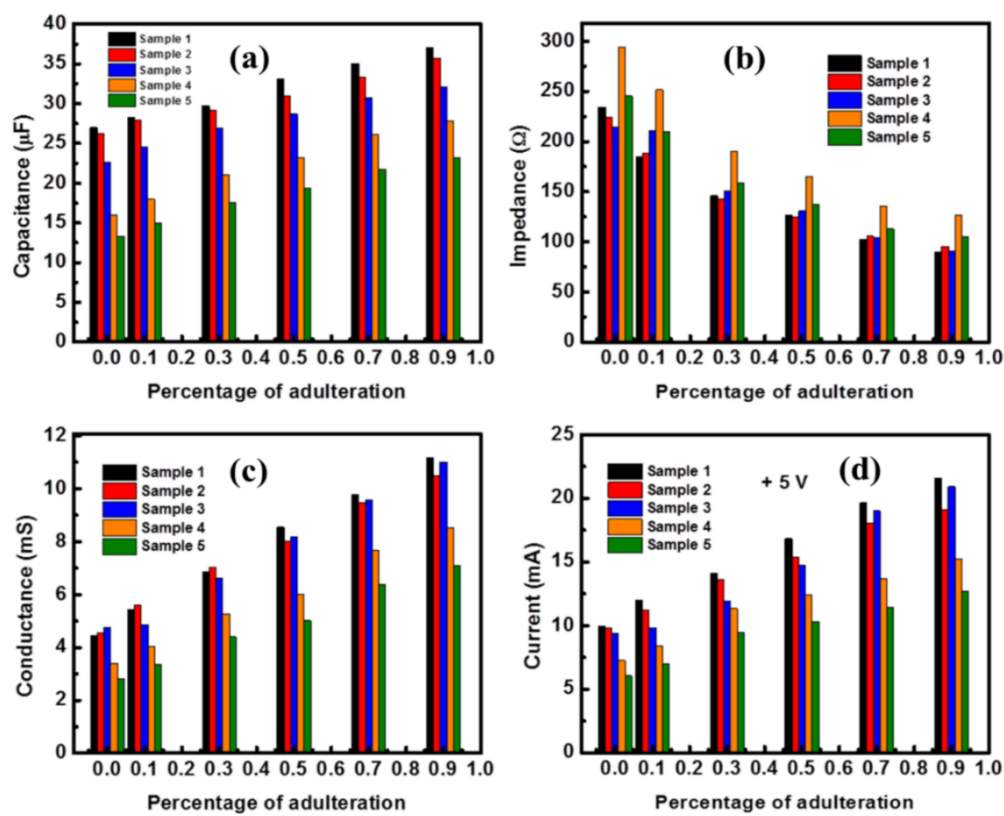

**Fig. 6.6**: Plots of (a) capacitance, (b) impedance, (c) conductance and (d) variation of current at +5 V for different percentages of detergent content in five milk samples.

Detergents are amphipathic molecules that contain both polar and non-polar groups. Their chemical structure constitutes a polar group (head) at the end of a long hydrophobic carbon chain (tail). In contrast to purely polar or non-polar molecules, detergents exhibit unique properties in water. Their polar group forms hydrogen bonds with water molecules, while the hydrocarbon chains aggregate due to non-polar interactions. When a polar molecule is added into a solution containing ions, the positive charge end of the polar molecule will be attracted towards the negative charge end (anion) of the ion. Similarly, the positive charge end (cation) of the ion will be attracted to the anion of the polar molecule. This ion-dipole interaction is stronger than the dipole-dipole interactions between polar molecules, but is much less weaker as compared to ion-ion interactions.

As discussed earlier, milk contains both polar and non-polar groups along with soluble salt fractions and detergents mainly consist of alkylbenzenesulphonates (polar) and a large number of water soluble sodium and calcium salts. Therefore, with the addition of percent weights of detergent in milk, all types of polar-polar, polar-non-polar, polar-ionic and ionic-ionic interactions will occur. Polar-polar and polar-non-polar dipole moment will form due to the interactions of both polar and non-polar groups present in milk and detergent. The interactions between ionic salts in detergents and polar groups in milk will contribute to the ionic-polar dipole moment. Finally, the interactions between ionic salts of detergents and soluble salt fractions in milk will result in ionic-ionic dipole moments. However, the ionic-ionic dipole interaction is the strongest among all other interactions and will dominate the effective polarization of the system. Therefore, gradual increase of detergent content in the adulterated milk-detergent solution increases the effective dipole moment which in turn increases the effective polarization of the system and thus results to an increment in the effective permittivity and the capacitance of the system.

**Fig. 6.6(d)** shows the variation of current with increasing percentages of detergent content in milk at +5 V. It is apparent from the plot that, with the increase in detergent content inside the solution, amount of charge flowing through the system increases effectively which in turn increases the overall current. This increment in current values is attributed to the formation of a large number of free ions inside the system due to the interactions between water soluble metallic salts present in detergents and water in milk.

## 6.7 Study of the FT-MIR spectral bands

### 6.7.1 FT-MIR spectroscopy analysis

FT-MIR spectroscopic measurements have been performed on the pure milk and adulterated samples with specific adulterants considered in the current study. The FT-MIR spectra of pure and such adulterated milk samples are shown in **Fig. 6.7**. IR-spectrum for pure milk and that for each adulterated milk sample exhibits distinct peaks at certain wavenumbers that correspond to the atomic stretching, bending, vibrations etc of different constituents of milk and adulterants. However in the adulterated milk, due to electronic interaction between the various constituents of milk and adulterants, new peaks originate from perturbation in the modes of atomic stretching, bending, and vibration. Such new peaks are observed in the present study, as shown in **Fig. 6.7**, in each of the adulterated samples.

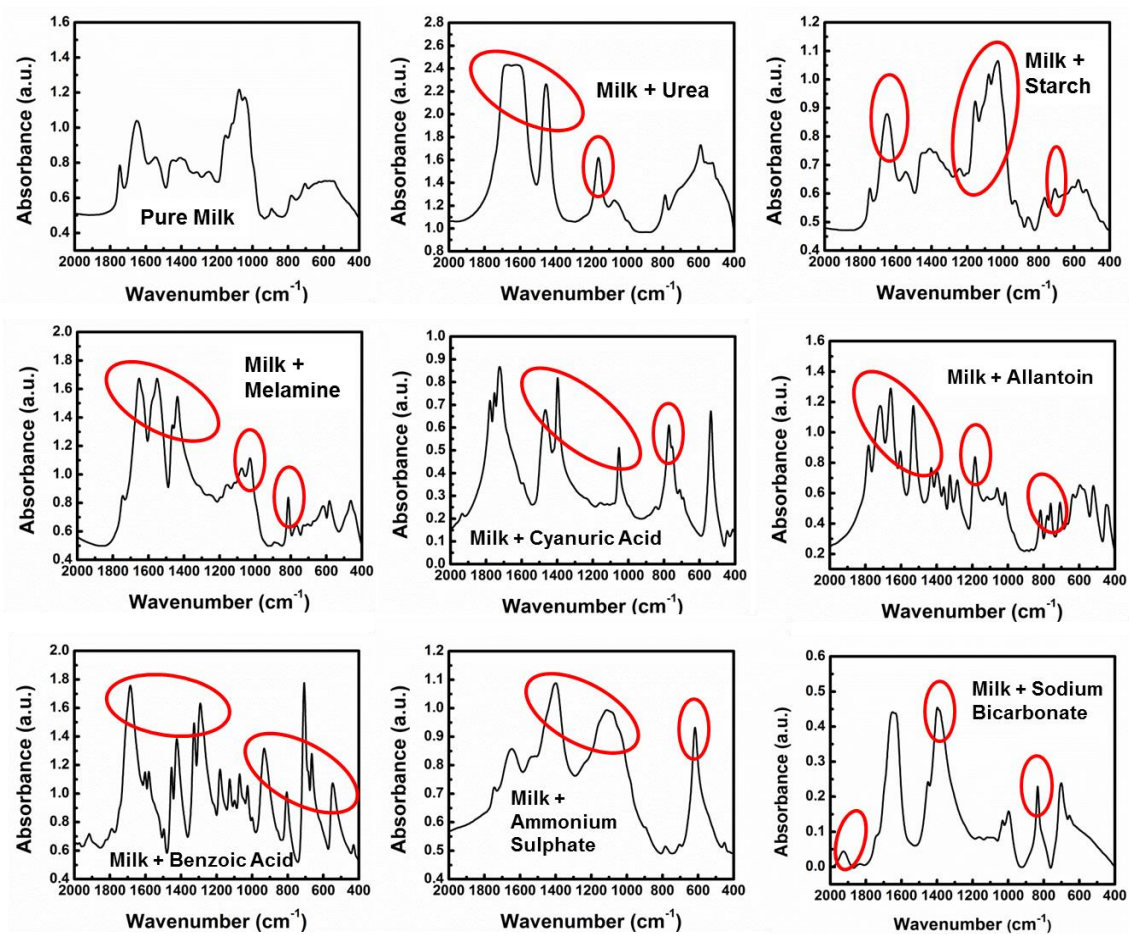

**Fig. 6.7**: FT-MIR spectra of pure and adulterated milk with circled area indicating the generation of new peaks due to adulteration.

The bond assignments along with the corresponding modes of vibrations in the FTIR spectra of adulterated milk are represented in **Table 6.5 and 6.6** respectively.

**Table 6.5** Summary of the FT-IR spectral bands and their assignments for milk adulterated with urea, melamine, starch and cyanuric acid.

| Milk adulterated with Urea | | Milk adulterated with Melamine | | Milk adulterated with Starch | | Milk adulterated with Cyanuric Acid | |
|---|---|---|---|---|---|---|---|
| IR (cm⁻¹) | Assignment | IR (cm⁻¹) | Assignment | IR (cm⁻¹) | Assignment | IR (cm⁻¹) | Assignment |
| 1675 | C=O stretching | 1744 | C=O stretching | 1745 | C=O stretching | 1724 | C=O stretching |
| 1622 | N-H stretching | 1653 | C=N bending | 1650 | C=O stretching | 1593 | N-H bending |
| 1161 | C-O stretching | 1551 | C=N stretching | 1546 | N-O stretching | 1466 | C-H bending / C-C stretching |
| 1072 | C–O stretching | 1437 | O-H stretching | 1412 | S=O stretching | 1397 | N-H stretching |
| 786 | C–Cl stretching | 1156 | C-F stretching | 1242 | C-O stretching | 1156 | C-O stretching |
|  |  | 1076 | C-O stretching | 1155 | C-O stretching | 1053 | O-H stretching |
|  |  | 1029 | C-N stretching | 1079 | C-O stretching | 772 | C-Cl stretching |
|  |  | 891 | C=C bending | 1028 | C-O stretching | 754 | C-Cl stretching |
|  |  | 814 | C-Cl stretching | 765 | C-Cl stretching |  |  |
|  |  | 765 | C-Cl stretching |  |  |  |  |
|  |  | 704 | C=C bending |  |  |  |  |

**Table 6.6** Summary of the FT-IR spectral bands and their assignments for milk sample adulterated with allontoin, benzoic acid, ammonium sulphate and sodium bicarbonate.

| Milk adulterated with Allantoin | | Milk adulterated with Benzoic Acid | | Milk adulterated with Ammonium Sulphate | | Milk adulterated with Sodium Bicarbonate | |
|---|---|---|---|---|---|---|---|
| IR (cm⁻¹) | Assignment | IR (cm⁻¹) | Assignment | IR (cm⁻¹) | Assignment | IR (cm⁻¹) | Assignment |
| 1782 | C=O stretching | 1685 | C=O stretching | 1744 | C=O stretching | 1926 | C=C=C stretching |
| 1717 | C=O stretching | 1424 | O-H bending | 1646 | C=C bending | 1650 | C=O stretching |
| 1658 | C=C stretching | 1292 | C-H stretching | 1399 | NO$_2$ stretching | 1635 | C=C stretching |
| 1604 | N-H bending | 1126 | C-O stretching | 1113 | S=O stretching | 1451 | C-H bending |
| 1531 | N-O stretching | 1070 | C-O stretching | 783 | C-Cl stretching | 1398 | S=O stretching |
| 1431 | C-C stretching | 1026 | C-O stretching | 703 | C-Cl stretching | 1031 | C-N stretching |
| 1398 | C-F stretching / S=O stretching | 932 | C-H bending/ O-H bending | 617 | C-Cl stretching | 997 | C=C bending |
| 1361 | N-O stretching | 805 | C-Cl stretching | | | 834 | C-H bending |
| 1184 | C-O stretching | | | | | 701 | Benzene derivative |
| 1060 | C-O stretching | | | | | | |
| 893 | C=C bending | | | | | | |
| 817 | C-Cl stretching | | | | | | |
| 779 | C-Cl stretching | | | | | | |
| 760 | C-Cl stretching | | | | | | |

The FT-IR spectra of all of the eight adulterated milk samples reveal distinct confirmatory peaks relevant to the milk constituents and corresponding adulterants, as shown in **Tables 6.5** and **6.6**. The distinct peaks at specific wavenumbers for each of the selected adulterants are listed below:

- The peaks at 1675 cm$^{-1}$, 1622 cm$^{-1}$, and 1161 cm$^{-1}$ signify the presence of urea in milk.

- The peaks at 1653 cm$^{-1}$, 1437 cm$^{-1}$, 1076 cm$^{-1}$, 1029 cm$^{-1}$ and 814 cm$^{-1}$ confirm the presence of melamine in milk.

- The peaks at 1650 cm$^{-1}$, 1412 cm$^{-1}$, 1242 cm$^{-1}$, 1155 cm$^{-1}$, 1079 cm$^{-1}$ and 765 cm$^{-1}$ confirm the presence of starch in milk.

- The peaks at 1466 cm$^{-1}$, 1397 cm$^{-1}$, 1053 cm$^{-1}$, 772 cm$^{-1}$ and 754 cm$^{-1}$ confirm the presence of cyanuric acid in milk.

- The peaks at 1782 cm$^{-1}$, 1717 cm$^{-1}$, 1658 cm$^{-1}$, 1604 cm$^{-1}$, 1531 cm$^{-1}$, 1184 cm$^{-1}$, 817 cm$^{-1}$, 779 cm$^{-1}$ and 760 cm$^{-1}$ confirm the presence of allantoin in milk.

- The peaks at 1685 cm$^{-1}$, 1424 cm$^{-1}$, 1292 cm$^{-1}$, 1026 cm$^{-1}$, 932 cm$^{-1}$ and 805 cm$^{-1}$ confirm the presence of benzoic acid in milk.

- The peaks at 1399 cm$^{-1}$, 1113 cm$^{-1}$ and 617 cm$^{-1}$ confirm the presence of ammonium sulphate in milk.

- The peaks at 1926 cm$^{-1}$, 1398 cm$^{-1}$, 1031 cm$^{-1}$, 997 cm$^{-1}$ and 834 cm$^{-1}$ confirm the presence of sodium bicarbonate in milk.

Additionally, as explained above, certain new peaks have originated in all the adulterated spectra that were not observed in any of the pure milk or pure adulterant IR spectra. Such peaks correspond to 787 cm$^{-1}$ (urea↔milk), 1115 cm$^{-1}$ (melamine↔milk), 899 cm$^{-1}$ (ammonium sulphate↔milk), 1379 cm$^{-1}$ (starch↔milk), 1117 cm$^{-1}$ (allantoin↔milk), 1181 cm$^{-1}$ (benzoic acid↔milk), 1162 cm$^{-1}$, 1117 cm$^{-1}$ and 1094 cm$^{-1}$ (sodium bicarbonate↔milk) and 1156 cm$^{-1}$, 1119 cm$^{-1}$ and 846 cm$^{-1}$ (cyanuric acid↔milk), respectively. The generation of such new peaks may be attributed to the formation of new bonds due to milk-adulterant molecular interactions.

## 6.8 EIS analysis (using on-wafer platform)

The electrical impedance (Z) of milk samples adulterated with the polar and non-polar/ionic compounds considered in the present work are measured for a frequency range of 50 Hz-1 MHz of the AC signal with amplitude being 50 mV (peak-to-peak). For sensing purpose, such Z values for varying concentration of adulterants are compared at an identical frequency of 1 kHz for all the samples. The corresponding C and G values are extracted from an equivalent circuit model, as described in our previous report [**13,16,17**]. The variation of impedance with percentages of adulteration (0% -0.9%) in milk are plotted in **Fig. 6.8(a) - (h)**, where the corresponding capacitance and conductance plots are provided at the insets. Among these, **Fig. 6.8(a) - (e)** represent the polar adulterants whereas **Fig. 6.8(f) - (h)** depict the non-polar/ionic ones.

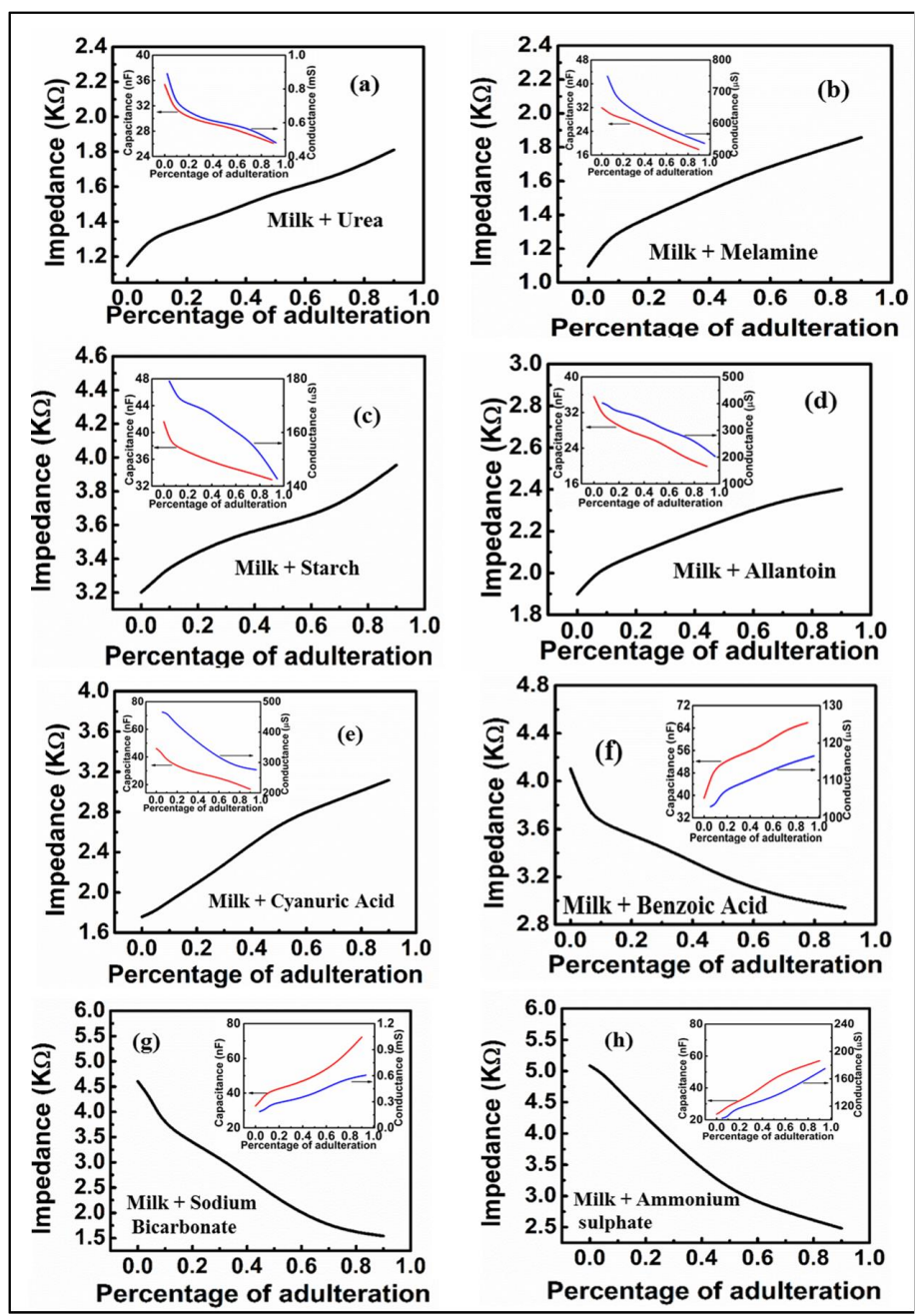

**Fig. 6.8**: Plots of capacitance, conductance and impedance of milk samples adulterated with polar adulterants - (a) Urea, (b) Melamine, (c) Starch, (d) Allantoin, (e) Cyanuric Acid and non-polar/ionic adulterants - (f) Benzoic Acid, (g) Sodium Bicarbonate and (h) Ammonium Sulphate.

It is interesting to note that the impedance values show an increasing trend for the polar adulterants whereas they are observed to decrease for the non-polar/ionic ones thereby emerging as a key parameter to differentiate qualitatively between polar and non-polar/ionic adulterants in milk. A reverse trend is observed in both the effective capacitance and conductance. However, in apparent contradiction to the conventional concept of leaky capacitance, the capacitance and conductance both are observed to follow similar trend with adulteration percentage for all the adulterants, whether polar or non-polar/ionic, as shown in **Fig. 6.8(a) - (h)**.

It is to be mentioned that the bulk content of pure milk is water which is a pure polar compound, and there are a few metallic salt fractions which contribute to its electrical conductivity. However, polar adulterants such as urea, melamine, starch, allantoin and cyanuric acid have permanent dipole moments as depicted in **Fig. 6.9(a) – (e)**.

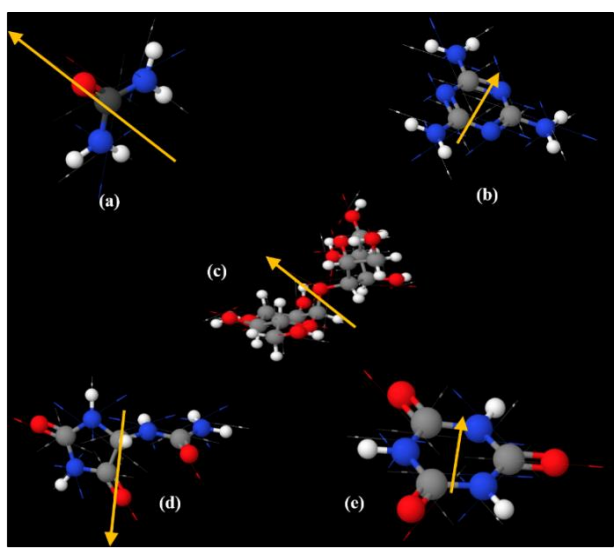

**Fig. 6.9**: Molecular structures of polar adulterants- (a) Urea, (b) Melamine, (c) Starch, (d) Allantoin and (e) Cyanuric acid showing individual bond dipoles and direction of their overall dipole moments. Multi-coloured spheres represent distinct elements - **Red:** Oxygen; **Blue:** Nitrogen; **Grey:** Carbon; **White:** Hydrogen.

When such polar compounds are added to milk, the partial charges of these polar molecules bound the free ions present in milk thereby reducing their effective concentration leading to the decrease in conductance. Simultaneously, such ions being bound to the polar compounds screen their effective polarity and reduce the effective capacitance.

Alternatively, the ionic compounds such as sodium bicarbonate and ammonium sulphate have free carriers, however, due to solvent sharing they may have very strong dipole moment as shown in **Fig. 6.10(a)** and **(b)**, respectively.

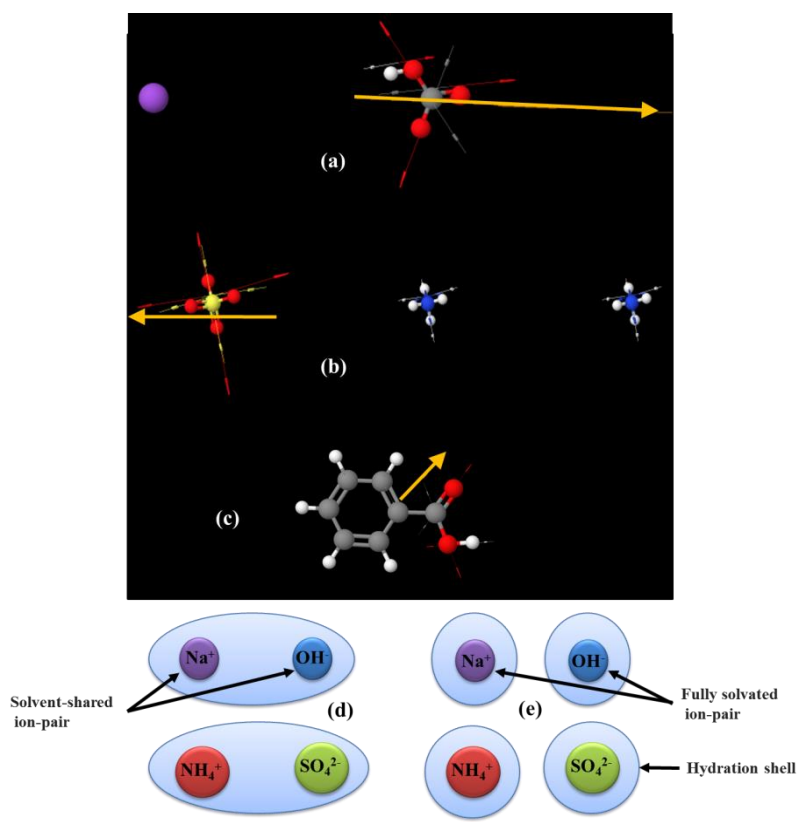

**Fig. 6.10**: Molecular structures of ionic and non-polar adulterants- (a) Sodium bicarbonate, (b) Ammonium sulphate and (c) Benzoic acid showing individual bond dipoles and direction of their overall dipole moments and schematic of (d) solvent-shared ion association and (e) fully solvated ion association resulting from ion dissociation in milk adulterated with ionic adulterants. Multi-coloured spheres represent distinct elements - **Red:** Oxygen; **Blue:** Nitrogen; **Grey:** Carbon; **White:** Hydrogen; **Purple:** Sodium; **Yellow:** Sulphur.

Since the water content in milk is ~85% - 90%, ionic adulterants in milk dissociate to generate a large number of ions inside the system and form solvent shared/separated ion pairs as well as fully solvated ion pairs, as shown in **Fig. 6.10(d)** and **(e)**, respectively. These bound solvent shared/separated ion pairs induces dipole moments which increases the effective polarization of the system with the gradual increment of adulterant content. This in turn increases the effective dielectric constant of the system and thereby the overall system capacitance, as observed in **Fig. 6.8(f) - (g)**. Simultaneously, gradual increment in the number of fully solvated ion pairs increases the net free carrier density of the system which increases its effective conductance.

Benzoic acid is partially non-polar (benzene ring part) and partially polar (carboxylic acid group) compound having a finite dipole moment as shown in **Fig. 6.10(c)**. In aqueous solution it forms hydronium ions and benzoate which almost doubles its dipole moment **[306,307]**. Thus in milk, it behaves similar to that in aqueous solution; and therefore due to the generation of benzoate dipoles and hydronium ions its electrical property shows similar trend to that of the ionic compounds.

## 6.9 Phase angle-Impedance analysis for sensory studies

It can be seen from the above study that the impedance changes with adulteration percentage in opposite manner for polar and non-polar/ionic adulterants, which indicates that such trend of the modulus of impedance can be used to detect whether an adulterant is polar or non-polar/ionic in nature. On the other hand, the ratio of conductance and capacitance, which represents the phase of impedance, changes with adulteration in similar trend for both polar and non-polar/ionic compounds, but in different ranges for different adulterants. Thus, such ranges of phase angle along with impedance value can be utilized to identify and quantify the polar and non-polar/ionic adulterants in milk. The percent change of impedance (ß=ΔZ/Z) and phase angle (2×argument) of impedance for all the individual adulterants considered in the present work in comparison to that of pure milk are plotted in **Fig. 6.11.** The changes in modulus and phase angle are plotted in the figure along radial and angular directions, respectively, where the adulterant concentrations are represented by linear colour gradient. In **Fig. 6.11**, the sample numbers are assigned in order of increasing phase angle, where SET 1 corresponds to all the polar adulterants (1,2,3,4,6), while SET 2 signifies the non-polar/ionic adulterants (5,7,8). The colour bars for SET 1 and SET 2 correspond to the linear increment from blue to red and increment from red to blue of the adulterant concentration, respectively.

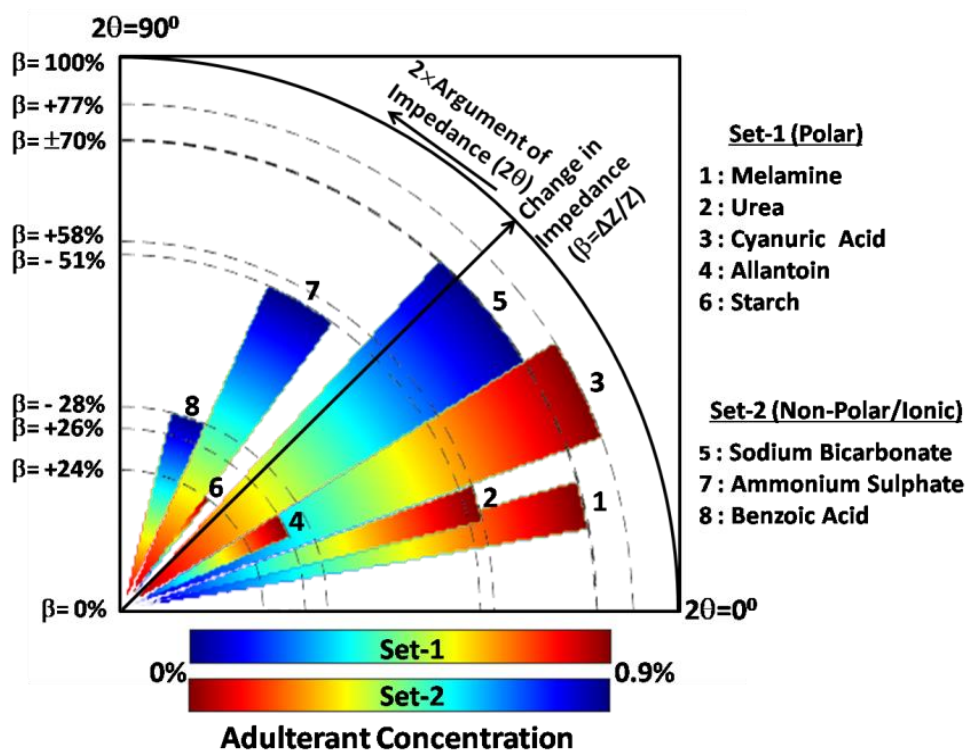

**Fig. 6.11**: Plot of the phase angle-impedance mapping for precise detection of eight adulterants in milk sample.

Using such phase angle-impedance plot, an unknown milk adulterant among the considered ones can be identified and quantified appropriately in the following steps:

1. The colour gradient for 'polar' and 'non-polar/ionic' adulterants of milk respond oppositely ('blue→red' or 'red→blue') in EIS measurement and thus such trend can identify whether an adulterant is polar or non-polar/ionic in nature.

2. Then a specific adulterant can be identified by finding the range of its angular coordinate. It is worthy to mention that in case of overlap of angular ranges, the adulterant(s) can be identified by comparing the respective colour gradients.

3. When the adulterant is identified, its quantification can be performed to precisely obtain its concentration in milk by comparing its radial coordinate with respect to the total radial length for the corresponding adulterant.

# 7. Summary

The present work attempted to detect and quantify the presence of different polar and non-polar/ionic adulterants added to common milk samples employing EIS and FT-MIR techniques by using conventional conductivity cell as well as on-chip device platform. From the measurements performed, EIS is observed to be a useful technique with highly sensitive to detect and quantify the adulterants. Such measurements have been performed on pure milk samples and when they are adulterated with tap water, detergent, melamine, urea, ammonium sulphate, cyanuric acid, benzoic acid, allantoin, sodium bicarbonate and starch, respectively. Primarily, for proof of concept, a parallel plate conductivity cell is dipped inside the sample solution to perform the measurements. EIS and I-V studies illustrate a systematic variation of the system impedance, capacitance, conductance and current values with gradual addition of tap water and detergent in milk. The generation of large number of free ions inside the system due to the interactions between the water soluble metallic salts present in detergent and that of water in milk is attributed to such variations. Analysis of both EIS and current measurements indicate that the method can be used to sense the presence of a minimum of ≥10% tap water and ≥0.05% of other adulterants in milk. For precise detection, a sensor chip (Pt/teflon/$SiO_2$/Si) has been fabricated on which all EIS measurements have been performed. A drastic reduction in sample volume is noticed when the measurement platform is changed from conductivity cell to LOC technique. Contact angle measurements have been performed to check the surface wettability of the liquid sample for different hydrophobic layer thicknesses. Spectroscopic ellipsometry measurements were performed to obtain different teflon layer thicknesses for hydrophobic layer optimization and lowering of contact angle hysteresis. EIS study illustrates a systematic variation of the system impedance, capacitance and conductance with controlled addition of adulterants. The nature of such variation as observed when milk is adulterated with polar adulterants has reversed for milk adulterated with non-polar/ionic compounds. Furthermore, a phase angle-impedance mapping based sensor plot has been developed which shows distinct range of phase angle and change in impedance values for all the adulterants considered. Thus, categorical determination of polar, non-polar and ionic adulterants in milk employing EIS technique has been achieved. FT-MIR spectroscopy has been performed on pure and adulterated milk samples and generation of new peaks corresponding to specific wavenumbers in the adulterated spectra have been observed. Analysis of both EIS and FT-MIR measurements indicate that the present method can detect a minimum of 0.05% adulterant in milk. Therefore, EIS can be successfully employed as an effective tool for rapid, precise and point-of-care detection and quantification of any varieties of adulterants present in any categories of milk as well as in other bio-consumables.

# Chapter 7

# Conclusion and Future Scopes

The present thesis provides an extensive study on the implementation of EIS technique supported by several optical spectroscopic techniques such as UV-Vis and FT-MIR for the detection and quantification of several toxic adulterants in foods and bio-consumables. Using such techniques, both the electrical and optical properties of several pure as well as percent adulterated bio-consumables have been investigated. Variation in such properties due to adulteration provides a systematic means by which the presence and quantification of an adulterant can be determined. Following such detailed investigations, the main findings for each conducted study are summarized below.

The first working chapter outlines the investigation of bio-dielectric properties of pure and adulterated sucrose solutions. $Na_2CO_3$ has been used as the adulterant in different weight percentages for different sucrose volume fractions in DI-water and their respective electrical properties have been observed. A comprehensive understanding on the theoretical background related to the study has been developed to analyse the overall polarization of the system and the effect of frequency on the complex permittivity of the system have been observed. All the corresponding measurements have been performed by dipping a custom-made parallel plate conductivity cell with unity cell constant inside the target solution. Electrical parameters such as impedance, capacitance and conductance of DI-water-sucrose solution with varying sucrose content reveals a quasi-oscillatory trend which is attributed to the random variations in the alignment of dipole moments at molecular level for corresponding sucrose content in the solution. Similar studies have been conducted with aqueous solution of glucose and results seem to validate with that of sucrose. However, with addition of $Na_2CO_3$ in selective weight percentages, such quasi-oscillatory behaviour gets altered to become linear. Such linear nature provides a plot for the sensor which determines the amount of adulteration in terms of the electrical parameters. Linear curve fitting technique was incorporated to check the linearity of the plots above 1% adulteration and the standard error of regression (S) with 95% prediction interval is calculated to be ±0.005. From the study, EIS spectroscopy can be exclusively used to sense a minimum amount of 1% $Na_2CO_3$ as an adulterant in sucrose. Therefore, this study provides a deterministic approach to analyse the electrical nature of any ionic compound present in sucrose solution as well as the quantitative estimation of such compounds based on the dielectric polarization of the system which can further be applied in the quality assessment of several bio-consumables.

Following the successful implementation of EIS technique on adulterated saccharide solutions, focus have been set to observe the electrical and optical properties of pure and adulterated honeys from different floral origin. Aqueous sucrose in varying percentages have been considered as the adulterant in this study. FT-MIR supported EIS analysis and characterization of the adulterated honey samples reveal that electrical and optical parameters investigated here significantly depend both on the varying floral species as well as adulterant content. EIS analysis has indicated distinct variation in the measured electrical parameters. Also, from this study, it is observed that the nature of variation of impedance, capacitance and conductance with percentage change in adulteration is linear. Such results have been validated by analyzing their corresponding I-V characteristics which show linear decrement in current values with the addition of SS. EIS and I-V results indicate that the system starts to sense the presence of 10% SS content and successfully continues till it is 70% adulterated. Simultaneously, FT-MIR study confirms the change in absorbance values for individual varieties of honey. Also, this study reflects the increase in absorbance values due to the increase in adulterant content. FWHM is calculated for the spectral peak at 1056 cm$^{-1}$ due to C-O, C-C and O-H stretching which tends to increase with increase in adulterant content. Finally, the electrical co-efficient of sensitivity have been extracted for individual varieties of honey in terms of the system conductance. Therefore, the current study reveals that EIS can be used as an efficient tool for rapid, cost-effective and accurate quantification of adulteration in different floral species of honey. Such technique can also be used for detecting and quantifying the presence of adulterants in other food products.

After two successful attempts in detecting and quantifying adulterants in saccharides and honey, a third attempt was made to detect MY in turmeric powder, both being a solid bio-consumable. In this study, EIS measurements have been performed on adulterated turmeric samples which were also characterized by employing UV-Vis and FT-MIR spectroscopy techniques. EIS study exhibited a steady variation of the system impedance, capacitance and conductance values with increasing adulterant percentage in the solution. I-V measurement indicated the increase of current values with the gradual increase of MY in the solution. This study also focusses on the occurrence of chemical ion transport phenomenon inside the system due to MY-DI-turmeric molecular interactions which is responsible for the variation of measured electrical and optical parameters. Analysis of both the EIS and I-V measurements indicate that the approach can be used to sense the presence of ≥5% MY in turmeric powder and it can be effectively applicable till the adulterant content is 50%. Alternatively, UV-Vis and FT-MIR studies also confirm a systematic increase of absorbance values for gradual addition of MY in the solution. Such observation corroborates the results obtained from the EIS measurement. Moreover, the UV-Vis spectroscopy data indicates a significant characteristic peak-shift (for MY) for absorbance from 420 nm for pure turmeric to 442 nm for the 50% MY-adulterated turmeric solution. Therefore, observations from the last three efforts make it very clear that EIS is the perfect tool to be implemented for rapid, cost-effective and precise quantification of adulterants in food samples.

In contrast to the conventional prototypes used in the previous works for measurement purposes, this work incorporates the entire processes of design and fabrication of sensor chips for detection and quantification of a variety of adulterants in milk samples. Adulterants used in this study were classified broadly into two different groups such as polar adulterants and non-polar/ionic adulterants with all of them having detrimental effects on consumers. Proper design of the sensor device is implemented by fabricating the chip on a Si wafer where $SiO_2$ is used as the dielectric layer, teflon AF 2400 as the hydrophobic layer with platinum (Pt) electrodes deposited on top of it for contact probing. Several characterizations have been performed on the device including spectroscopic ellipsometry measurements to obtain different teflon layer thicknesses for hydrophobic layer optimization and lowering of contact angle hysteresis. Contact angle measurements have been performed to check the surface wettability of the liquid sample for different hydrophobic layer thicknesses. EIS study illustrates a systematic variation of the system impedance, capacitance and conductance with controlled addition of adulterants. The nature of such variation as observed when milk is adulterated with polar adulterants has reversed for milk adulterated with non-polar/ionic compounds. Furthermore, a phase angle-impedance mapping based sensor plot has been developed which shows distinct range of Ψ-Δ values for all the adulterants considered which categorically determines the respective adulterants in milk. However, generation of new peaks corresponding to specific wavenumbers in the adulterated spectra have been observed which is neither observed in the pure milk FT-MIR spectrum nor found in the adulterant spectra and are attributed to the generation of new bonds due to milk-adulterant molecular interactions. Analysis of both EIS and FT-MIR measurements indicate that the present method can detect at least 0.05% adulterant in milk.

The present work, therefore concludes with the fact that EIS technique can be successfully employed as an effective tool for rapid, precise, cost-effective and point-of-care detection and quantification of any varieties of adulterants present in any categories of foods and bio-consumables using both conventional prototype systems as well as state-of-the-art miniaturized bio-sensor chips.

# Key contributions of the research work

Key contributions of the research work include:

- Development of background theory, electrical equivalent circuit and experimental setup for the detection and quantification of adulterants in food by employing EIS technique.

- Development of a parallel plate conductivity cell (prototype) based measurement setup for the detection of food adulterants.

- Successful implementation of EIS and optical spectroscopic techniques such as UV-Vis and FT-MIR for the detection of adulterants in sucrose, honey and turmeric by using such prototypes.

- Design, fabrication and characterization of novel miniaturized on-chip portable devices for bio-sensing applications.

- Categorical detection and quantification of polar and non-polar/ionic adulterants in milk by developing a novel phase angle-impedance plot for the sensor.

# Future Scopes

The present thesis sought to explore some of the potential applications of EIS technique offers in food quality monitoring. The technique is applied to adulterated saccharides, honey, turmeric and milk samples through a prototype sensing device. Measurements on pure and adulterated solutions provide a clear understanding of the variation of impedance, capacitance, conductance and current. However, for a complex solution with multiple constituents, it is difficult to detect each one of those as well as their impact on the electrical properties of the system which limits its success. Although, this problem can be solved by performing step by step inclusion of each known constituents along with their concentrations and electrical monitoring after every step. This will generate a statistical database for a particular adulterant from which one can compare during working on an unknown sample.

Continuation of this process for other known adulterants will eventually generate a databank which may be a useful source for the researchers in the near future.

As discussed in Chapter 6, a bio-sensor device should be robust, cost-effective and easy to operate. All the measurements performed in this research work is within a frequency range of 50 Hz - 5 MHz, which indicates that low frequency analyzers can be implemented in food quality monitoring rather than using in high frequency setups. Furthermore, it would simplify the design of LCR meters or impedance analyzers at a much lower cost.

Digital microfluidic lab-on-chip devices are transforming the shape of modern bio-sensing, bio-medical and agricultural domains by their novel state-of-the-art designs and ground-breaking applications. One of the key future scopes of this research work is to find and provide applications of EIS technique employing digital microfluidic lab-on-chip devices or bio-chips as a platform for novel applications in the bio-medical research arena.

The present work provides a systematic study to design and fabricate on-wafer devices for food adulteration sensing with its possible applications in real-time food safety issues. However, targeted design and fabrication of high precision multi-layered LOC devices have not been performed due to unavailability of appropriate infrastructure; though it can be developed in future. Furthermore, the synergy of EIS and microfluidic lab-on-chips in bio-sensing arena makes it a unique platform which is yet to revolutionize the entire world of food and bio-sensors.

# References

1. Kennedy, G., Nantel, G., & Shetty, P. (2004). Globalization of food systems in developing countries: impact on food security and nutrition.

2. Olayiwola, K., Soyibo, A., & Atinmo, T. (2004). Impact of globalization on food consumption, health and nutrition in Nigeria. Globalization of Food Systems in Developing Countries: impact on food security and nutrition, 83, 99.

3. Di Renzo, L et al. (2015). Food safety and nutritional quality for the prevention of non-communicable diseases: the Nutrient, hazard Analysis and Critical Control Point process (NACCP). *Journal of translational medicine*, *13*(1), 128.

4. https://www.europarl.europa.eu/factsheets/en/sheet/51/food-safety

5. http://www.fao.org/3/V2890T/v2890t05.htm

6. Institute of Medicine, National Research Council. (1998). Ensuring safe food from production to consumption.

7. Paré, J. R. J., & Bélanger, J. M. (Eds.). (1997). *Instrumental methods in food analysis* (Vol. 18). Elsevier.

8. Otles, S., & Ozyurt, V. H. (2015). Instrumental Food Analysis. *Handbook of Food Chemistry*, 1-19.

9. Müller, A., & Steinhart, H. (2007). Recent developments in instrumental analysis for food quality. *Food Chemistry*, *102*(2), 436-444.

10. MacLeod, A. J. (1973). *Instrumental methods of food analysis*. Elek Science.

11. Mishra, G. K., Barfidokht, A., Tehrani, F., & Mishra, R. K. (2018). Food safety analysis using electrochemical biosensors. *Foods*, *7*(9), 141.

12. Raz, S. R., & Haasnoot, W. (2011). Multiplex bioanalytical methods for food and environmental monitoring. *TrAC Trends in Analytical Chemistry*, *30*(9), 1526-1537.

13. Das, C., Chakraborty, S., Bera, N. K., Chattopadhyay, D., Karmakar, A., & Chattopadhyay, S. (2019). Quantitative estimation of soda ash as an adulterant in aqueous sucrose solution by employing electrical impedance and capacitance spectroscopy. *Measurement*, *148*, 106937.

14. Das, C., Chakraborty, S., Acharya, K., Bera, N. K., Chattopadhyay, D., Karmakar, A., & Chattopadhyay, S. (2017). FT-MIR supported Electrical Impedance Spectroscopy based study of sugar adulterated honeys from different floral origin. *Talanta*, *171*, 327-334.

15. Das, C., Chakraborty, S., Bera, N. K., Acharya, K., Chattopadhyay, D., Karmakar, A., & Chattopadhyay, S. (2019). Impedimetric Approach for Estimating the Presence of Metanil Yellow in Turmeric Powder from Tunable Capacitance Measurement. *Food Analytical Methods*, *12*(4), 1017-1027.

16. Chakraborty, S., Das, C., Bera, N. K., Chattopadhyay, D., Karmakar, A., & Chattopadhyay, S. (2017). Analytical modelling of electrical impedance based adulterant sensor for aqueous sucrose solutions. *Journal of Electroanalytical Chemistry*, *784*, 133-139.

17. Chakraborty, S., Das, C., Saha, R., Das, A., Bera, NK., Chattopadhyay, D., Karmakar, A., Chattopadhyay, D & Chattopadhyay, S. (2015). Investigating the quasi-oscillatory behaviour of electrical parameters with the concentration of D-glucose in its aqueous solution at room temperature by employing impedance spectroscopy technique. *Journal of Electrical Bioimpedance*. 6(1):10.

18. Tripathy, S., Ghole, A. R., Deep, K., Vanjari, S. R. K., & Singh, S. G. (2017). A comprehensive approach for milk adulteration detection using inherent bio-physical properties as 'Universal Markers': Towards a miniaturized adulteration detection platform. *Food chemistry*, *217*, 756-765.

19. Durante, G., Becari, W., Lima, F. A., & Peres, H. E. (2015). Electrical impedance sensor for real-time detection of bovine milk adulteration. *IEEE Sensors Journal*, *16*(4), 861-865.

20. Grossi, M., & Riccò, B. (2017). Electrical impedance spectroscopy (EIS) for biological analysis and food characterization: A review.

21. Nakonieczna, A., Paszkowski, B., Wilczek, A., Szypłowska, A., & Skierucha, W. (2016). Electrical impedance measurements for detecting artificial chemical additives in liquid food products. *Food Control*, *66*, 116-129.

22. Dean, D. A., Ramanathan, T., Machado, D., & Sundararajan, R. (2008). Electrical impedance spectroscopy study of biological tissues. *Journal of electrostatics*, *66*(3-4), 165-177.

23. Chakraborty, S., Das, C., Ghoshal, K., Bhattacharyya, M., Karmakar, A., & Chattopadhyay, S. (2019). Low Frequency Impedimetric Cell Counting: Analytical Modeling and Measurements. *IRBM*.

24. Bera, T. K. (2018, March). Bioelectrical Impedance and The Frequency Dependent Current Conduction Through Biological Tissues: A Short Review. In *IOP Conference Series: Materials Science and Engineering* (Vol. 331, No. 1, p. 012005). IOP Publishing.

25. Ramírez-Chavarría, R. G., Sanchez-Perez, C., Matatagui, D., Qureshi, N., Pérez-García, A., & Hernandez-Ruiz, J. (2018). Ex-vivo biological tissue differentiation by the Distribution of Relaxation Times method applied to Electrical Impedance Spectroscopy. *Electrochimica Acta*, *276*, 214-222.

26. Chao, P. J., Huang, E. Y., Cheng, K. S., & Huang, Y. J. (2013). Electrical impedance spectroscopy as electrical biopsy for monitoring radiation sequelae of intestine in rats. *BioMed research international*.

27. Da Silva, J. E., De Sá, J. M., & Jossinet, J. (2000). Classification of breast tissue by electrical impedance spectroscopy. *Medical and Biological Engineering and Computing*, *38*(1), 26-30.

28. Casas, O et al. (1999). In Vivo and In Situ Ischemic Tissue Characterization Using Electrical Impedance Spectroscopy a. *Annals of the New York Academy of Sciences*, *873*(1), 51-58.

29. Ayza, A., & Belete, E. (2015). Food adulteration: its challenges and impacts. *Food Science and Quality Management*, *41*, 50-56.

30. Paiva, C. L. (2013). Quality management: Important aspects for the food industry. In *Food industry*. IntechOpen.

31. Mizrach, A. (2000). Determination of avocado and mango fruit properties by ultrasonic technique. *Ultrasonics*, *38*(1-8), 717-722.

32. Alander, J. T., Bochko, V., Martinkauppi, B., Saranwong, S., & Mantere, T. (2013). A review of optical nondestructive visual and near-infrared methods for food quality and safety. *International Journal of Spectroscopy*, *2013*.

33. Zhang, B., Peng, B., Zhang, C., Song, Z., & Ma, R. (2017). Determination of fruit maturity and its prediction model based on the pericarp index of absorbance difference (IAD) for peaches. *PloS one*, *12*(5), e0177511.

34. Aboonajmi, M., Jahangiri, M., & Hassan-Beygi, S. R. (2015). A Review on Application of Acoustic Analysis in Quality Evaluation of Agro-food Products. *Journal of food processing and preservation*, *39*(6), 3175-3188.

35. García-Ramos, F. J., Valero, C., Homer, I., Ortiz-Cañavate, J., & Ruiz-Altisent, M. (2005). Non-destructive fruit firmness sensors: a review. *Spanish Journal of Agricultural Research*, *3*(1), 61-73.

36. Taniwaki, M., & Sakurai, N. (2010) .Evaluation of the internal quality of agricultural products using acoustic vibration techniques. Journal *of the Japanese Society for Horticultural Science* , *79* (2), 113-128.

37. Magwaza, L. S., & Opara, U. L. (2015). Analytical methods for determination of sugars and sweetness of horticultural products—A review. *Scientia Horticulturae*, *184*, 179-192.

38. Wang, H., Peng, J., Xie, C., Bao, Y., & He, Y. (2015). Fruit quality evaluation using spectroscopy technology: a review. *Sensors*, *15*(5), 11889-11927.

39. Rongtong, B., Suwonsichon, T., Ritthiruangdej, P., & Kasemsumran, S. (2018). Determination of water activity, total soluble solids and moisture, sucrose, glucose and fructose contents in osmotically dehydrated papaya using near-infrared spectroscopy. *Agriculture and Natural Resources*, *52*(6), 557-564.

40. Brattoli, M., Cisternino, E., Dambruoso, P. R., De Gennaro, G., Giungato, P., Mazzone, A., ... & Tutino, M. (2013). Gas chromatography analysis with olfactometric detection (GC-O) as a useful methodology for chemical characterization of odorous compounds. *Sensors*, *13*(12), 16759-16800.

41. Di, Y., Liu, J., Liu, J., Liu, S., & Yan, L. (2013). Characteristic analysis for odor gas emitted from food waste anaerobic fermentation in the pretreatment workshop. *Journal of the Air & Waste Management Association*, *63*(10), 1173-1181.

42. Tian, H., Shen, Y., Yu, H., & Chen, C. (2018). Aroma features of honey measured by sensory evaluation, gas chromatography-mass spectrometry, and electronic nose. *International journal of food properties*, *21*(1), 1755-1768.

43. Chen, P., McCarthy, M. J., & Kauten, R. (1989). NMR for internal quality evaluation of fruits and vegetables. *Trans. ASAE*, *32*(5), 1747-1753.

44. Spyros, A., & Dais, P. (2012). *NMR spectroscopy in food analysis*. Royal Society of Chemistry.

45. Torrecilla, J. S., Rojo, E., Dominguez, J. C., & Rodríguez, F. (2010). A novel method to quantify the adulteration of extra virgin olive oil with low-grade olive oils by UV−vis. *Journal of agricultural and food chemistry*, *58*(3), 1679-1684.

46. Srivastava, L. P., Khanna, S. K., & Singh, G. B. (1978). Spectrophotometric estimation of metanil yellow in foodstuffs. *International Journal of Environmental Analytical Chemistry*, *5*(2), 119-124.

47. Nath, P. P., Sarkar, K., Tarafder, P., & Paul, G. (2013). Development of a visible spectrophotometric method for the quantitative determination of metanil yellow in different food samples. *Int J Pharma Bio Sci*, *4*(2), 685-692.

48. Li, S., Zhang, X., Shan, Y., Su, D., Ma, Q., Wen, R., & Li, J. (2017). Qualitative and quantitative detection of honey adulterated with high-fructose corn syrup and maltose syrup by using near-infrared spectroscopy. *Food chemistry*, *218*, 231-236.

49. Jaiswal, P., Jha, S. N., Borah, A., Gautam, A., Grewal, M. K., & Jindal, G. (2015). Detection and quantification of soymilk in cow–buffalo milk using Attenuated Total Reflectance Fourier Transform Infrared spectroscopy (ATR–FTIR). *Food chemistry*, *168*, 41-47.

50. Sharifi, M., & Young, B. (2012). Milk total solids and fat content soft sensing via electrical resistance tomography and temperature measurement. *Food and Bioproducts Processing*, *90*(4), 659-666.

51. Grossi, M., Di Lecce, G., Toschi, T. G., & Riccò, B. (2014). Fast and accurate determination of olive oil acidity by electrochemical impedance spectroscopy. *IEEE Sensors Journal*, *14*(9), 2947-2954.

52. Vasilescu, A., Vezeanu, A., & Badea, M. (2014). Electrochemical impedance spectroscopy investigations focused on food allergens. *Sensing in Electroanalysis.(K. Kalcher, R. Metelka, I. Švancara, K. Vytřas; Eds.). 2013/2014, Volume 8.*

53. Ahmed, M. U., Hossain, M. M., & Tamiya, E. (2008). Electrochemical biosensors for medical and food applications. *Electroanalysis: An International Journal Devoted to Fundamental and Practical Aspects of Electroanalysis*, *20*(6), 616-626.

54. Schell, L. M., Gallo, M. V., & Cook, K. (2012). What's NOT to eat—food adulteration in the context of human biology. *American Journal of Human Biology*, *24*(2), 139-148.

55. Bansal, S., Singh, A., Mangal, M., Mangal, A. K., & Kumar, S. (2017). Food adulteration: Sources, health risks, and detection methods. *Critical reviews in food science and nutrition*, *57*(6), 1174-1189.

56. Chowdhury, M. A. J., & Islam, M. T. (2014). Look for the Flies-the Future Ahead!. *Bangladesh Journal of Medicine*, *25*(2), 40-41.

57. Dhanya, K., & Sasikumar, B. (2010). Molecular marker based adulteration detection in traded food and agricultural commodities of plant origin with special reference to spices. *Current Trends in Biotechnology & Pharmacy*, *4*(1).

58. https://old.fssai.gov.in/Portals/0/Pdf/Draft_Manuals/MILK_AND_MILK_PRODUCTS.pdf

59. Franco-Duarte, R., Černáková, L., Kadam, S., S Kaushik, K., Salehi, B., Bevilacqua, A., ... & Relison Tintino, S. (2019). Advances in Chemical and Biological Methods to Identify Microorganisms—From Past to Present. *Microorganisms*, *7*(5), 130.

60. Diba, K., Kordbacheh, P., Mirhendi, S. H., Rezaie, S., & Mahmoudi, M. (2007). Identification of Aspergillus species using morphological characteristics. *Pakistan journal of medical sciences*, *23*(6), 867.

61. Bharathi, M. J., Ramakrishnan, R., Vasu, S., Meenakshi, R., & Palaniappan, R. (2003). Epidemiological characteristics and laboratory diagnosis of fungal keratitis. A three-year study. *Indian journal of ophthalmology*, *51*(4), 315.

62. https://old.fssai.gov.in/Portals/0/Pdf/Draft_Manuals/SPICES_AND_CONDIMENTS.pdf

63. Zábrodská, B., & Vorlová, L. (2015). Adulteration of honey and available methods for detection–a review. *Acta Veterinaria Brno*, *83*(10), 85-102.

64. Paudayal, K. N., & Gautam, I. (2012). SEM Investigation of pollen taxa in honeys from Autochtone Apis cerana in Godavari, Lalitpur District, Nepal. *Journal of Natural History Museum*, *26*, 29-67.

65. Hassan, Z. A. (2018). Application of Scanning Electron Microscope in Palynology Study of Floral Resources By Indo-Malayan Stingless Bees Genus Tetragonula. *Malaysian Journal of Microscopy*, *14*(1).

66. Dustmann, J. H., & Von Der Ohe, K. (1993). Scanning electron microscopic studies on pollen from honey. IV. Surface pattern of pollen of Sapium sebiferum and Euphorbia spp (Euphorbiaceae). *Apidologie*, *24*(1), 59-66.

67. Jones, G. D., & Bryant Jr, V. M. (2007). A comparison of pollen counts: Light versus scanning electron microscopy. *Grana*, *46*(1), 20-33.

68. Silva, B. M., Andrade, P. B., Seabra, R. M., & Ferreira, M. A. (2001). Determination of selected phenolic compounds in quince jams by solid-phase extraction and HPLC. *Journal of liquid chromatography & related technologies*, *24*(18), 2861-2872.

69. Andrade, P. B., Carvalho, A. R. F., Seabra, R. M., & Ferreira, M. A. (1998). A previous study of phenolic profiles of quince, pear, and apple purees by HPLC diode array detection for the evaluation of quince puree genuineness. *Journal of Agricultural and Food Chemistry*, *46*(3), 968-972.

70. Dionisi, F., Prodolliet, J., & Tagliaferri, E. (1995). Assessment of olive oil adulteration by reversed-phase high-performance liquid chromatography/amperometric detection of tocopherols and tocotrienols. *Journal of the American Oil Chemists' Society*, *72*(12), 1505-1511.

71. González-Domínguez, R., Sayago, A., Morales, M. T., & Fernández-Recamales, Á. (2019). Assessment of Virgin Olive Oil Adulteration by a Rapid Luminescent Method. *Foods*, *8*(8), 287.

72. Arlorio, M., Coisson, J. D., Bordiga, M., Travaglia, F., Garino, C., Zuidmeer, L., ... & Martelli, A. (2010). Olive oil adulterated with hazelnut oils: simulation to identify possible risks to allergic consumers. *Food Additives and Contaminants*, *27*(1), 11-18.

73. Cautela, D., Laratta, B., Santelli, F., Trifirò, A., Servillo, L., & Castaldo, D. (2008). Estimating bergamot juice adulteration of lemon juice by high-performance liquid chromatography (HPLC) analysis of flavanone glycosides. *Journal of agricultural and food chemistry*, *56*(13), 5407-5414.

74. Feng, X., Zhang, Q., Cong, P., & Zhu, Z. (2012). Simultaneous determination of flavonoids in different citrus fruit juices and beverages by high-performance liquid chromatography and analysis of their chromatographic profiles by chemometrics. *Analytical Methods*, *4*(11), 3748-3753.

75. Mouly, P., Gaydou, E. M., & Estienne, J. (1993). Column liquid chromatographic determination of flavanone glycosides in Citrus: Application to grapefruit and sour orange juice adulterations. *Journal of Chromatography A*, *634*(1), 129-134.

76. Caristi, C., Bellocco, E., Panzera, V., Toscano, G., Vadalà, R., & Leuzzi, U. (2003). Flavonoids detection by HPLC-DAD-MS-MS in lemon juices from Sicilian cultivars. *Journal of agricultural and food chemistry*, *51*(12), 3528-3534.

77. Del Rio, D., Stewart, A. J., Mullen, W., Burns, J., Lean, M. E., Brighenti, F., & Crozier, A. (2004). HPLC-MSn analysis of phenolic compounds and purine alkaloids in green and black tea. *Journal of agricultural and food chemistry*, *52*(10), 2807-2815.

78. Calabrese, M., Stancher, B., & Riccobon, P. (1995). High-performance liquid chromatography determination of proline isomers in Italian wines. *Journal of the Science of Food and Agriculture*, *69*(3), 361-366.

79. Dasenaki, M. E., & Thomaidis, N. S. (2019). Quality and Authenticity Control of Fruit Juices-A Review. *Molecules*, *24*(6), 1014.

80. Haughey, S. A., Galvin-King, P., Malechaux, A., & Elliott, C. T. (2015). The use of handheld near-infrared reflectance spectroscopy (NIRS) for the proximate analysis of poultry feed and to detect melamine adulteration of soya bean meal. *Analytical Methods*, *7*(1), 181-186.

81. Graham, S. F., Haughey, S. A., Ervin, R. M., Cancouët, E., Bell, S., & Elliott, C. T. (2012). The application of near-infrared (NIR) and Raman spectroscopy to detect adulteration of oil used in animal feed production. *Food Chemistry*, *132*(3), 1614-1619.

82. Rodriguez-Saona, L. E., & Allendorf, M. E. (2011). Use of FTIR for rapid authentication and detection of adulteration of food. *Annual review of food science and technology*, *2*, 467-483.

83. Coitinho, T. B., Cassoli, L. D., Cerqueira, P. H. R., da Silva, H. K., Coitinho, J. B., & Machado, P. F. (2017). Adulteration identification in raw milk using Fourier transform infrared spectroscopy. *Journal of food science and technology*, *54*(8), 2394-2402.

84. Hansen, P. W., & Holroyd, S. E. (2019). Development and application of Fourier transform infrared spectroscopy for detection of milk adulteration in practice. *International Journal of Dairy Technology*.

85. Jaiswal, P., Jha, S. N., Borah, A., Gautam, A., Grewal, M. K., & Jindal, G. (2015). Detection and quantification of soymilk in cow–buffalo milk using Attenuated Total Reflectance Fourier Transform Infrared spectroscopy (ATR–FTIR). *Food chemistry*, *168*, 41-47.

86. Jawaid, S., Talpur, F. N., Sherazi, S. T. H., Nizamani, S. M., & Khaskheli, A. A. (2013). Rapid detection of melamine adulteration in dairy milk by SB-ATR–Fourier transform infrared spectroscopy. *Food chemistry*, *141*(3), 3066-3071.

87. Jha, S. N., Jaiswal, P., Borah, A., Gautam, A. K., & Srivastava, N. (2015). Detection and quantification of urea in milk using attenuated total reflectance-Fourier transform infrared spectroscopy. *Food and bioprocess technology*, *8*(4), 926-933.

88. Ghisi, M., Chaves, E. S., Quadros, D. P., Marques, E. P., Curtius, A. J., & Marques, A. L. (2011). Simple method for the determination of Cu and Fe by electrothermal atomic absorption spectrometry in biodiesel treated with tetramethylammonium hydroxide. *Microchemical Journal*, *98*(1), 62-65.

89. Bucci, R., Magrì, A. D., Magrì, A. L., & Marini, F. (2003). Comparison of three spectrophotometric methods for the determination of γ-oryzanol in rice bran oil. *Analytical and bioanalytical chemistry*, *375*(8), 1254-1259.

90. Ruiz-Matute, A. I., Soria, A. C., Martínez-Castro, I., & Sanz, M. L. (2007). A new methodology based on GC⁻MS to detect honey adulteration with commercial syrups. *Journal of agricultural and food chemistry*, *55*(18), 7264-7269.

91. Ruiz-Matute, A. I., Rodríguez-Sánchez, S., Sanz, M. L., & Martínez-Castro, I. (2010). Detection of adulterations of honey with high fructose syrups from inulin by GC analysis. *Journal of Food Composition and Analysis*, *23*(3), 273-276.

92. Oliveira, R. C., Oliveira, L. S., Franca, A. S., & Augusti, R. (2009). Evaluation of the potential of SPME-GC-MS and chemometrics to detect adulteration of ground roasted coffee with roasted barley. *Journal of Food Composition and analysis*, *22*(3), 257-261.

93. Liang, Q., Qu, J., Luo, G., & Wang, Y. (2006). Rapid and reliable determination of illegal adulterant in herbal medicines and dietary supplements by LC/MS/MS. *Journal of pharmaceutical and biomedical analysis*, *40*(2), 305-311.

94. Bogusz, M. J., Hassan, H., Al-Enazi, E., Ibrahim, Z., & Al-Tufail, M. (2006). Application of LC–ESI–MS–MS for detection of synthetic adulterants in herbal remedies. *Journal of pharmaceutical and biomedical analysis*, *41*(2), 554-564.

95. Galvin-King, P., Haughey, S. A., Montgomery, H., & Elliott, C. T. (2019). The Rapid Detection of Sage Adulteration Using Fourier Transform Infra-Red (FTIR) Spectroscopy and Chemometrics. *Journal of AOAC International*, *102*(2), 354-362.

96. Briandet, R., Kemsley, E. K., & Wilson, R. H. (1996). Approaches to adulteration detection in instant coffees using infrared spectroscopy and chemometrics. *Journal of the Science of Food and Agriculture*, *71*(3), 359-366.

97. Souza, E. M. T. D., Arruda, S. F., Brandão, P. O., & Siqueira, E. M. D. A. (2000). Electrophoretic analisys to detect and quantify additional whey in milk and dairy beverages. *Food Science and Technology*, *20*(3), 314-317.

98. Cartoni, G., Coccioli, F., Jasionowska, R., & Masci, M. (1999). Determination of cows' milk in goats' milk and cheese by capillary electrophoresis of the whey protein fractions. *Journal of Chromatography A*, *846*(1-2), 135-141.

99. R. Vemireddy, L., Archak, S., & Nagaraju, J. (2007). Capillary electrophoresis is essential for microsatellite marker based detection and quantification of adulteration of Basmati rice (Oryza sativa). *Journal of agricultural and food chemistry*, *55*(20), 8112-8117.

100. Ren, Q. R., Zhang, H., Guo, H. Y., Jiang, L., Tian, M., & Ren, F. Z. (2014). Detection of cow milk adulteration in yak milk by ELISA. *Journal of dairy science*, *97*(10), 6000-6006.

101. Pizzano, R., & Salimei, E. (2014). Isoelectric focusing and ELISA for detecting adulteration of donkey milk with cow milk. *Journal of agricultural and food chemistry*, *62*(25), 5853-5858.

102. Olsen, J. E. (2000). DNA-based methods for detection of food-borne bacterial pathogens. *Food Research International*, *33*(3-4), 257-266.

103. Prasad, D., & Vidyarthi, A. S. (2009). DNA based methods used for characterization and detection of food borne bacterial pathogens with special consideration to recent rapid methods. *African Journal of Biotechnology*, *8*(9).

104. Alberts, B., Johnson, A., Lewis, J., Raff, M., Roberts, K., & Walter, P. (2002). Isolating, cloning, and sequencing DNA. In *Molecular Biology of the Cell. 4th edition*. Garland Science.

105. Mafra, I., Ferreira, I. M., & Oliveira, M. B. P. (2008). Food authentication by PCR-based methods. *European Food Research and Technology*, *227*(3), 649-665.

106. Vemireddy, L. R., Satyavathi, V. V., Siddiq, E. A., & Nagaraju, J. (2015). Review of methods for the detection and quantification of adulteration of rice: Basmati as a case study. *Journal of food science and technology*, *52*(6), 3187-3202.

107. Li, T. T., Jalbani, Y. M., Zhang, G. L., Zhao, Z. Y., Wang, Z. Y., Zhao, X. Y., & Chen, A. L. (2019). Detection of goat meat adulteration by real-time PCR based on a reference primer. *Food chemistry*, *277*, 554-557.

108. Kumari, R., Rank, D. N., Kumar, S., Joshi, C. G., & Lal, S. V. (2015). Real time PCR-an approach to detect meat adulteration. *Buffalo Bulletin*, *34*(1), 124-129.

109. Wang, Q., Cai, Y., He, Y., Yang, L., & Pan, L. (2018). Collaborative ring trial of two real-time PCR assays for the detection of porcine-and chicken-derived material in meat products. *PloS one*, *13*(10), e0206609.

110. Kaltenbrunner, M., Hochegger, R., & Cichna-Markl, M. (2018). Red deer (Cervus elaphus)-specific real-time PCR assay for the detection of food adulteration. *Food control*, *89*, 157-166.

111. Soares, S., Amaral, J. S., Mafra, I., & Oliveira, M. B. P. (2010). Quantitative detection of poultry meat adulteration with pork by a duplex PCR assay. *Meat science*, *85*(3), 531-536.

112. Kotowicz, M., Adamczyk, E., & Bania, J. (2007). Application of a duplex-PCR for detection of cows' milk in goats' milk. *Annals of Agricultural and Environmental Medicine*, *14*(2).

113. Dhanya, K., Syamkumar, S., Jaleel, K., & Sasikumar, B. (2008). Random amplified polymorphic DNA technique for detection of plant based adul-terants in chilli powder (Capsicum annuum). *Journal of Spices and Aromatic Crops*, *17*(2), 75-81.

114. Dhanya, K., Syamkumar, S., Siju, S., & Sasikumar, B. (2011). SCAR markers for adulterant detection in ground chilli. *British Food Journal*, *113*(5), 656-668.

115. Cunha, J. T., Ribeiro, T. I., Rocha, J. B., Nunes, J., Teixeira, J. A., & Domingues, L. (2016). RAPD and SCAR markers as potential tools for detection of milk origin in dairy products: Adulterant sheep breeds in Serra da Estrela cheese production. *Food chemistry*, *211*, 631-636.

116. Dhanya, K., Syamkumar, S., & Sasikumar, B. (2009). Development and application of SCAR marker for the detection of papaya seed adulteration in traded black pepper powder. *Food Biotechnology*, *23*(2), 97-106.

117. Deb, R., Sengar, G. S., Singh, U., Kumar, S., Alyethodi, R. R., Alex, R., ... & Prakash, B. (2016). Application of a loop-mediated isothermal amplification assay for rapid detection of cow components adulterated in buffalo milk/meat. *Molecular biotechnology*, *58*(12), 850-860.

118. Wang, D., Wang, Y., Zhu, K., Shi, L., Zhang, M., Yu, J., & Liu, Y. (2019). Detection of Cassava Component in Sweet Potato Noodles by Real-Time Loop-mediated Isothermal Amplification (Real-time LAMP) Method. *Molecules*, *24*(11), 2043.

119. Zhao, M., Shi, Y., Wu, L., Guo, L., Liu, W., Xiong, C., ... & Chen, S. (2016). Rapid authentication of the precious herb saffron by loop-mediated isothermal amplification (LAMP) based on internal transcribed spacer 2 (ITS2) sequence. *Scientific reports*, *6*, 25370.

120. Park, M., Tsai, S. L., & Chen, W. (2013). Microbial biosensors: engineered microorganisms as the sensing machinery. *Sensors*, *13*(5), 5777-5795.

121. Mehrotra, P. (2016). Biosensors and their applications–A review. *Journal of oral biology and craniofacial research*, *6*(2), 153-159.

122. Bunney, J., Williamson, S., Atkin, D., Jeanneret, M., Cozzolino, D., & Chapman, J. (2017). The use of electrochemical biosensors in food analysis. *Current Research in Nutrition and Food Science Journal*, *5*(3), 183-195.

123. Sforza, S., Corradini, R., Tedeschi, T., & Marchelli, R. (2011). Food analysis and food authentication by peptide nucleic acid (PNA)-based technologies. *Chemical Society Reviews*, *40*(1), 221-232.

124. Germini, A., Mezzelani, A., Lesignoli, F., Corradini, R., Marchelli, R., Bordoni, R., ... & De Bellis, G. (2004). Detection of genetically modified soybean using peptide nucleic acids (PNAs) and microarray technology. *Journal of agricultural and food chemistry*, *52*(14), 4535-4540.

125. Dominguez-Benetton, X., Sevda, S., Vanbroekhoven, K., & Pant, D. (2012). The accurate use of impedance analysis for the study of microbial electrochemical systems. *Chemical Society Reviews*, *41*(21), 7228-7246.

126. Yang, L. (2008). Electrical impedance spectroscopy for detection of bacterial cells in suspensions using interdigitated microelectrodes. *Talanta*, *74*(5), 1621-1629.

127. Hamlaoui, Y., Pedraza, F., & Tifouti, L. (2008). Corrosion monitoring of galvanised coatings through electrochemical impedance spectroscopy. *Corrosion Science*, *50*(6), 1558-1566.

128. Mansfeld, F. (1995). Use of electrochemical impedance spectroscopy for the study of corrosion protection by polymer coatings. *Journal of Applied Electrochemistry*, *25*(3), 187-202.

129. Nishikata, A., Ichihara, Y., & Tsuru, T. (1995). An application of electrochemical impedance spectroscopy to atmospheric corrosion study. *Corrosion science*, *37*(6), 897-911.

130. Mansfeld, F., & Kendig, M. W. (1985). Electrochemical impedance spectroscopy of protective coatings. *Materials and Corrosion*, *36*(11), 473-483.

131. Jayaraj, B., Vishweswaraiah, S., Desai, V. H., & Sohn, Y. H. (2004). Electrochemical impedance spectroscopy of thermal barrier coatings as a function of isothermal and cyclic thermal exposure. *Surface and Coatings Technology*, *177*, 140-151.

132. Elsener, B., Rota, A., & Böhni, H. (1989). Impedance study on the corrosion of PVD and CVD titanium nitride coatings. In *Materials Science Forum* (Vol. 44, pp. 29-38). Trans Tech Publications.

133. Hilpert, T., & Ivers-Tiffee, E. (2004). Correlation of electrical and mechanical properties of zirconia based thermal barrier coatings. *Solid State Ionics*, *175*(1-4), 471-476.

134. Cabeza, M., Merino, P., Miranda, A., Nóvoa, X. R., & Sanchez, I. (2002). Impedance spectroscopy study of hardened Portland cement paste. *Cement and Concrete Research*, *32*(6), 881-891.

135. Andrade, C., Blanco, V. M., Collazo, A., Keddam, M., Novoa, X. R., & Takenouti, H. (1999). Cement paste hardening process studied by impedance spectroscopy. *Electrochimica acta*, *44*(24), 4313-4318.

136. Iqbal, M. Z. (2016). Preparation, characterization, electrical conductivity and dielectric studies of Na2SO4 and V2O5 composite solid electrolytes. *Measurement*, *81*, 102-112.

137. Wendler, F., Büschel, P., Kanoun, O., Schadewald, J., Bufon, C. B., & Schmidt, O. G. (2012, March). Impedance spectroscopy in solid state electrolyte characterization. In *International Multi-Conference on Systems, Signals & Devices* (pp. 1-4). IEEE.

138. Lukaski, H. C. (1999). Requirements for clinical use of bioelectrical impedance analysis (BIA). *ANNALS-NEW YORK ACADEMY OF SCIENCES*, *873*, 72-76.

139. Baarends, E. M., Lichtenbelt, W. V. M., Wouters, E. F. M., & Schols, A. M. W. J. (1998). Body-water compartments measured by bio-electrical impedance spectroscopy in patients with chronic obstructive pulmonary disease. *Clinical Nutrition*, *17*(1), 15-22.

140. Min, M., Ollmar, S., & Gersing, E. (2003). Electrical impedance and cardiac monitoring-technology, potential and applications. *International Journal of Bioelectromagnetism*, *5*(1), 53-56.

141. Lukaski, H. C. (1996). Biological indexes considered in the derivation of the bioelectrical impedance analysis. *The American journal of clinical nutrition*, *64*(3), 397S-404S.

142. Scandurra, G., Tripodi, G., & Verzera, A. (2013). Impedance spectroscopy for rapid determination of honey floral origin. *Journal of Food Engineering*, *119*(4), 738-743.

143. Paszkowski, B., Wilczek, A., Szypłowska, A., Nakonieczna, A., & Skierucha, W. (2014). A low-frequency sensor for determination of honey electrical properties in varying temperature conditions. *Journal of food engineering*, *138*, 17-22.

144. KITAMURA, Y., TOYODA, K., & PARK, B. (2000). Electric impedance spectroscopy for yogurt processing. *Food science and technology research*, *6*(4), 310-313.

145. Funke, K. (2013). Solid state ionics: from Michael Faraday to green energy—the European dimension. *Science and technology of advanced materials*, *14*(4), 043502.

146. Raistrick, I. D. (1986). Application of impedance spectroscopy to materials science. *Annual Review of Materials Science*, *16*(1), 343-370.

147. Joshi, M. (2017). Importance of Impedance Spectroscopy Technique in Materials Characterization: A Brief Review. *Mechanics, Materials Science & Engineering MMSE Journal. Open Access*, *9*.

148. Lvovich, V. F. (2012). *Impedance spectroscopy: applications to electrochemical and dielectric phenomena*. John Wiley & Sons.

149. Lasia, A. (2002). Electrochemical impedance spectroscopy and its applications. In *Modern aspects of electrochemistry* (pp. 143-248). Springer, Boston, MA.

150. Lackermeier, A. H., McAdams, E. T., Moss, G. P., & Woolfson, A. D. (1999). In vivo ac impedance spectroscopy of human skin: theory and problems in monitoring of passive percutaneous drug delivery. *Annals of the New York Academy of Sciences*, *873*(1), 197-213.

151. Clemente, F., Arpaia, P., & Manna, C. (2013). Characterization of human skin impedance after electrical treatment for transdermal drug delivery. *Measurement*, *46*(9), 3494-3501.

152. Nover, G., Heikamp, S., & Freund, D. (2000). Electrical impedance spectroscopy used as a tool for the detection of fractures in rock samples exposed to either hydrostatic or triaxial pressure conditions. In *Natural Hazards* (pp. 317-330). Springer, Dordrecht.

153. Kahraman, S., & Alber, M. (2006). Predicting the physico-mechanical properties of rocks from electrical impedance spectroscopy measurements. *International Journal of Rock Mechanics and Mining Sciences*, *43*(4), 543-553.

154. Ahmed, R., & Reifsnider, K. (2011). Study of influence of electrode geometry on impedance spectroscopy. *Int. J. Electrochem. Sci*, *6*, 1159-1174.

155. Holder, D. (Ed.). (1993). *Clinical and physiological applications of electrical impedance tomography*. CRC Press.

156. Adler, A., & Guardo, R. (1994). A neural network image reconstruction technique for electrical impedance tomography. *IEEE Transactions on Medical Imaging*, *13*(4), 594-600.

157. Dong, Y., Ma, Y., Wang, H., Zhang, J., Zhang, G., & Yang, M. S. (2013). Assessment of tolerance of willows to saline soils through electrical impedance measurements. *Forest Science and Practice*, *15*(1), 32-40.

158. Zhang, X., Yu, N., Xi, G., & Meng, X. (2012). Changes in the power spectrum of electrical signals in maize leaf induced by osmotic stress. *Chinese Science Bulletin*, *57*(4), 413-420.

159. Zhao, X., Zhuang, H., Yoon, S. C., Dong, Y., Wang, W., & Zhao, W. (2017). Electrical impedance spectroscopy for quality assessment of meat and fish: A review on basic principles, measurement methods, and recent advances. *Journal of Food Quality*, *2017*.

160. Jha, S.N., Narsaiah, K., Basediya, A.L., Sharma, R., Jaiswal, P., Kumar, R. and Bhardwaj, R., 2011. Measurement techniques and application of electrical properties for nondestructive quality evaluation of foods—a review. *Journal of food science and technology*, *48*(4), pp.387-411.

161. Mishra, G. K., Barfidokht, A., Tehrani, F., & Mishra, R. K. (2018). Food safety analysis using electrochemical biosensors. *Foods*, *7*(9), 141.

162. Vasilescu, A., & Marty, J. L. (2016). Electrochemical aptasensors for the assessment of food quality and safety. *TrAC Trends in Analytical Chemistry*, *79*, 60-70.

163. Mészáros, P., Vozáry, E., & B. Funk, D. (2005). Connection between moisture content and electrical parameters of apple slices during drying. *Progress in Agricultural Engineering Sciences*, *1*(1), 95-121.

164. Kertész, Á., Hlaváčová, Z., Vozáry, E., & Staroňová, L. (2015). Relationship between moisture content and electrical impedance of carrot slices during drying. *International Agrophysics*, *29*(1), 61-66.

165. Toyoda, K., Tsenkova, R. N., & Nakamura, M. (2001). Characterization of osmotic dehydration and swelling of apple tissues by bioelectrical impedance spectroscopy. *Drying Technology*, *19*(8), 1683-1695.

166. Ando, Y., Maeda, Y., Mizutani, K., Wakatsuki, N., Hagiwara, S., & Nabetani, H. (2016). Effect of air-dehydration pretreatment before freezing on the electrical impedance characteristics and texture of carrots. *Journal of Food Engineering*, *169*, 114-121.

167. Repo, T., Paine, D. H., & Taylor, A. G. (2002). Electrical impedance spectroscopy in relation to seed viability and moisture content in snap bean (Phaseolus vulgaris L.). *Seed Science Research*, *12*(1), 17-29.

168. Chowdhury, A., Bera, T. K., Ghoshal, D., & Chakraborty, B. (2015, February). Studying the electrical impedance variations in banana ripening using electrical impedance spectroscopy (EIS). In *Proceedings of the 2015 Third International Conference on Computer, Communication, Control and Information Technology (C3IT)* (pp. 1-4). IEEE.

169. Chowdhury, A., Kanti Bera, T., Ghoshal, D., & Chakraborty, B. (2017). Electrical impedance variations in banana ripening: an analytical study with electrical impedance spectroscopy. *Journal of food process engineering*, *40*(2), e12387.

170. Liu, X., Fang, Q., Zheng, S., & Cosic, I. (2008). Electrical impedance spectroscopy investigation on banana ripening. In *Europe-Asia Symposium on Quality Management in Postharvest Systems-Eurasia 2007*. International Society for Horticultural Science.

171. Neto, A. F., Olivier, N. C., Cordeiro, E. R., & de Oliveira, H. P. (2017). Determination of mango ripening degree by electrical impedance spectroscopy. *Computers and Electronics in Agriculture*, *143*, 222-226.

172. Watanabe, T., Ando, Y., Orikasa, T., Shiina, T., & Kohyama, K. (2017). Effect of short time heating on the mechanical fracture and electrical impedance properties of spinach (Spinacia oleracea L.). *Journal of food engineering*, *194*, 9-14.

173. Ando, Y., Mizutani, K., & Wakatsuki, N. (2014). Electrical impedance analysis of potato tissues during drying. *Journal of Food Engineering*, *121*, 24-31.

174. Bai, X et al. (2018). Electrical impedance analysis of pork tissues during storage. *Journal of Food Measurement and Characterization*, *12*(1), 164-172.

175. Nguyen, H. B., & Nguyen, L. T. (2015). Rapid and non-invasive evaluation of pork meat quality during storage via impedance measurement. *International Journal of Food Science & Technology*, *50*(8), 1718-1725.

176. Damez, J. L., Clerjon, S., Abouelkaram, S., & Lepetit, J. (2008). Beef meat electrical impedance spectroscopy and anisotropy sensing for non-invasive early assessment of meat ageing. *Journal of Food Engineering*, *85*(1), 116-122.

177. Damez, J. L., Clerjon, S., Abouelkaram, S., & Lepetit, J. (2007). Dielectric behavior of beef meat in the 1–1500 kHz range: Simulation with the Fricke/Cole–Cole model. *Meat Science*, *77*(4), 512-519.

178. Guermazi, M., Kanoun, O., & Derbel, N. (2014). Investigation of long time beef and veal meat behavior by bioimpedance spectroscopy for meat monitoring. *IEEE Sensors Journal*, *14*(10), 3624-3630.

179. Ćurić, T., Marušić Radovčić, N., Janči, T., Lacković, I., & Vidaček, S. (2017). Salt and moisture content determination of fish by bioelectrical impedance and a needle-type multi-electrode array. *International journal of food properties*, *20*(11), 2477-2486.

180. Bertemes-Filho, P., Valicheski, R., Pereira, R. M., & Paterno, A. S. (2010). Bioelectrical impedance analysis for bovine milk: preliminary results. In *Journal of Physics: Conference Series* (Vol. 224, No. 1, p. 012133). IOP Publishing.

181. Khaled, A. Y., Aziz, S. A., & Rokhani, F. Z. (2014). Development and evaluation of an impedance spectroscopy sensor to assess cooking oil quality. *International Journal of Environmental Science and Development*, *5*(3), 299.

182. Valli, E., Bendini, A., Berardinelli, A., Ragni, L., Riccò, B., Grossi, M., & Gallina Toschi, T. (2016). Rapid and innovative instrumental approaches for quality and authenticity of olive oils. *European journal of lipid science and technology*, *118*(11), 1601-1619.

183.  He, F. (2014). Development of capillary-driven microfludic biosensors for food safety and quality assurance.

184.  Gupta, S., Ramesh, K., Ahmed, S., & Kakkar, V. (2016). Lab-on-chip technology: A review on design trends and future scope in biomedical applications. *Int. J. Bio-Sci. Bio-Technol, 8*, 311-322.

185.  Kelly, J. D., & Downey, G. (2005). Detection of sugar adulterants in apple juice using Fourier transform infrared spectroscopy and chemometrics. *Journal of Agricultural and Food Chemistry, 53*(9), 3281-3286.

186.  Dixit S, Consumer Guidance Society of India (CGSI). http://cgsiindia.org/wp-content/uploads/2015/01/Identifying-common-food-adulterants.pdf (2015).

187.  Rodriguez-Saona, L. E., & Allendorf, M. E. (2011). Use of FTIR for rapid authentication and detection of adulteration of food. *Annual review of food science and technology, 2*, 467-483.

188.  Kumaravelu, C., & Gopal, A. (2015). Detection and quantification of adulteration in honey through near infrared spectroscopy. *International Journal of Food Properties, 18*(9), 1930-1935.

189.  Kim, J. M., Kim, H. J., & Park, J. M. (2015). Determination of milk fat adulteration with vegetable oils and animal fats by gas chromatographic analysis. *Journal of food science, 80*(9), C1945-C1951.

190.  Domingues, D. S., Pauli, E. D., de Abreu, J. E., Massura, F. W., Cristiano, V., Santos, M. J., & Nixdorf, S. L. (2014). Detection of roasted and ground coffee adulteration by HPLC by amperometric and by post-column derivatization UV–Vis detection. *Food chemistry, 146*, 353-362.

191.  Rivellino, S. R., Hantao, L. W., Risticevic, S., Carasek, E., Pawliszyn, J., & Augusto, F. (2013). Detection of extraction artifacts in the analysis of honey volatiles using comprehensive two-dimensional gas chromatography. *Food chemistry, 141*(3), 1828-1833.

192.  Woodbury, S. E., Evershed, R. P., Rossell, J. B., Griffith, R. E., & Farnell, P. (1995). Detection of vegetable oil adulteration using gas chromatography combustion/isotope ratio mass spectrometry. *Analytical Chemistry, 67*(15), 2685-2690.

193. Al-Jowder, O., Kemsley, E. K., & Wilson, R. H. (2002). Detection of adulteration in cooked meat products by mid-infrared spectroscopy. *Journal of agricultural and food chemistry*, *50*(6), 1325-1329.

194. Bertelli, D., Lolli, M., Papotti, G., Bortolotti, L., Serra, G., & Plessi, M. (2010). Detection of honey adulteration by sugar syrups using one-dimensional and two-dimensional high-resolution nuclear magnetic resonance. *Journal of agricultural and food chemistry*, *58*(15), 8495-8501.

195. Rios-Corripio, M. A., Rojas-López*, M., & Delgado-Macuil, R. (2012). Analysis of adulteration in honey with standard sugar solutions and syrups using attenuated total reflectance-Fourier transform infrared spectroscopy and multivariate methods. *CyTA-Journal of Food*, *10*(2), 119-122.

196. Kamal, M. A., & Klein, P. (2011). Determination of sugars in honey by liquid chromatography. *Saudi journal of biological sciences*, *18*(1), 17-21.

197. Sharma, R., Rajput, Y. S., Dogra, G., & TOMAR, S. K. (2009). Estimation of sugars in milk by HPLC and its application in detection of adulteration of milk with soymilk. *International journal of dairy technology*, *62*(4), 514-519.

198. Guo, W., Zhu, X., Liu, Y., & Zhuang, H. (2010). Sugar and water contents of honey with dielectric property sensing. *Journal of Food Engineering*, *97*(2), 275-281.

199. Guo, W., Liu, Y., Zhu, X., & Wang, S. (2011). Temperature-dependent dielectric properties of honey associated with dielectric heating. *Journal of Food Engineering*, *102*(3), 209-216.

200. Macdonald JR, (1992). Impedance spectroscopy. Annals of Biomedical Engineering, *20*, 289-305.

201. Zhu, Z., Frey, O., Haandbaek, N., Franke, F., Rudolf, F., & Hierlemann, A. (2015). Time-lapse electrical impedance spectroscopy for monitoring the cell cycle of single immobilized S. pombe cells. *Scientific reports*, *5*, 17180.

202. Zhang, M. I. N., Repo, T., Willison, J. H. M., & Sutinen, S. (1995). Electrical impedance analysis in plant tissues: on the biological meaning of Cole-Cole α in Scots pine needles. *European Biophysics Journal*, *24*(2), 99-106.

203. Kuralay, F., Erdem, A., Abacı, S., Özyörük, H., & Yıldız, A. (2009). Characterization of redox polymer based electrode and electrochemical behavior for DNA detection. *Analytica chimica acta*, *643*(1-2), 83-89.

204. Cho, S., & Thielecke, H. (2008). Electrical characterization of human mesenchymal stem cell growth on microelectrode. *Microelectronic Engineering*, *85*(5-6), 1272-1274.

205. Silva, M. G., Helali, S., Esseghaier, C., Suarez, C. E., Oliva, A., & Abdelghani, A. (2008). An impedance spectroscopy method for the detection and evaluation of Babesia bovis antibodies in cattle. *Sensors and Actuators B: Chemical*, *135*(1), 206-213.

206. Yang, W., Butler, J. E., Russell Jr, J. N., & Hamers, R. J. (2007). Direct electrical detection of antigen–antibody binding on diamond and silicon substrates using electrical impedance spectroscopy. *Analyst*, *132*(4), 296-306.

207. Repo, T., Zhang, G. A. N. G., Ryyppö, A., & Rikala, R. (2000). The electrical impedance spectroscopy of Scots pine (Pinus sylvestris L.) shoots in relation to cold acclimation. *Journal of Experimental Botany*, *51*(353), 2095-2107.

208. Spinelli, F. R., Dutra, S. V., Carnieli, G., Leonardelli, S., Drehmer, A. P., & Vanderlinde, R. (2016). Detection of addition of apple juice in purple grape juice. *Food Control*, *69*, 1-4.

209. Das, S., Sivaramakrishna, M., Biswas, K., & Goswami, B. (2011). Performance study of a 'constant phase angle based'impedance sensor to detect milk adulteration. *Sensors and Actuators A: Physical*, *167*(2), 273-278.

210. Guo, W., Zhu, X., Liu, H., Yue, R., & Wang, S. (2010). Effects of milk concentration and freshness on microwave dielectric properties. *Journal of Food Engineering*, *99*(3), 344-350.

211. Halambre, S. N., & Badhe, S. G. (2013). Characterization of milk using impedance analysis technique. *Deccean Current Science International Research Journal*, *8*, 132-136.

212. Seidlová, R., Poživil, J., Seidl, J., Ďaďo, S., Průšová, P., & Malec, L. (2015). Electrode and electrodeless impedance measurement for determination of orange juices parameters. *Chemical Papers*, *69*(7), 938-949.

213. Yang, L., & Bashir, R. (2008). Electrical/electrochemical impedance for rapid detection of foodborne pathogenic bacteria. *Biotechnology advances*, *26*(2), 135-150.

214. Griffiths DJ, (2012). Introduction to Electrodynamics. (4$^{th}$ ed.). Prentice-Hall Inc. (Chapter 4)

215. Kaškonienė, V., & Venskutonis, P. R. (2010). Floral markers in honey of various botanical and geographic origins: a review. *Comprehensive reviews in food science and food safety*, *9*(6), 620-634.

216. Buba, F., Gidado, A., & Shugaba, A. (2013). Physicochemical and microbiological properties of honey from North East Nigeria. *Biochem Anal Biochem*, *2*(142), 61-67.

217. Boussaid, A., Chouaibi, M., Rezig, L., Hellal, R., Donsì, F., Ferrari, G., & Hamdi, S. (2018). Physicochemical and bioactive properties of six honey samples from various floral origins from Tunisia. *Arabian Journal of Chemistry*, *11*(2), 265-274.

218. El Sohaimy, S. A., Masry, S. H. D., & Shehata, M. G. (2015). Physicochemical characteristics of honey from different origins. *Annals of Agricultural Sciences*, *60*(2), 279-287.

219. Rebiai, A., & Lanez, T. (2014). Comparative study of honey collected from different flora of Algeria. *Journal of Fundamental and Applied Sciences*, *6*(1), 48-55.

220. IS 4941:1994 Extracted Honey – Specification, 01 April (2011)

221. Kelly, J. D., Downey, G., & Fouratier, V. (2004). Initial study of honey adulteration by sugar solutions using midinfrared (MIR) spectroscopy and chemometrics. *Journal of agricultural and food chemistry*, *52*(1), 33-39.

222. Anjos, O., Campos, M. G., Ruiz, P. C., & Antunes, P. (2015). Application of FTIR-ATR spectroscopy to the quantification of sugar in honey. *Food chemistry*, *169*, 218-223.

223. Chen, L., Xue, X., Ye, Z., Zhou, J., Chen, F., & Zhao, J. (2011). Determination of Chinese honey adulterated with high fructose corn syrup by near infrared spectroscopy. *Food Chemistry*, *128*(4), 1110-1114.

224. Sivakesava, S., & Irudayaraj, J. (2001). A rapid spectroscopic technique for determining honey adulteration with corn syrup. *Journal of Food Science*, *66*(6), 787-791.

225. Guo, W., Liu, Y., Zhu, X., & Wang, S. (2011). Dielectric properties of honey adulterated with sucrose syrup. *Journal of Food Engineering*, *107*(1), 1-7.

226. Li, S., Shan, Y., Zhu, X., Zhang, X., & Ling, G. (2012). Detection of honey adulteration by high fructose corn syrup and maltose syrup using Raman spectroscopy. *Journal of Food Composition and Analysis*, *28*(1), 69-74.

227. Monakhova, Y. B., Kuballa, T., & Lachenmeier, D. W. (2013). Chemometric methods in NMR spectroscopic analysis of food products. *Journal of Analytical Chemistry*, *68*(9), 755-766.

228. Lenhardt, L., Zeković, I., Dramićanin, T., Tešić, Ž., Milojković-Opsenica, D., & Dramićanin, M. D. (2014). Authentication of the botanical origin of unifloral honey by infrared spectroscopy coupled with support vector machine algorithm. *Physica Scripta*, *2014*(T162), 014042.

229. Pataca, L. C., Neto, W. B., Marcucci, M. C., & Poppi, R. J. (2007). Determination of apparent reducing sugars, moisture and acidity in honey by attenuated total reflectance-Fourier transform infrared spectrometry. *Talanta*, *71*(5), 1926-1931.

230. Wang, J., Kliks, M. M., Jun, S., Jackson, M., & Li, Q. X. (2010). Rapid analysis of glucose, fructose, sucrose, and maltose in honeys from different geographic regions using Fourier transform infrared spectroscopy and multivariate analysis. *Journal of food science*, *75*(2), C208-C214.

231. Tewari, J., & Irudayaraj, J. (2004). Quantification of saccharides in multiple floral honeys using fourier transform infrared microattenuated total reflectance spectroscopy. *Journal of agricultural and food chemistry*, *52*(11), 3237-3243.

232. Qiu, P. Y., Ding, H. B., Tang, Y. K., & Xu, R. J. (1999). Determination of chemical composition of commercial honey by near-infrared spectroscopy. *Journal of agricultural and food chemistry*, *47*(7), 2760-2765.

233. Rybak-Chmielewska, H. (2007). High performance liquid chromatography (HPLC) study of sugar composition in some kinds of natural honey and winter stores processed by bees from starch syrup. *Journal of Apicultural Science*, *51*(1), 23-38.

234. Makawi, S. Z. A., Gadkariem, E. A., & Ayoub, S. M. H. (2009). Determination of antioxidant flavonoids in Sudanese honey samples by solid phase extraction and high performance liquid chromatography. *Journal of Chemistry*, *6*(S1), S429-S437.

235. Jerković, I., Tuberso, C. I., Gugić, M., & Bubalo, D. (2010). Composition of sulla (Hedysarum coronarium L.) honey solvent extractives determined by GC/MS: norisoprenoids and other volatile organic compounds. *Molecules*, *15*(9), 6375-6385.

236. Tosun, M. (2013). Detection of adulteration in honey samples added various sugar syrups with 13C/12C isotope ratio analysis method. *Food chemistry*, *138*(2-3), 1629-1632.

237. Basak, S. P., Kanjilal, B., Sarkar, P., & Turner, A. P. (2013). Application of electrical impedance spectroscopy and amperometry in polyaniline modified ammonia gas sensor. *Synthetic metals*, *175*, 127-133.

238. Lin, C. M., Chen, L. H., & Chen, T. M. (2012). The development and application of an electrical impedance spectroscopy measurement system for plant tissues. *Computers and electronics in agriculture*, *82*, 96-99.

239. Galvis-Sánchez, A. C., Barros, A., & Delgadillo, I. (2007). FTIR-ATR infrared spectroscopy for the detection of ochratoxin A in dried vine fruit. *Food additives and contaminants*, *24*(11), 1299-1305.

240. D.W.Sun, (2009). Infrared spectroscopy for food quality analysis and control, Academic Press.

241. Hafner, J. (1985). Bond-angle distribution functions in metallic glasses. *Le Journal de Physique Colloques*, *46*(C9), C9-69.

242. Cronin, J. R. (2003). Curcumin: Old spice is a new medicine. *Alternative & Complementary Therapies*, *9*(1), 34-38.

243. Zorofchian Moghadamtousi, S., Abdul Kadir, H., Hassandarvish, P., Tajik, H., Abubakar, S., & Zandi, K. (2014). A review on antibacterial, antiviral, and antifungal activity of curcumin. *BioMed research international*, *2014*.

244. Duvoix, A et al. (2005). Chemopreventive and therapeutic effects of curcumin. *Cancer letters*, *223*(2), 181-190.

245. Hishikawa, N et al. (2012). Effects of turmeric on Alzheimer's disease with behavioral and psychological symptoms of dementia. *Ayu*, *33*(4), 499.

246. Sasikumar, B., Syamkumar, S., Remya, R., & John Zachariah, T. (2004). PCR based detection of adulteration in the market samples of turmeric powder. *Food Biotechnology*, *18*(3), 299-306.

247. Sen, A. R., Gupta, P. S., & Dastidar, N. G. (1974). Detection of Curcuma zedoaria and Curcuma aromatica in Curcuma longa (turmeric) by thin-layer chromatography. *Analyst*, *99*(1176), 153-155.

248. Gupta, S., Sundarrajan, M., & Rao, K. V. K. (2003). Tumor promotion by metanil yellow and malachite green during rat hepatocarcinogenesis is associated with dysregulated expression of cell cycle regulatory proteins. *Teratogenesis, carcinogenesis, and mutagenesis*, *23*(S1), 301-312.

249. Nagaraja, T. N., & Desiraju, T. (1993). Effects of chronic consumption of metanil yellow by developing and adult rats on brain regional levels of noradrenaline, dopamine and serotonin, on acetylcholine esterase activity and on operant conditioning. *Food and chemical toxicology*, *31*(1), 41-44.

250. Fernandes, C., & Rao, K. V. (1994). Dose related promoter effect of metanil yellow on the development of hepatic pre-neoplastic lesions induced by N-nitrosodiethylamine in rats. *The Indian journal of medical research*, *100*, 140-149.

251. Prasad, O. M., & Rastogi, P. B. (1983). Haematological changes induced by feeding a common food colour, metanil yellow, in albino mice. *Toxicology letters*, *16*(1-2), 103-107.

252. Nallappan, K., Dash, J., Ray, S., & Pesala, B. (2013, September). Identification of adulterants in turmeric powder using terahertz spectroscopy. In *2013 38th International Conference on Infrared, Millimeter, and Terahertz Waves (IRMMW-THz)* (pp. 1-2). IEEE.

253. Dhanya, K., Syamkumar, S., Siju, S., & Sasikumar, B. (2011). Sequence characterized amplified region markers: A reliable tool for adulterant detection in turmeric powder. *Food research international*, *44*(9), 2889-2895.

254. Tiwari, M., Agrawal, R., Pathak*, A. K., Rai, A. K., & Rai, G. K. (2013). Laser-induced breakdown spectroscopy: an approach to detect adulteration in turmeric. *Spectroscopy Letters*, *46*(3), 155-159.

255. Dhakal, S., Chao, K., Schmidt, W., Qin, J., Kim, M., & Chan, D. (2016). Evaluation of turmeric powder adulterated with metanil yellow using FT-Raman and FT-IR spectroscopy. *Foods*, *5*(2), 36.

256. Guo, X., Wei, Q., Du, B., Zhang, Y., Xin, X., Yan, L., & Yu, H. (2013). Removal of Metanil Yellow from water environment by amino functionalized graphenes (NH2-G)–Influence of surface chemistry of NH2-G. *Applied Surface Science*, *284*, 862-869.

257. Eisenschitz R, London F (1930) On the relationship of van der Waals forces to homopolar bonding forces. FZ Physik 60(7):491–527

258. Krishnaji, & Srivastava, S. L. (1964). First-Order London Dispersion Forces and Microwave Spectral Linewidth. *The Journal of Chemical Physics*, *41*(8), 2266-2270.

259. Bergmann, E., & Weizmann, A. (1936). Dipole moment and molecular structure. Part XVII. The dipole moments of azo-dyes and some similar substances. *Transactions of the Faraday Society*, *32*, 1318-1326.

260. Ghanadzadeh, A., Shahzamanian, M. A., Shoarinejad, S., Zakerhamidi, M. S., & Moghadam, M. (2007). Guest–host interaction of some aminoazobenzene dyes doped in liquid crystalline matrix. *Journal of Molecular Liquids*, *136*(1-2), 22-28.

261. Zakerhamidi, M. S., Ahmadi-Kandjani, S., Moghadam, M., Ortyl, E., & Kucharski, S. (2012). Solvatochromism effects on the dipole moments and photo-physical behavior of some azo sulfonamide dyes. *Spectrochimica Acta Part A: Molecular and Biomolecular Spectroscopy*, *85*(1), 105-110.

262. Grossi, M., & Riccò, B. (2017). An automatic titration system for oil concentration measurement in metalworking fluids. *Measurement*, *97*, 8-14.

263. Grossi, M., Di Lecce, G., Toschi, T. G., & Riccò, B. (2014). A novel electrochemical method for olive oil acidity determination. *Microelectronics Journal*, *45*(12), 1701-1707.

264. Swinehart, D. F. (1962). The beer-lambert law. *Journal of chemical education*, *39*(7), 333.

265. Singh, V., & Mishra, A. K. (2015). White light emission from vegetable extracts. *Scientific reports*, *5*, 11118.

266. Mainreck, N., Brézillon, S., Sockalingum, G. D., Maquart, F. X., Manfait, M., & Wegrowski, Y. (2011). Rapid characterization of glycosaminoglycans using a combined approach by infrared and Raman microspectroscopies. *Journal of pharmaceutical sciences*, *100*(2), 441-450.

267. Samiei, E., Tabrizian, M., & Hoorfar, M. (2016). A review of digital microfluidics as portable platforms for lab-on a-chip applications. *Lab on a Chip*, *16*(13), 2376-2396.

268. Ho, T. Y., Zeng, J., & Chakrabarty, K. (2010, November). Digital microfluidic biochips: A vision for functional diversity and more than Moore. In *Proceedings of the International Conference on Computer-Aided Design* (pp. 578-585). IEEE Press.

269. Oosterbroek, R. E., Oosterbroek, E., & van den Berg, A. (Eds.). (2003). *Lab-on-a-chip: miniaturized systems for (bio) chemical analysis and synthesis*. Elsevier.

270. Chakrabarty, K., & Zeng, J. (2005). Design automation for microfluidics-based biochips. *ACM Journal on Emerging Technologies in Computing Systems (JETC)*, *1*(3), 186-223.

271. Ho, T. Y., Chakrabarty, K., & Pop, P. (2011, October). Digital microfluidic biochips: recent research and emerging challenges. In *Proceedings of the seventh IEEE/ACM/IFIP international conference on Hardware/software codesign and system synthesis* (pp. 335-344). ACM.

272. Roy, P., Sohid, M., Chakraborty, S., Rahaman, H., & Dasgupta, P. (2012, December). System on Biochips: A new design for integration of multiple DMFBs. In *2012 International Symposium on Electronic System Design (ISED)* (pp. 256-260). IEEE.

273. Jokerst, J. V., & McDevitt, J. T. (2010). Programmable nano-bio-chips: multifunctional clinical tools for use at the point-of-care. *Nanomedicine*, *5*(1), 143-155.

274. Zhou, X., Gao, M., & Gui, L. (2017). A Liquid-Metal Based Spiral Magnetohydrodynamic Micropump. *Micromachines*, *8*(12), 365.

275. Peng, C., Zhang, Z., & Ju, Y. S. (2014). EWOD (electrowetting on dielectric) digital microfluidics powered by finger actuation. *Lab on a Chip*, *14*(6), 1117-1122.

276. Wang, H., Chen, L., & Sun, L. (2017). Digital microfluidics: A promising technique for biochemical applications. *Frontiers of Mechanical Engineering*, *12*(4), 510-525.

277. Cui, W., Zhang, M., Duan, X., Pang, W., Zhang, D., & Zhang, H. (2015). Dynamics of electrowetting droplet motion in digital microfluidics systems: from dynamic saturation to device physics. *Micromachines*, *6*(6), 778-789.

278. Laben, R. C. (1963). Factors responsible for variation in milk composition. *Journal of Dairy Science*, *46*(11), 1293-1301.

279. Meurant, G. (1995). *Handbook of milk composition*. Elsevier.

280. Handford, C. E., Campbell, K., & Elliott, C. T. (2016). Impacts of milk fraud on food safety and nutrition with special emphasis on developing countries. *Comprehensive Reviews in Food Science and Food Safety*, *15*(1), 130-142.

281. Poonia, A., Jha, A., Sharma, R., Singh, H. B., Rai, A. K., & Sharma, N. (2017). Detection of adulteration in milk: A review. *International journal of dairy technology*, *70*(1), 23-42.

282. Ma, Y., Dong, W., Bao, H., Fang, Y., & Fan, C. (2017). Simultaneous determination of urea and melamine in milk powder by nonlinear chemical fingerprint technique. *Food chemistry*, *221*, 898-906.

283. Azad, T., & Ahmed, S. (2016). Common milk adulteration and their detection techniques. *International Journal of Food Contamination*, *3*(1), 22.

284. Lipp, M. (1995). Review of methods for the analysis of triglycerides in milk fat: Application for studies of milk quality and adulteration. *Food Chemistry*, *54*(2), 213-221.

285. Singh, P., & Gandhi, N. (2015). Milk preservatives and adulterants: processing, regulatory and safety issues. *Food Reviews International*, *31*(3), 236-261.

286. Cohen, S. M. (2018). Crystalluria and Chronic Kidney Disease. *Toxicologic pathology*, *46*(8), 949-955.

287. Kasemsumran, S., Thanapase, W., & Kiatsoonthon, A. (2007). Feasibility of near-infrared spectroscopy to detect and to quantify adulterants in cow milk. *Analytical Sciences*, *23*(7), 907-910.

288. Jawaid, S., Talpur, F. N., Afridi, H. I., Nizamani, S. M., Khaskheli, A. A., & Naz, S. (2014). Quick determination of melamine in infant powder and liquid milk by Fourier transform infrared spectroscopy. *Analytical Methods*, *6*(14), 5269-5273.

289. Zhang, X. F., Zou, M. Q., Qi, X. H., Liu, F., Zhu, X. H., & Zhao, B. H. (2010). Detection of melamine in liquid milk using surface-enhanced Raman scattering spectroscopy. *Journal of Raman Spectroscopy*, *41*(12), 1655-1660.

290. Santos, P. M., Pereira-Filho, E. R., & Colnago, L. A. (2016). Detection and quantification of milk adulteration using time domain nuclear magnetic resonance (TD-NMR). *Microchemical Journal*, *124*, 15-19.

291. Sharma, R., Rajput, Y. S., Dogra, G., & TOMAR, S. K. (2009). Estimation of sugars in milk by HPLC and its application in detection of adulteration of milk with soymilk. *International journal of dairy technology*, *62*(4), 514-519.

292. Motta, T. C., Hoff, R. B., Barreto, F., Andrade, R. B. S., Lorenzini, D. M., Meneghini, L. Z., & Pizzolato, T. M. (2014). Detection and confirmation of milk adulteration with cheese whey using proteomic-like sample preparation and liquid chromatography–electrospray–tandem mass spectrometry analysis. *Talanta*, *120*, 498-505.

293. Gutiérrez, R., Vega, S., Díaz, G., Sánchez, J., Coronado, M., Ramírez, A., ... & Schettino, B. (2009). Detection of non-milk fat in milk fat by gas chromatography and linear discriminant analysis. *Journal of dairy science*, *92*(5), 1846-1855.

294. Das, C., Chakraborty, S., Karmakar, A., & Chattopadhyay, S. (2018, March). On-chip detection and quantification of soap as an adulterant in milk employing electrical impedance spectroscopy. In *2018 International Symposium on Devices, Circuits and Systems (ISDCS)* (pp. 1-4). IEEE.

295. Jossinet, J., & Schmitt, M. (1999). A review of parameters for the bioelectrical characterization of breast tissue. *Annals of the New York academy of sciences*, *873*(1), 30-41.

296. El Hasni, A., Schmitz, C., Bui-Göbbels, K., Bräunig, P., Jahnen-Dechent, W., & Schnakenberg, U. (2017). Electrical impedance spectroscopy of single cells in hydrodynamic traps. *Sensors and Actuators B: Chemical*, *248*, 419-429.

297. Gomila, G et al. (2006). Advances in the production, immobilization, and electrical characterization of olfactory receptors for olfactory nanobiosensor development. *Sensors and Actuators B: Chemical*, *116*(1-2), 66-71.

298. Mukherjee, A., Chakraborty, S., Das, C., Karmakar, A., & Chattopadhyay, S. (2019, March). Study of Optical and Electrical Characteristics of chemically extracted Lotus and Taro Bio-Wax for Hydrophobic Surface Engineering. In *2019 International Conference on Opto-Electronics and Applied Optics (Optronix)* (pp. 1-4). IEEE.

299. Chakraborty, S et al. (2018, March). Bio-dielectric variation as a signature of shape alteration and lysis of human erythrocytes: An on-chip analysis. In *2018 International Symposium on Devices, Circuits and Systems (ISDCS)* (pp. 1-4). IEEE.

300. Åberg, P., Geladi, P., Nicander, I., Hansson, J., Holmgren, U., & Ollmar, S. (2005). Non-invasive and microinvasive electrical impedance spectra of skin cancer–a comparison between two techniques. *Skin research and technology*, *11*(4), 281-286.

301. Casas, O et al. (1999). In Vivo and In Situ Ischemic Tissue Characterization Using Electrical Impedance Spectroscopy a. *Annals of the New York Academy of Sciences*, *873*(1), 51-58.

302. Seoane, F., Ferreira, J., Sanchez, J. J., & Bragós, R. (2008). An analog front-end enables electrical impedance spectroscopy system on-chip for biomedical applications. *Physiological measurement*, *29*(6), S267.

303. Yun, J., Hong, Y. T., Hong, K. H., & Lee, J. H. (2018). Ex vivo identification of thyroid cancer tissue using electrical impedance spectroscopy on a needle. *Sensors and Actuators B: Chemical*, *261*, 537-544.

304. Van Der Wal, P., & Steiner, U. (2007). Super-hydrophobic surfaces made from Teflon. *Soft Matter*, *3*(4), 426-429.

305. Yang, M. K., French, R. H., & Tokarsky, E. W. (2008). Optical properties of Teflon® AF amorphous fluoropolymers. *Journal of Micro/Nanolithography, MEMS, and MOEMS*, *7*(3), 033010.

306. Sanjeev, R., Jagannadham, V., & Vrath, R. V. (2012). Dipole moments and melting points and their unsolved miracles on the application of hammett equation. *Education Journal*, *1*(1), 1-4.

307. Parungo, F. P., & Lodge Jr, J. P. (1965). Molecular structure and ice nucleation of some organics. *Journal of the Atmospheric Sciences*, *22*(3), 309-313.